\documentclass[useAMS,usenatbib]{mn2e}

\usepackage{color}
\usepackage{colortbl}
\usepackage{multirow}
\usepackage{dcolumn}
\usepackage{amsmath}
\usepackage{amssymb}
\usepackage{graphicx}
\usepackage{epsfig}
\usepackage{changebar}
\usepackage{natbib}
\usepackage{multirow}
\usepackage{subfigure}
\usepackage{txfonts}
\usepackage{rotating}

\usepackage{longtable,lscape}
\usepackage{pdflscape}

\newcolumntype{d}[1]{D{.}{\cdot}{#1}}

\newcolumntype{.}{D{.}{.}{-1}}

\newcommand{\lsun}{L$_\odot$}
\newcommand{\msun}{M$_\odot$}

\newcommand{\mum}{$\umu$m}

\newcommand{\kms}{km\,s$^{-1}$}

\newcommand{\hi}{H~{\sc i}}
\newcommand{\hii}{H~{\sc ii}}
\newcommand{\uchii}{UC\,H~{\sc ii}}
\newcommand{\hchii}{HC\,H~{\sc ii}}

\newcommand{\sex}{\texttt{SExtractor}}
\newcommand{\rms}{r.m.s.}
\newcommand{\submm}{submillimetre}

\newcommand{\uchiinum}{213}

\title[ATLASGAL-CORNISH \uchii\ regions]{ATLASGAL --- properties of compact \hii\ regions and their natal clumps\thanks{The full version of Tables\,3 and 4 are only available in electronic form at the CDS via anonymous ftp to cdsarc.u-strasbg.fr (130.79.125.5) or via http://cdsweb.u-strasbg.fr/cgi-bin/qcat?J/MNRAS/, while Figs\,11 and 12 are only available in the electronic version of the journal.}}
\author[J. S. Urquhart et al.]{J.\,S.\,Urquhart$^{1}$\thanks{E-mail:
jurquhart@mpifr-bonn.mpg.de (MPIfR)}, M.\,A.\,Thompson$^{2}$,  T.\,J.\,T.\,Moore$^{3}$, C.\,R.\,Purcell${^4}$, M.\,G.\,Hoare${^4}$,  F.\,Schuller$^{5}$, \newauthor F.\,Wyrowski$^{1}$, T.\,Csengeri$^{1}$, K.\,M.\,Menten$^{1}$, S.\,L.\,Lumsden${^4}$, S.\,Kurtz${^6}$, C.\,M.\,Walmsley$^{7, 8}$, \newauthor L.\,Bronfman$^{9}$,  L.\,K.\,Morgan$^{3}$, D.\,J.\,Eden$^{3}$, D.\,Russeil$^{10}$\\ \\
$^{1}$
 Max-Planck-Institut f\"ur Radioastronomie, Auf dem H\"ugel
  69, D-53121 Bonn, Germany \\
  $^{2}$ Centre for Astrophysics Research, Science and Technology Research Institute, University of Hertfordshire, College Lane, Hatfield, AL10 9AB, UK \\
 $^{3}$Astrophysics Research Institute, Liverpool John Moores University, 146 Brownlow Hill, Liverpool, L3\,5RF, UK\\ 
  $^{4}$School of Physics and Astrophysics, University of Leeds, Leeds, LS2\,9JT, UK \\
  $^{5}$European Southern Observatory, Alonso de Cordova 3107, Vitacura, Santiago, Chile\\
    $^{6}$Centro de Radioastronomia y Astrofisica, Universidad Nacional Aut\'onoma de M\'exico, Antigua Carretera a Pátzcuaro \# 8701 Morelia, 58089 Michoac\'an, M\'exico\\
  $^{7}$Osservatorio Astrofisico di Arcetri, Largo E. Fermi, 5, 50125 Firenze, Italy\\
$^{8}$Dublin Institute for Advanced Studies, Burlington Road 10, Dublin 4, Ireland\\
$^{9}$Departamento de Astronom\'{i}a, Universidad de Chile, Casilla 36-D, Santiago, Chile\\
$^{10}$Aix Marseille Universit\'e, CNRS, LAM (Laboratoire d'Astrophysique de Marseille) UMR 7326, 13388, Marseille, France}

\begin{document}

\date{Accepted ??. Received ??; in original form ??}

\pagerange{\pageref{firstpage}--\pageref{lastpage}} \pubyear{2009}

\maketitle

\label{firstpage}

\begin{abstract}

We present a complete sample of molecular clumps containing compact and ultra-compact (UC) \hii\ regions between $\ell=10\degr$ and 60\degr\ and $|b|<1$\degr, identified by combining the the ATLASGAL sub-mm and CORNISH radio continuum surveys with visual examination of archival infrared data. Our sample is complete to optically thin, compact and \uchii\ regions driven by a zero age main sequence star of spectral type B0 or earlier embedded within a 1,000\,\msun\ clump. In total we identify \uchiinum\ compact and \uchii\ regions, associated with 170 clumps. Unambiguous kinematic distances are derived for these clumps and used to estimate their masses and physical sizes, as well as the Lyman continuum fluxes and sizes of their embedded \hii\ regions. We find a clear lower envelope for the surface density of molecular clumps hosting massive star formation of 0.05 g\,cm$^{-2}$, which is consistent with a similar sample of clumps associated with 6.7\,GHz masers. The mass of the most massive embedded stars is closely correlated with the mass of their natal clump. Young B stars appear to be significantly more luminous in the ultraviolet than predicted by current stellar atmosphere models. The properties of clumps associated with compact and \uchii\ regions are very similar to those associated with 6.7\,GHz methanol masers and we speculate that there is little evolution in the structure of the molecular clumps between these two phases. Finally, we identify a significant peak in the surface density of compact and \uchii\ regions associated with the W49A star-forming complex, noting that this complex is truly one of the most massive and intense regions of star formation in the Galaxy.

\end{abstract}
\begin{keywords}
Stars: formation -- Stars: early-type -- ISM: clouds -- ISM: radio continuum -- ISM: \hii\ regions -- Galaxy: structure.
\end{keywords}

\section{Introduction}
\label{sect:intro}
Ultracompact \hii\  (hereafter \uchii) regions represent one of the earliest stages in the development of an \hii\ region, where a massive star (or stars) has begun to ionise the surrounding gas and produce a small (diameter, $D$, $<$ 0.1\,pc) photoionised bubble of gas embedded within a dense molecular cloud clump. \uchii\ are particularly important tracers of \emph{recent} massive star formation that are readily detectable across most of the Milky Way via their centimetre-wavelength free-free emission or their strong far-infrared emission \citep[e.g.][]{wood1989b, becker1990, giveon2005a, hoare2012}. As such, \uchii\ regions give us a reliable snapshot of  massive star formation activity within the last $\sim$10$^{5}$ years (e.g., \citealt{davies2011,mottram2011b}), enabling a census of massive star formation on the scale of individual complexes throughout the entire Galaxy. \uchii\ regions are a particularly important signpost of the dense conditions within molecular clouds that give rise to massive star formation, and also as a direct tracer of the luminosity of the recently formed stars via their free-free emission. 

Most \uchii\ regions have been identified by targeted radio interferometry of objects selected by their mid- or far-infrared colours \citep[e.g.,][]{wood1989b, kurtz1994, walsh1998, urquhart_radio_south,urquhart2009}.  The drawback of such targeted studies is that their infrared selection techniques can be contaminated by other IR-bright objects such as planetary nebulae and intermediate mass YSOs \citep{ramesh1997}, introducing biases into conclusions based on larger samples drawn solely from  infrared catalogues. Wide area radio interferometric surveys remove this constraint and have also been utilised to identify \uchii\ regions (\citealt{becker1990, zoonematkermani1990, becker1994, giveon2005a, murphy2010, hindson2012}). Most of these surveys were carried out in snapshot mode, which limits the  $uv$-coverage of the observations. It is thus difficult to unambiguously identify \uchii\ regions due to the limited range of angular scales traced by snapshot observations;  ultracompact sources detected in snapshot surveys may simply be bright and compact components of more extended emission \citep{kurtz1999, kim2001, ellingsen2005}.

Another compounding problem in the identification of \uchii\ regions is the fact that at their typical 5\,GHz flux densities (a few mJy to tens mJy) the source counts are dominated by background radio galaxies \citep{anglada1998}. In order to conclusively identify \uchii\ regions one must also show that the ultracompact radio sources  have spectra consistent with thermal free-free continuum, are clearly associated with thermal infrared emission from hot dust heated within the ionized nebula or polycyclic aromatic hydrocarbon (PAH) emission from the photo-dominated region (PDR)
surrounding the nebula, or show signs of being embedded within molecular cloud clumps.  Existing studies have mainly used the IRAS or MSX surveys to fulfil the infrared criterion \citep{walsh1998, giveon2005a}. The main drawback with these surveys is their low angular resolution compared to the $\sim$ arcsecond resolution of the radio images, which makes it difficult to unambiguously determine the correct infrared counterpart. While the angular resolution of the ground based mm and \submm\ observations used to trace the star forming molecular clumps is higher, wide-area surveys have not come into being until relatively recently. This means that the number of \uchii\ hosting clumps studied is limited by targeted studies to typically a few tens to a hundred objects \citep[see e.g.,][]{thompson2006}. 

We take advantage of three recent Galactic plane surveys to address these issues and compile a robust well-selected sample of molecular clumps associated with deeply embedded \hii\ regions. The surveys are the CORNISH 5\,GHz survey \citep{hoare2012,purcell2013}, the GLIMPSE mid-infrared survey \citep{benjamin2003_ori}, and the ATLASGAL 870 $\umu$m survey \citep{schuller2009}. Each of these surveys covers a common field (10\degr\ $\le \ell \le$ 60\degr, $|b| \le$ 1\degr), with matched angular resolution between the  5\,GHz radio and GLIMPSE mid-infrared data sets ($\sim$2\arcsec). By combining these three surveys we are able to use multi-wavelength identification techniques to identify \hii\ regions and confirm their embedded nature via the coincidence of radio and mid-infrared emission \citep[e.g.][]{hoare2007} and \submm\ calorimetry \citep{thompson2006}.  Another concern is the possibility of comtamination from stellar winds of massive young stellar objects (MYSOs). However, the radio continuum emission arising from these winds is relatively weak ($S_\nu<$ few \,mJy at 1\,kpc; e.g., \citealt{tonfani1995,hoare2002}) and it is therefore unlikely that CORNISH is sensitive enough for this to be an issue.  We also determine unambiguous kinematic distances for the majority of our sample. This results in a complete catalogue of molecular clumps with embedded  \hii\ regions over the common GLIMPSE, CORNISH \& ATLASGAL survey region. This sample will include a mixture of ultracompact ($D\lesssim$0.1\,pc and particle densities, $n$, $\gtrsim$ 10$^4$\,cm$^{-3}$) and compact ($D\lesssim$0.5\,pc and $n\gtrsim$ $5\times10^3$\,cm$^{-3}$) \hii\ regions (\citealt{kurtz2005a}).

This catalogue is used to investigate the physical nature of the \uchii\ region hosting clumps, the mass-size relationship for massive star-forming clumps \citep{kauffmann2010c}, the Galactic distribution of \uchii\ regions and their statistical lifetimes. Finally we explore correlations between the \uchii\ region properties and those of their embedding clumps in order to give insights into the accretion history and star formation efficiency.

We describe the individual surveys used in this study in Sect.\,2, the multi-wavelength identification techniques and observed properties of the \uchii\ hosting clumps in Sect.\,3, and their kinematic properties in Sect.\,4. Sect.\,5 details the derived kinematic distances and the \hi\ self absorption techniques that we have used to break the near-far distance ambiguity. In Sect.\,6 and 7 we derive the physical properties of the molecular clumps and \uchii\ regions, respectively. We discuss these properties in the context of current star formation thresholds and with the properties of methanol maser associated clumps discussed in \citet[][hereafter Paper\,I]{urquhart2013}. Finally, in Sect.\,9 we summarise our findings and discuss their implications.  

This paper is the second in a series of three papers focusing on the properties of molecular clumps from the ATLASGAL survey that are associated with massive star formation. The first paper in the series investigates the properties of clumps associated with 6.7\,GHz methanol masers (Paper\,I) identified by the Methanol Multibeam (MMB; \citealt{caswell2010b}) survey. In this paper we deal with \uchii\ region hosting clumps, and the third will investigate the properties of clumps associated with a sample of mid-infrared-selected massive YSOs (Urquhart et al 2013, in prep). We also note that there are two forthcoming CORNISH papers focusing on the radio-derived parameters of the \uchii\ regions (Purcell et al., in prep.) and their morphologies (Hoare et al., in prep.).

\section{Survey descriptions}

\subsection{ATLASGAL Survey}

The APEX Telescope Large Area Survey of the Galaxy (ATLASGAL; \citealt{schuller2009}) is the first systematic survey of the inner Galactic plane in the \submm\ band. The survey was carried out with the Large APEX Bolometer Camera (LABOCA; \citealt{siringo2009}), an array of 295 bolometers observing at 870\,$\umu$m (345\,GHz). At this wavelength the APEX Telescope has a full-width at half-maximum (FWHM) beam size of 19.2\arcsec. The initial survey region covered a Galactic longitude region of $300\degr < \ell < 60\degr$ and $|b| < 1.5\degr$, but this was extended to include $280\degr < \ell < 300\degr$, however, the latitude range was shifted to $-2\degr < b < 1\degr$ to take account of the Galactic warp in this region of the plane.

\citet{contreras2013} produced a compact source catalogue for the central part of the survey region (i.e., 330\degr\ $ <\ell <$ 21\degr) using the source extraction algorithm \sex\ (\citealt{bertin1996}). This catalogue consists of 6,774 sources and is 99\,per\,cent complete at $\sim$6$\sigma$, which corresponds to a flux sensitivity of 0.3-0.4\,Jy\,beam$^{-1}$. We have used the same source extraction algorithm and method described by Contreras et al. (2013) to produce a catalogue for the currently unpublished 280\degr\ $ <\ell <$ 330\degr\ and 21\degr\ $ <\ell <$ 60\degr\ regions of the survey. When the sources identified in these regions are combined with those identified by  \citet{contreras2013} we obtain a final compact source catalogue consisting of $\sim$10,000 sources (the full catalogue will be presented in Csengeri et al. in prep.). The observations typically had an \rms\ pointing accuracy of $\sim$4\arcsec\ (\citealt{schuller2009}), which we adopted as the positional accuracy for the catalogue. This catalogue provides a complete census of dense dust clumps located in the inner Galaxy and includes all potential massive star forming clumps with masses greater than 1,000\,\msun\ out to 20\,kpc (Paper\,I).

\begin{figure*}
\begin{center}

\includegraphics[width=0.33\textwidth, trim= 0 0 0 0]{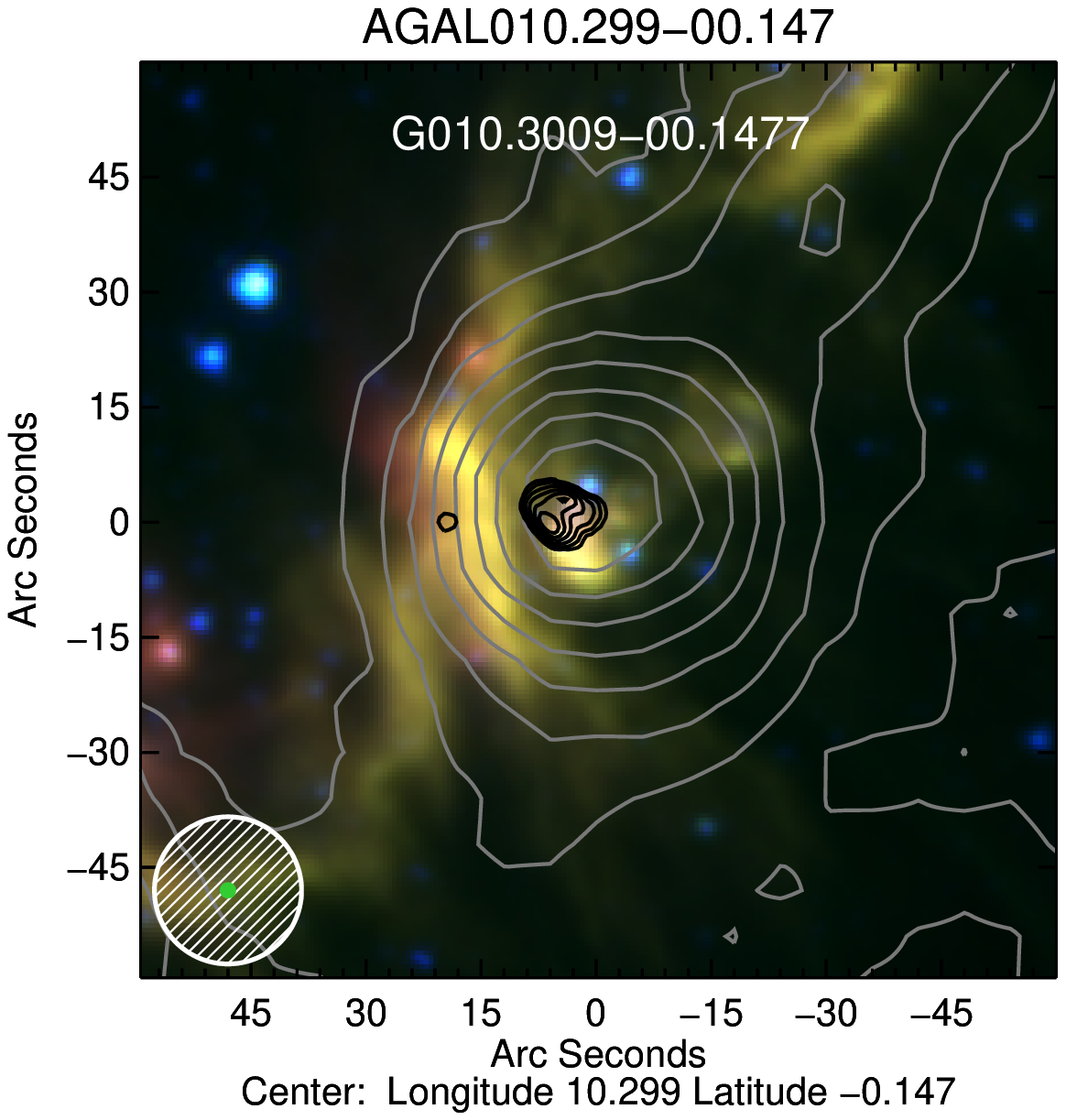}
\includegraphics[width=0.33\textwidth, trim= 0 0 0 0]{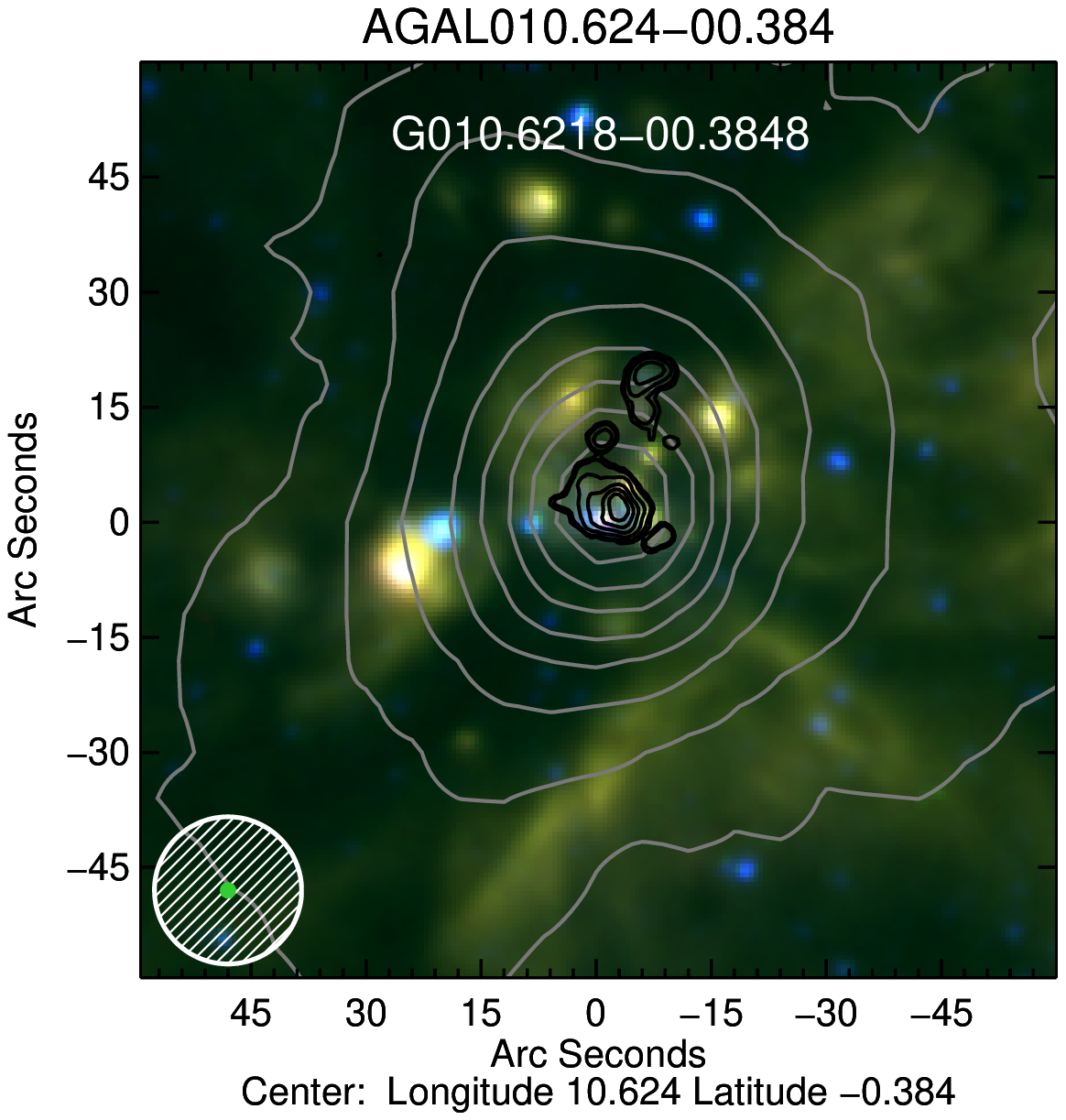}
\includegraphics[width=0.33\textwidth, trim= 0 0 0 0]{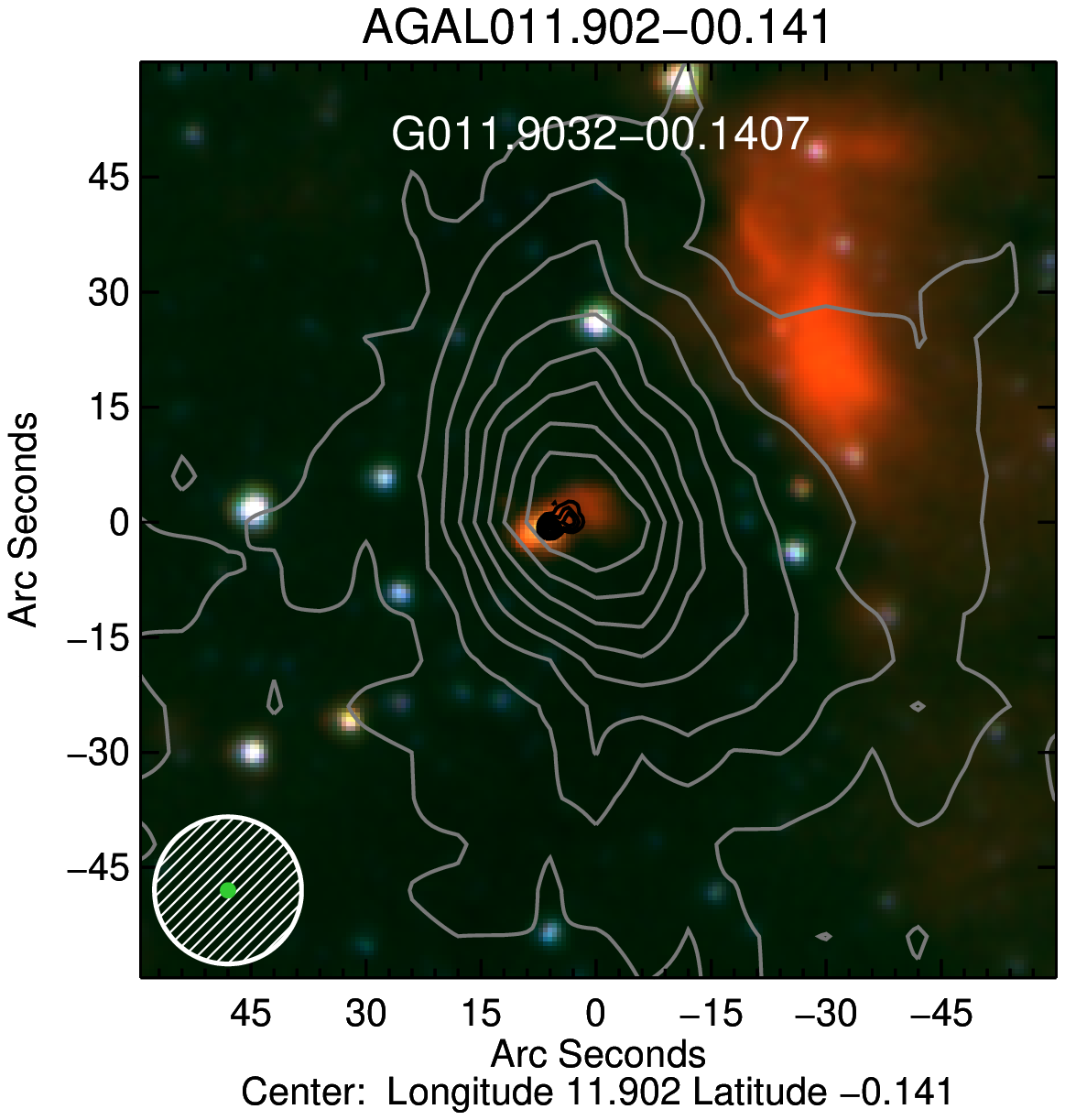}
\includegraphics[width=0.33\textwidth, trim= 0 0 0 0]{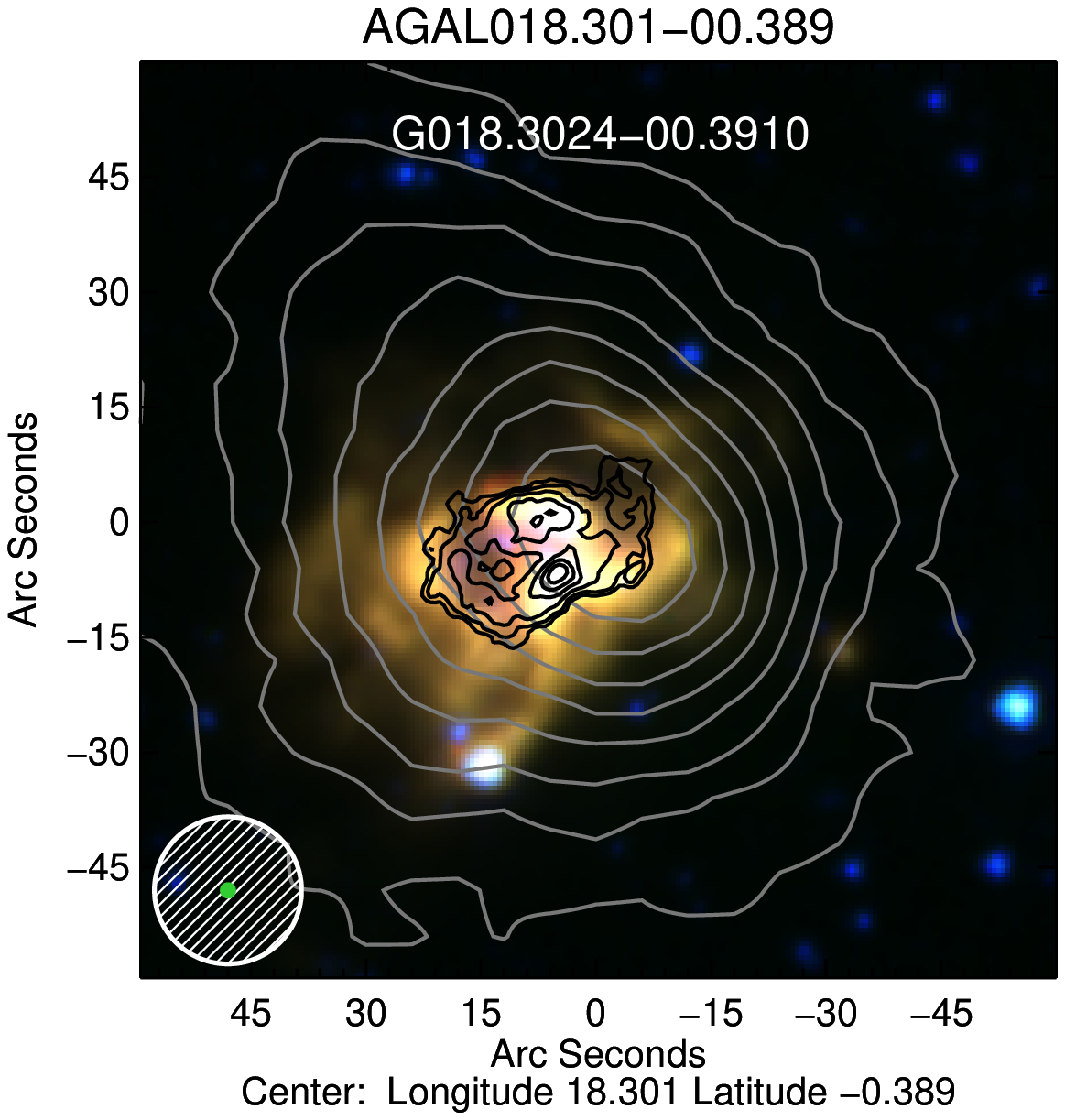}
\includegraphics[width=0.33\textwidth, trim= 0 0 0 0]{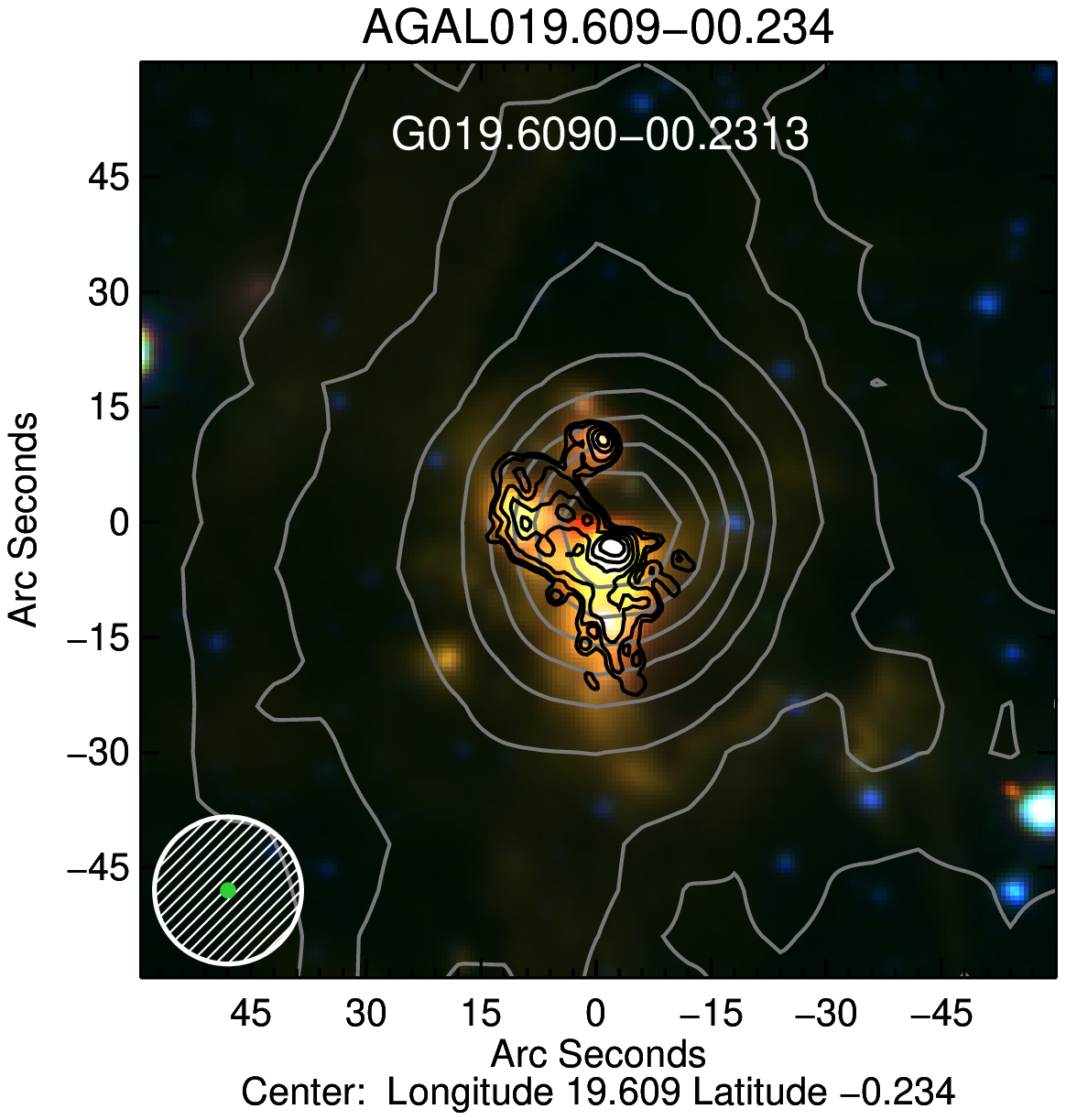}
\includegraphics[width=0.33\textwidth, trim= 0 0 0 0]{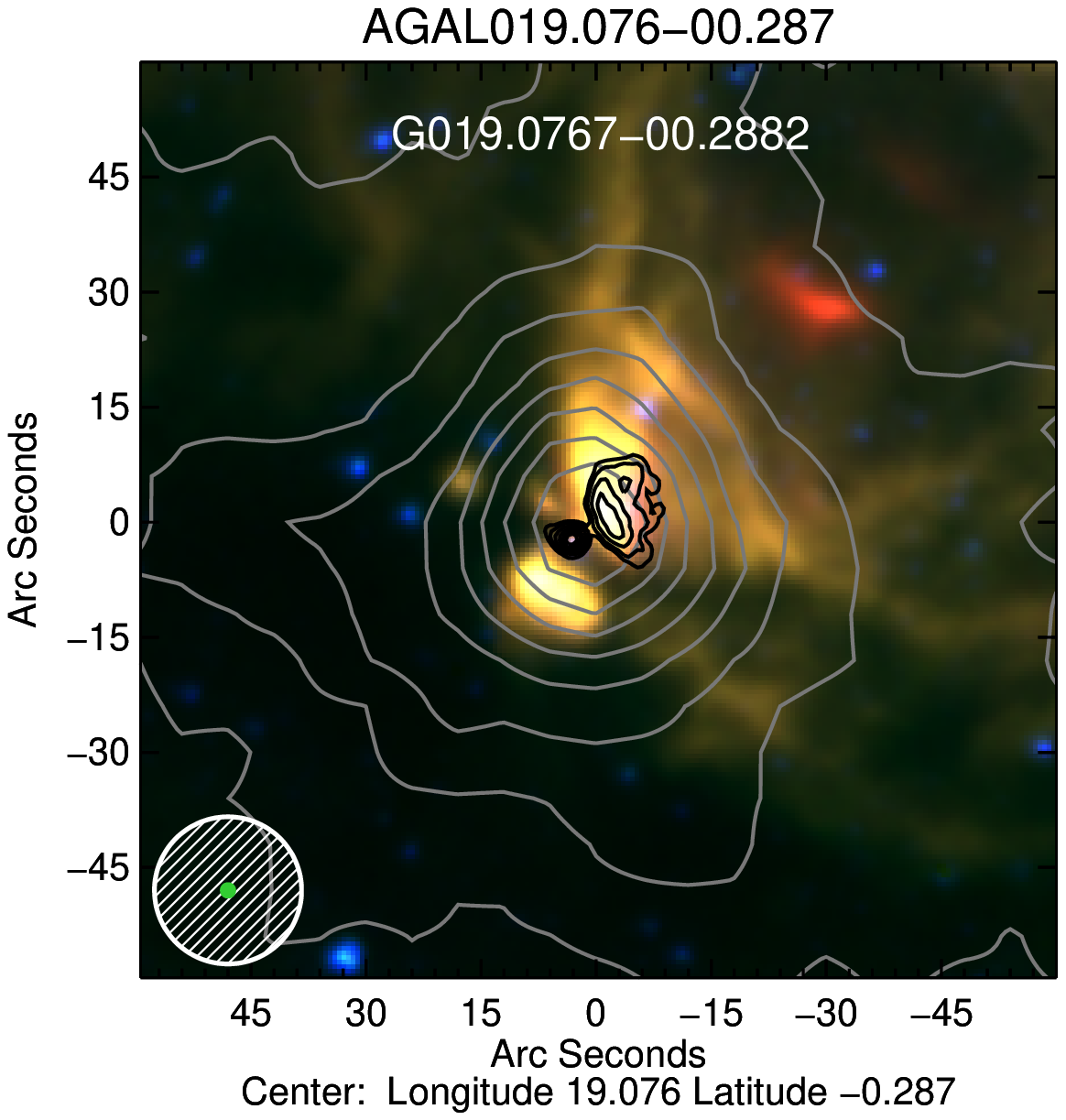}

\caption{\label{fig:irac_images_uchiis}  Examples of the local mid-infrared environments found towards the ATLASGAL-CORNISH identified compact and \uchii\ regions and included in the final catalogue (see text for details). These images are composed of the GLIMPSE 3.6, 4.5 and 8.0\,\mum\ IRAC bands (coloured blue, green and red respectively) which are overlaid with grey contours showing the ATLASGAL 870\,$\umu$m thermal dust emission, and black contours showing the CORNISH 5\,GHz radio continuum emission. The ATLASGAL source name is given above the image, while the CORNISH source name is given in white in the top part of the image. The CORNISH and ATLASGAL survey resolutions are indicated by the green circles and the white hatched circle in the lower left corner of each image, respectively. The contour levels start at 2$\sigma$ and increase in steps set by a dynamically determined power-law of the form $D=3\times N^i+2$, where $D$ is the dynamic range of the \submm\ emission map (defined as the peak brightness divided by the local r.m.s. noise), $N$ is the number of contours used (8 in this case), and $i$ is the contour power-law index. The lowest power-law index used was one, which results in linearly spaced contours starting at 2$\sigma$ and increasing in steps of 3$\sigma$ (see \citealt{thompson2006} for more details). The advantage of this scheme over a linear scheme is its ability to emphasise both emission from diffuse extended structures with low surface brightness and emission from bright compact sources.}

\end{center}
\end{figure*}

\subsection{CORNISH Survey}
\label{sect:cornish_description}

The CORNISH project has mapped the northern Galactic plane between $\ell=10\degr$ and 65\degr\ and $|b|<1$\degr\ for 5\,GHz radio-continuum emission. The survey was designed to detect \uchii\ regions across the Galactic disk (\citealt{hoare2012}) and to form a radio counterpart to recent arcsecond-resolution infrared surveys (e.g., UKIDSS, GLIMPSE, and MIPSGAL). Utilising the Karl Jansky Very Large Array (VLA) in B and BnA array configurations, it is possible to resolve radio emission on angular scales ranging from $\sim$1.5\arcsec\ to 20\arcsec. The r.m.s. noise level in the CORNISH images is better than 0.4\,mJy\,beam$^{-1}$, sufficient to detect free-free emission from an optically thin \hii\ region around a B0 star on the other side of the Galaxy. The high-reliability CORNISH catalogue (\citealt{purcell2013}) contains 2637 distinct sources detected above a 7-sigma intensity cutoff, $\sim$250 of which have been identified as possible \hii\ regions via a visual inspection of infrared images (Purcell et al. 2013, in prep.).

The catalogue is 90\,per\,cent complete for point sources at a flux density of 3.9\,mJy, however, this value decreases with increasing source size as the incomplete \emph{uv}-coverage leads to missing flux in extended objects. \citet{purcell2013} have used the GLIMPSE 8\,$\umu$m data to identify and re-integrate over-resolved radio emission which has been broken up by the source finder.

\section{ATLASGAL-CORNISH associations}
\label{sect:atlas-cornish}

\begin{figure*}
\begin{center}

\includegraphics[width=0.33\textwidth, trim= 0 0 0 0]{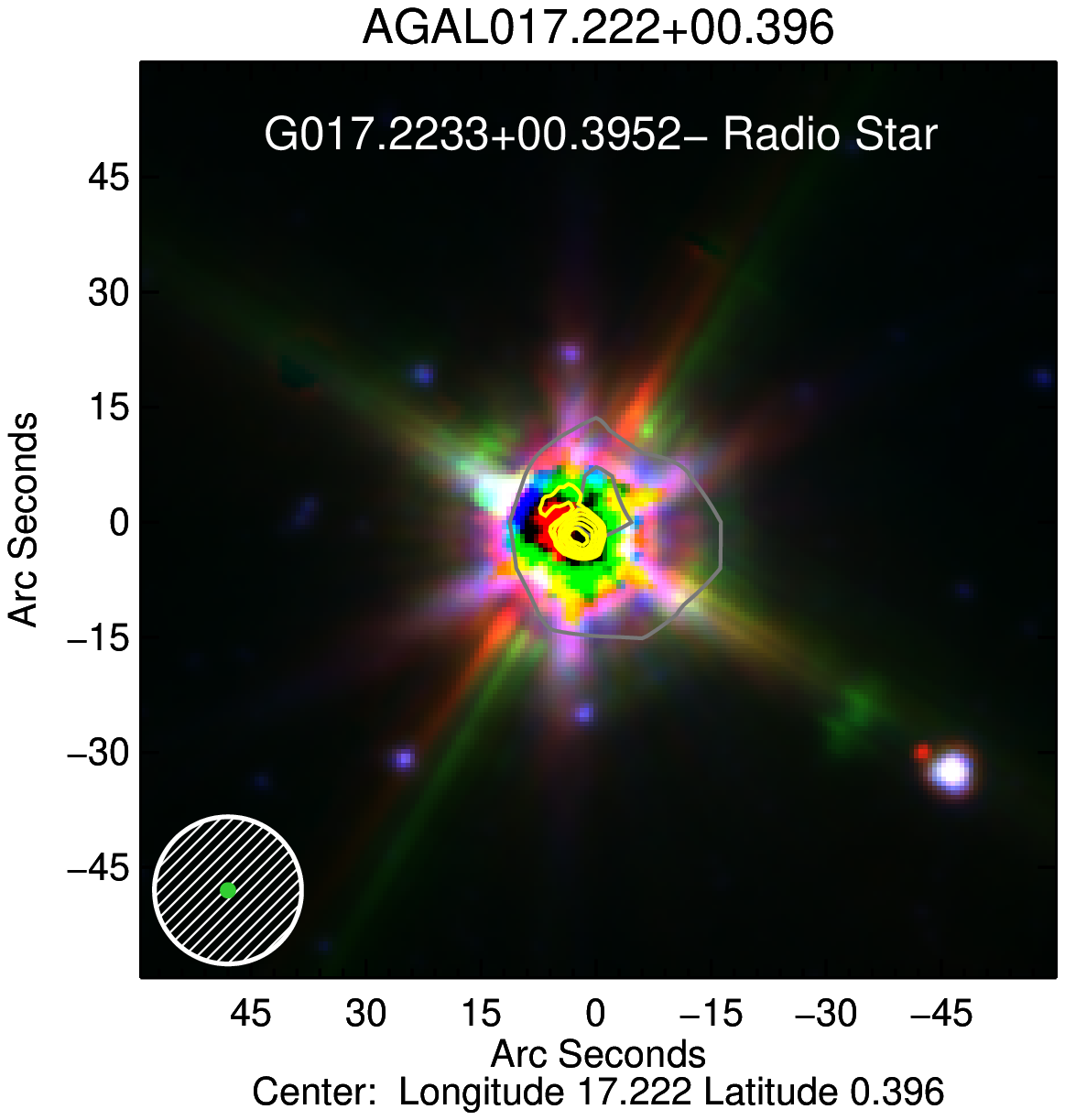}
\includegraphics[width=0.33\textwidth, trim= 0 0 0 0]{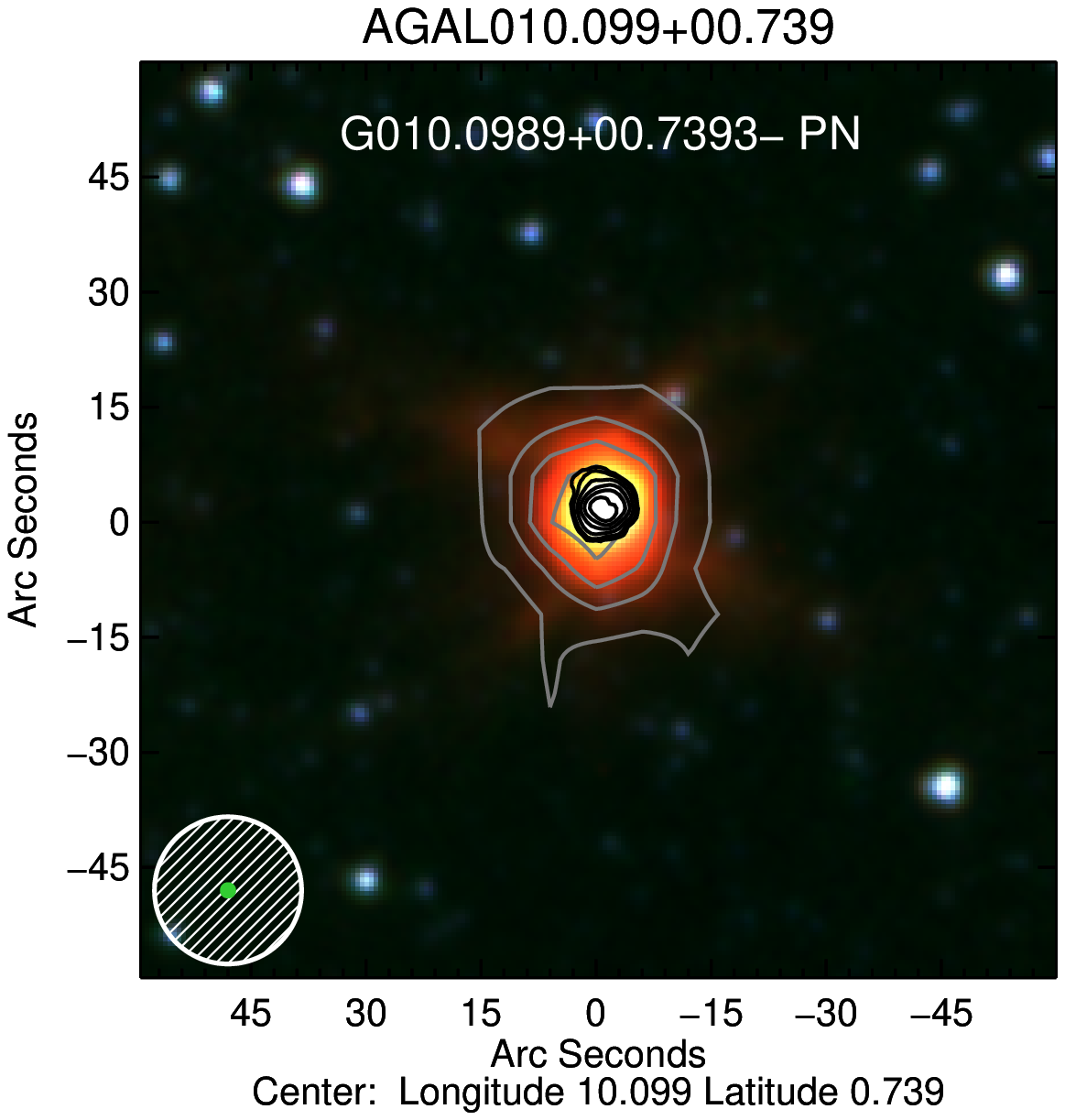}
\includegraphics[width=0.33\textwidth, trim= 0 0 0 0]{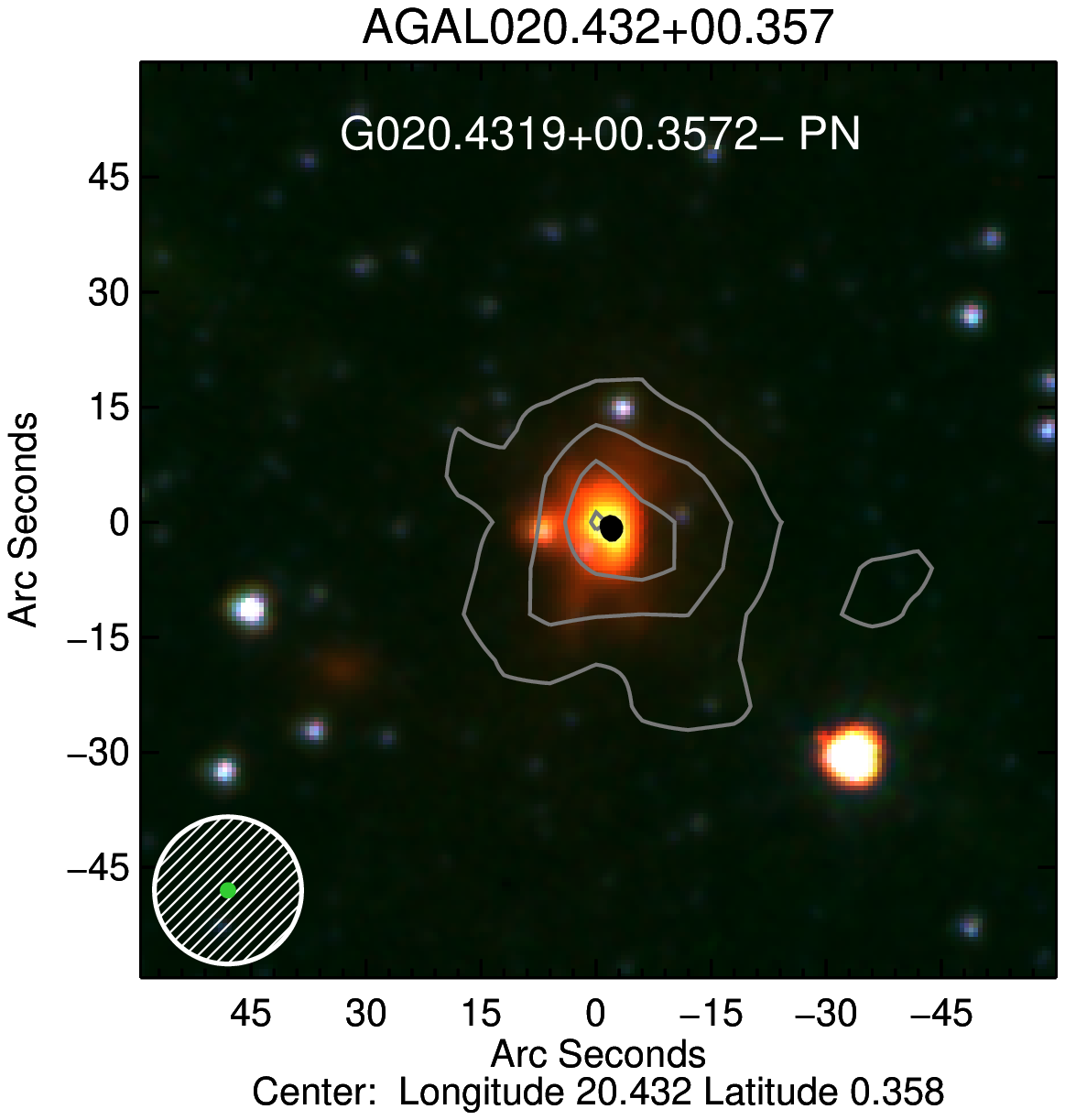}
\\
\includegraphics[width=0.33\textwidth, trim= 0 0 0 0]{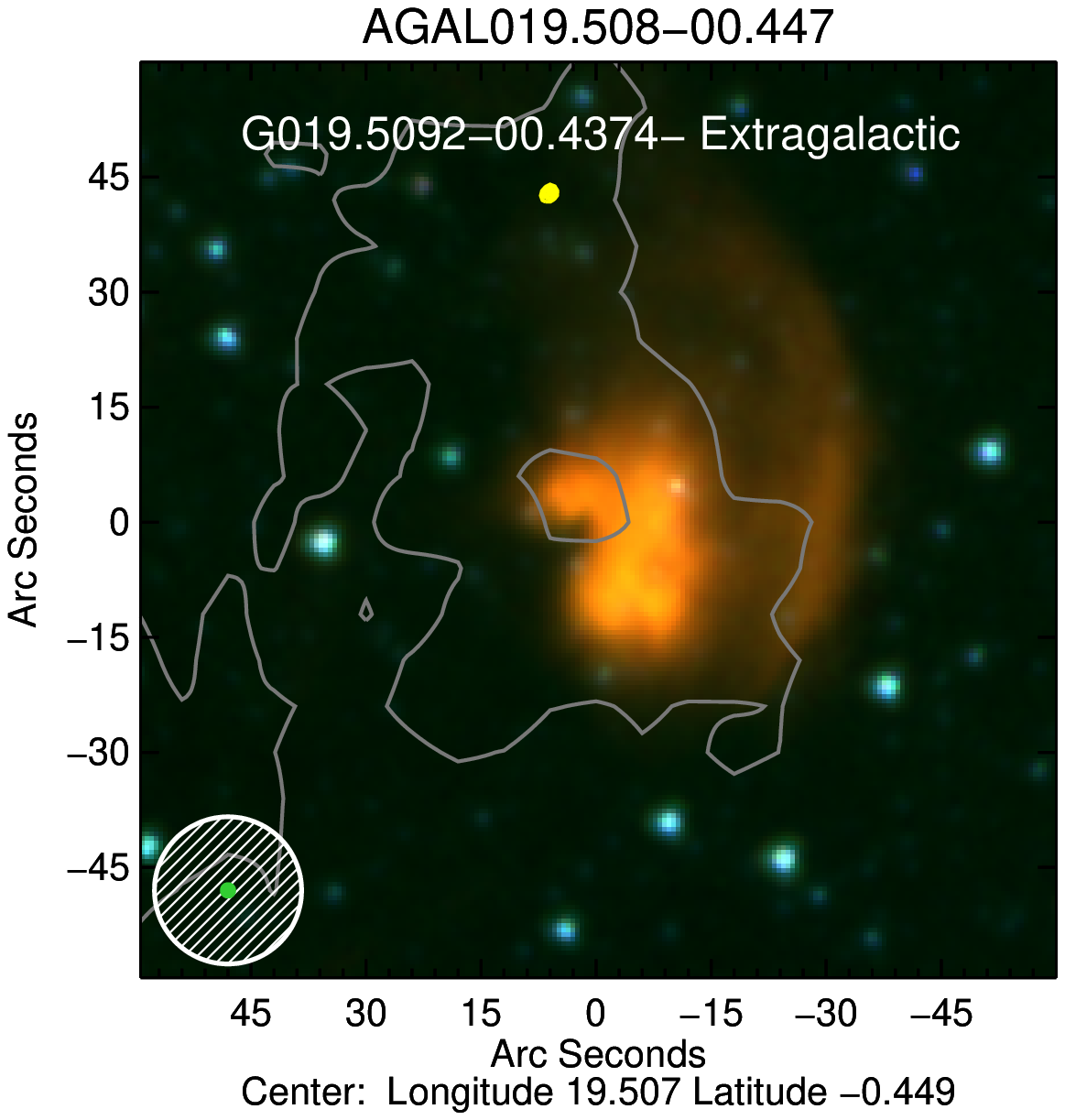}
\includegraphics[width=0.33\textwidth, trim= 0 0 0 0]{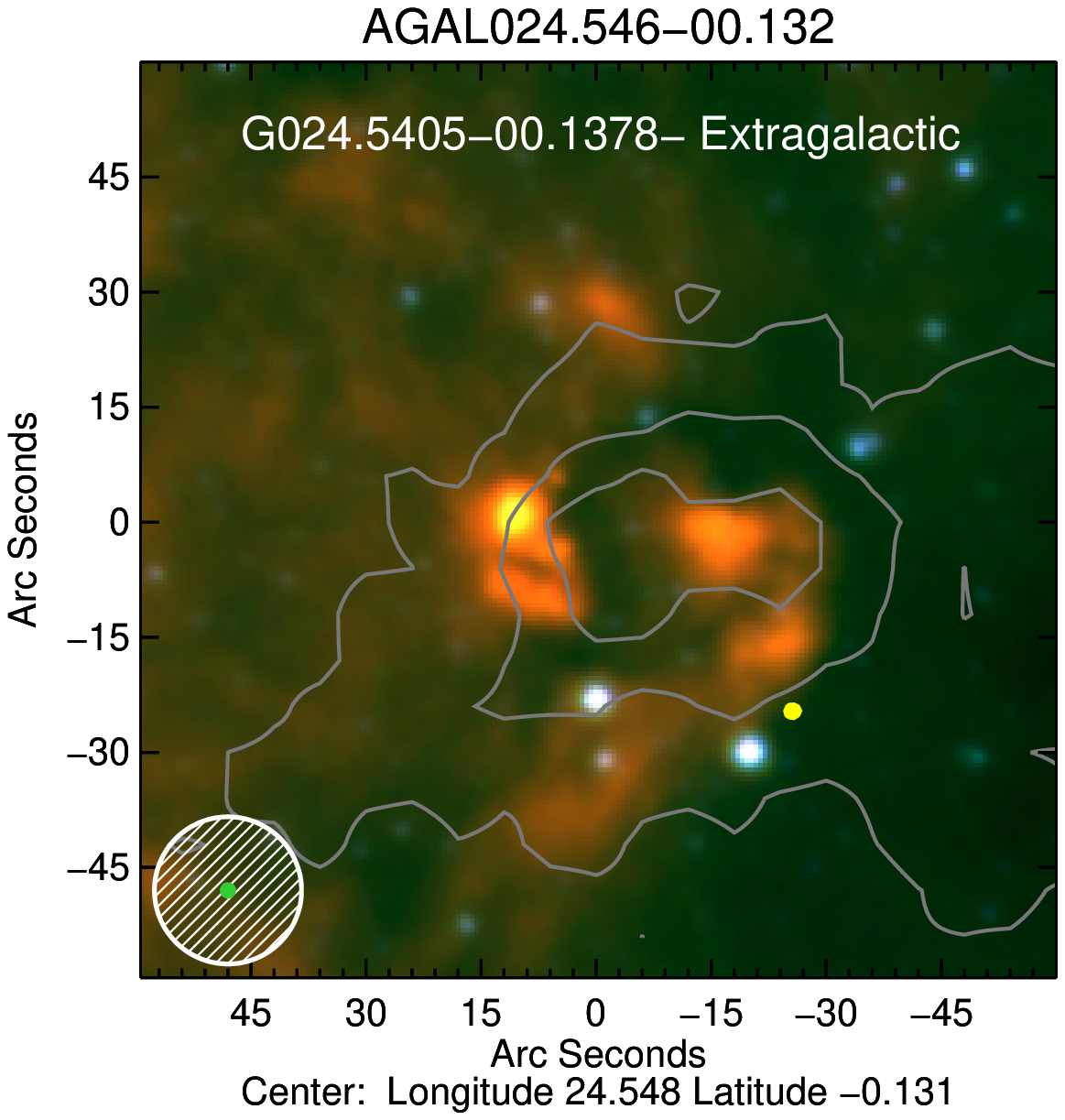}
\includegraphics[width=0.33\textwidth, trim= 0 0 0 0]{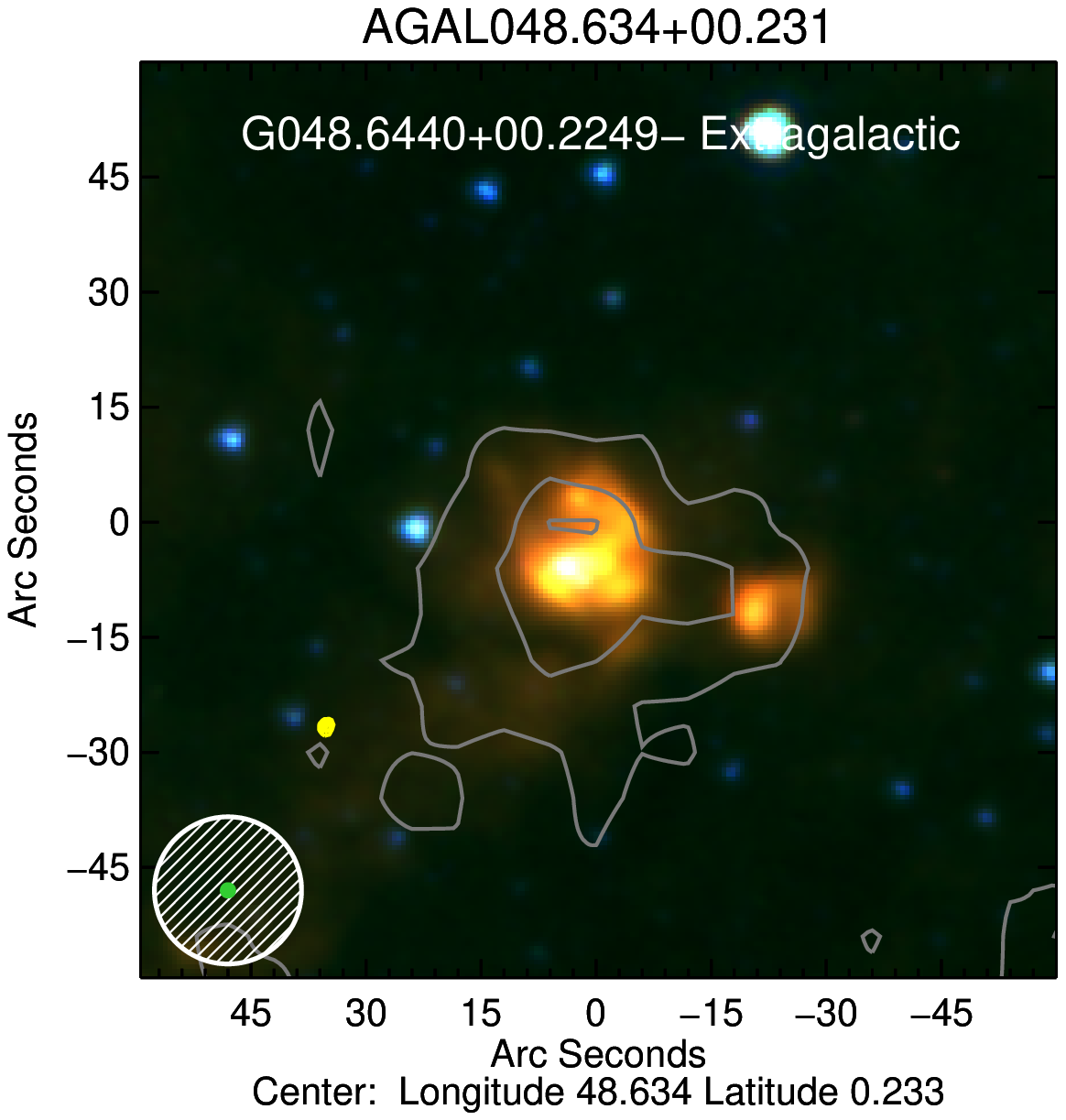}\\
\includegraphics[width=0.33\textwidth, trim= 0 0 0 0]{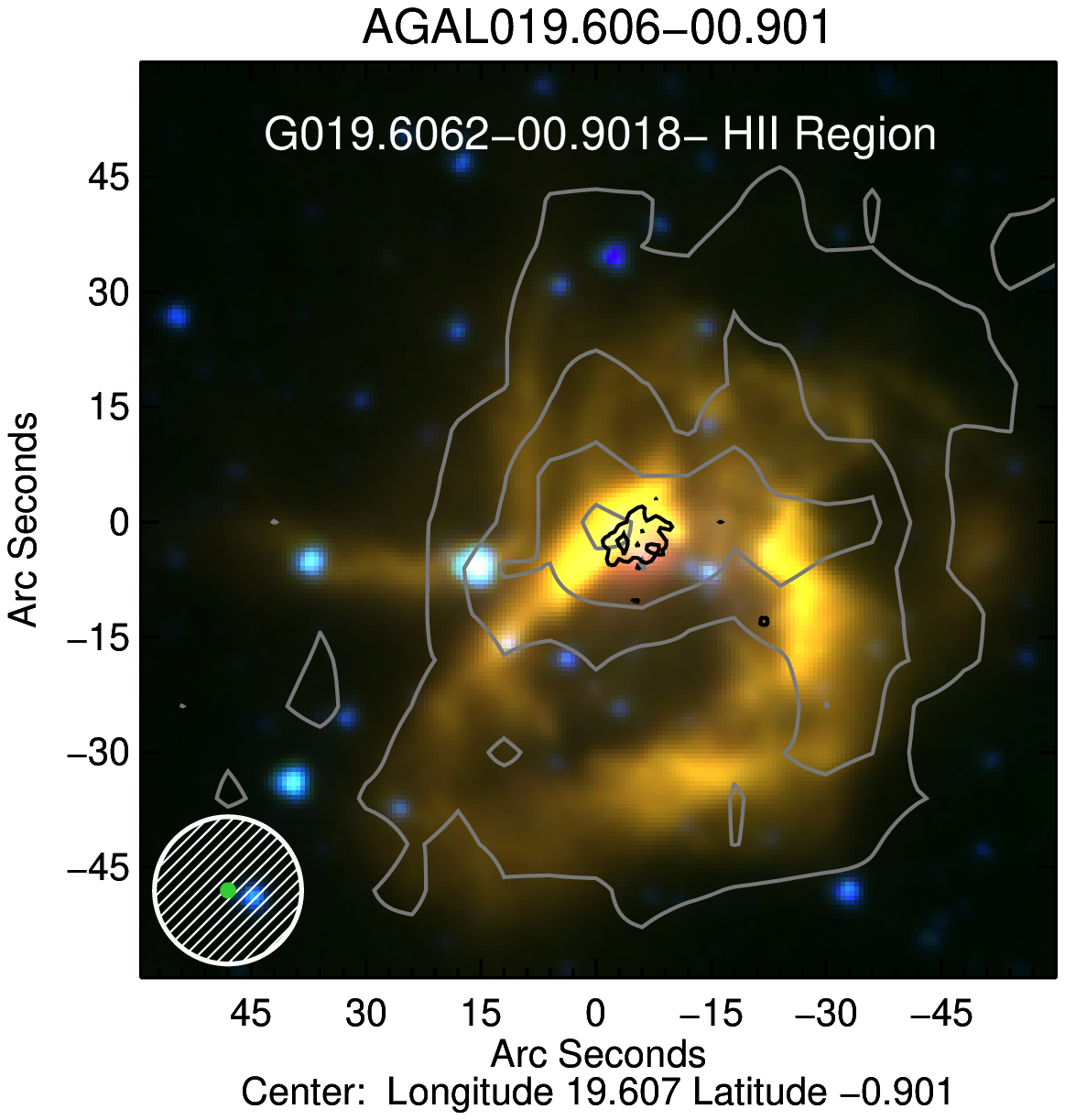}
\includegraphics[width=0.33\textwidth, trim= 0 0 0 0]{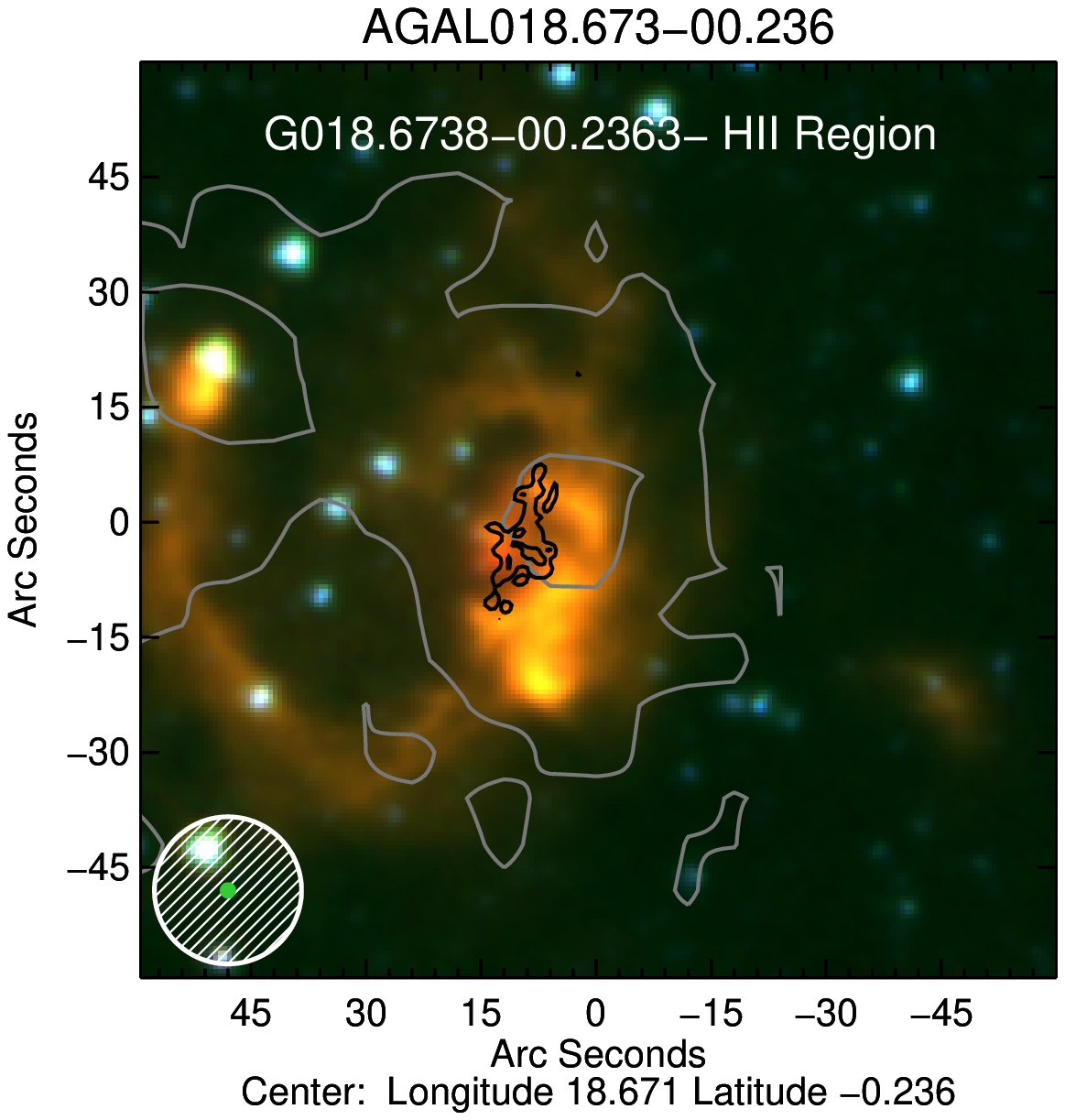}
\includegraphics[width=0.33\textwidth, trim= 0 0 0 0]{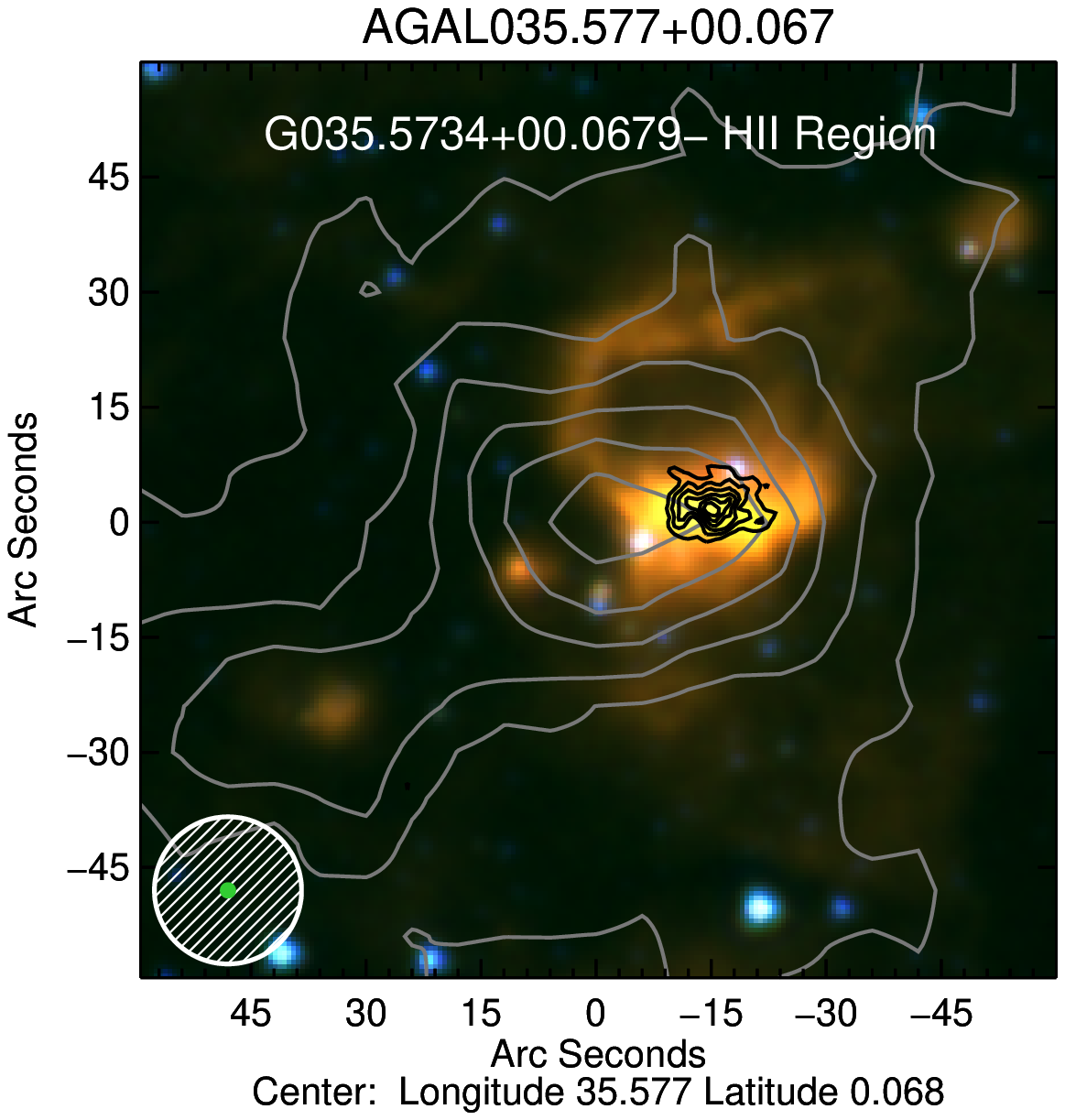}

\caption{\label{fig:irac_images_contaminating_sources}  Examples of the local mid-infrared environments found towards the ATLASGAL-CORNISH matches that have been excluded from the final catalogue (see text for details). The image details are as described in Fig.\,\ref{fig:irac_images_uchiis} with the exception of the contours overlaid on the images presented in the middle and upper left panels; these are shown in yellow to make the radio emission more prominent.}

\end{center}
\end{figure*}

There are 2337 CORNISH sources located within the region common to both surveys, which corresponds to approximately 90\,per\,cent of the CORNISH catalogue. In this section we will describe the matching procedure and classification of the matched sources and discuss the properties of the associated clumps. 

\subsection{Matching statistics and classification}
\label{sect:matching_stats}

In addition to the source catalogue, \sex\ also produced image masks marking the location and extent of each extracted ATLASGAL source. We have used these masks to identify positional matches with the CORNISH radio catalogue. This procedure identified 285 ATLASGAL-CORNISH matches. This initial sample includes chance alignments of Galactic dust emission with extragalactic background sources, planetary nebulae (PNe) and radio emission from more extended \hii\ regions. The first two forms of contamination are relatively straightforward to identify and remove, but emission from extended \hii\ regions can be more difficult to identify. Snapshot interferometric observations, such as those used for CORNISH, filter out the larger-scale emission ($>$14\arcsec), but are still sensitive to more compact, bright spots of emission commonly found towards the edges of more extended \hii\ regions where the ionisation front is interacting with the molecular material.

To assist in the classification of radio emission we have created false-colour, mid-infrared images of each of the ATLASGAL-CORNISH matches, using data extracted from the GLIMPSE survey (\citealt{benjamin2003_ori}). These images are $5\arcmin\times5$\arcmin\ in size and combine the 4.5, 5.8 and 8.0\,$\umu$m IRAC bands (coloured blue, green and red, respectively) centred on the peak of the \submm\ emission. Examples are presented in Figs.\,\ref{fig:irac_images_uchiis} and \ref{fig:irac_images_contaminating_sources} with radio and \submm\ emission contours overlaid (see figure captions for more details). Fig.\,\ref{fig:irac_images_uchiis} shows examples of the compact and \uchii\ regions that make up our final sample, while in Fig.\,\ref{fig:irac_images_contaminating_sources} we present examples of the contaminating sources that have been excluded.

The correlation of the mid-infrared emission with tracers of the molecular gas and dust has been used in a number of previous studies to aid in the classification of radio emission (e.g., \citealt{urquhart_radio_north, hindson2012, purcell2013}). Here we use the false-colour images to classify all of the ATLASGAL-CORNISH matches into one of five distinct types using the following criteria: 

\begin{enumerate}

\item Radio Star: radio emission is ubiquitous from stars across the HR diagram \citep{guedel2002} and is thought to originate mainly from dynamo effects in convective envelopes or magnetospheric structures. At the sensitivity of CORNISH these sources are expected to be detected out to distances of several kpc. Stellar sources are easily identified by their point source morphology in GLIMPSE images and their generally blue mid-infrared colours, as can be seen in the upper left panel of Fig.\,\ref{fig:irac_images_contaminating_sources}.

\item PNe: approximately 1,000 PNe are expected to be detected by CORNISH. These are expected to have typical radio continuum flux densities of 5-50\,mJy and sizes of a few arcseconds (\citealt{hoare2012}). These objects have similar mid-infrared colours to the embedded \hii\ regions in which we are interested but their spectral energy distributions (SEDs) tend to peak at shorter wavelengths than the latter and, thus, the majority will not be detected by ATLASGAL. The dusty envelopes of some nearby PNe will be detected. However, these objects are generally isolated in near- and mid-infrared images, having dispersed their natal clump, and have a relatively simple infrared morphology making them relatively easy to identify (\citealt{purcell2012}). Two examples are presented in the upper middle and right panels of Fig.\,\ref{fig:irac_images_contaminating_sources}.

\item Extragalactic background sources: these are usually located away from the peak of the \submm\ continuum and are not typically associated with any mid-infrared emission. Their radio emission also tends to be weaker than for \uchii\ regions and they are generally only marginally resolved and so can be well fitted with a Gaussian profile. Three examples of some of these extragalactic background sources are presented in the middle panels of Fig.\,\ref{fig:irac_images_contaminating_sources}.

\item Evolved \hii\ regions: for these sources the radio emission is found to be coincident with diffuse PAH emission at 8$\umu$m, associated with the ionization front where UV photons are interacting with the surrounding dust. The radio emission is typically a factor of three weaker (see Table\,1 for average values), often has an irregular morphology, and tends to be associated with only part of the shell structure surrounding the \hii\ region, as viewed in mid-infrared images. Any compact radio emission is therefore more likely to arise from localised dense parts of the shell rather than being a genuine compact or \uchii\ region. Three examples of these more extended \hii\ regions are presented in the lower panels of Fig.\,\ref{fig:irac_images_contaminating_sources} (see also \citealt{deharveng2010} for detailed study of ATLASGAL sources associated with these more evolved \hii\ regions).

\item Compact and \uchii\ regions: radio sources that appear to be self-contained regions of radio continuum emission that are coincident with compact mid-infrared sources. These sources are usually found towards the peak of the \submm\ emission seen in the ATLASGAL maps. In cases where the structure of the radio emission is resolved their morphologies are also found to be correlated with emission features seen in the mid-infrared images (\citealt{hoare2007}). A selection of these \hii\ regions is presented in Fig.\,\ref{fig:irac_images_uchiis}.

\end{enumerate}

It is the last of these classifications that will be the focus of this study. In a small number of radio sources there is some ambiguity in the classification. We have opted to identify such cases as being extended \hii\ regions in order to avoid the possibility of contaminating our target sample of compact and \uchii\ regions. This may result in a small number ($<$6) of genuine sources being wrongly excluded but these represent only a few per\,cent of our final sample and are therefore unlikely to affect the results significantly. In Table\,\ref{tbl:type} we present a summary of the classifications, the number of each type identified and the average observed radio continuum parameters.

Since the CORNISH survey is sensitive to radio emission with angular scales up to $\sim$20\arcsec, it is likely that our final sample will consist of a mixture of both compact and \uchii\ regions (the physical sizes will be investigated in Sect.\,\ref{sect:hii_size}). For brevity, we will refer to the contents of this sample collectively as \hii\ regions in the discussion that follows.

\setlength{\tabcolsep}{2pt}

\begin{table}
%select id_type,'\&',count(*),'\&',round(avg(radio_flux_p),1),'\&',round(avg(radio_flux_i),1),'\&', round(avg(radio_angular_size),1),'\&', round(avg(radio_flux_p/submm_peak_at_cornish_pos),1),'\\\\' from atlas_cornish_matches_sex where photSigma > 7 and reject_flag is null group by left(id_type,2)

\begin{center}\caption{Summary of ATLASGAL-CORNISH radio source classifications and average radio continuum parameters.}
\label{tbl:type}
\begin{minipage}{\linewidth}
\scriptsize
\begin{tabular}{lc...}
\hline \hline
  \multicolumn{1}{l}{Classification}&  \multicolumn{1}{c}{Number}&  \multicolumn{1}{c}{Peak Flux}&  \multicolumn{1}{c}{Int. Flux}&  \multicolumn{1}{c}{Angular Diameter}\\
   \multicolumn{1}{c}{}&  \multicolumn{1}{c}{}&  \multicolumn{1}{c}{(mJy\,beam$^{-1}$)}&  \multicolumn{1}{c}{(mJy)}&  \multicolumn{1}{c}{(\arcsec)}\\
\hline

Radio Star	&	1	&	54.2	&	199.9	&	3.6	\\
PNe	&	7	&	19.2	&	84.3	&	2.2	\\
Background galaxies	&	8	&	5.7	&	9.5	&	2.0	\\
\hii\ regions	&	56	&	13.4	&	312.1	&	9.7	\\
Compact/\uchii\ regions	&	\uchiinum	&	44.4	&	373.0	&	4.4	\\
\hline
Total &  285  & 36.6 & 343.1 & 5.3\\
\hline
\end{tabular}\\
\end{minipage}

\end{center}
\end{table}

 \setlength{\tabcolsep}{6pt}

\subsubsection{Contamination from background radio sources}

There are approximately 2,600 compact radio sources in the CORNISH catalogue, most of which ($\sim$80\,per\,cent) are likely to be extragalactic in origin. With such a large number of background sources there is a non-negligible probability of contamination from the chance alignment between these background sources and the dust emission detected by ATLASGAL. In our visual inspection of the mid-infrared images we have identified 8 of the ATLASGAL-CORNISH matches as being due to background source contamination. However, we need to check this number against the probability of such alignments occurring.

To estimate the number of chance alignments we have conducted a Monte-Carlo simulation in which 2,000 point sources were randomly distributed in longitude and latitude over the CORNISH survey region. The resulting catalogue was then compared with the ATLASGAL emission masks and the number of matches noted. After one hundred repeats the number of chance alignments ranged from 9 to 20 with a mean value of $14\pm4$, where the error is the standard deviation.

The number of expected chance alignments of background sources with the Galactic dust emission is in reasonable agreement with the actual number identified. It is therefore likely that we have identified the majority of the extragalactic background sources that would otherwise contaminate our sample. There may still be a small number of extragalactic interlopers that have made it into our final \hii\ region sample but contamination from these is at most few per\,cent. Furthermore, it is likely that these interlopers will be weeded out at a later stage of the analysis as we expect their derived parameters will deviate from those of the larger sample of \hii\ regions.

\subsection{Reliability of the \hii\ region catalogue}

As a consistency check on our classifications we have compared our \hii\ region catalogue to the samples identified from the targeted surveys of \citet{wood1989b} and \citet{kurtz1994}. These studies identified candidate \uchii\ regions by selecting sources with similar infrared colours to those of known \uchii\ regions.  The infrared-selected sample was then observed in the radio continuum with the VLA.

\citet{wood1989b} identified 75 \uchii\ regions, 60 of which are located in the ATLASGAL-CORNISH region. We recover 48 of these; however, of these we classify four as extended, rather than compact or \uchii\ regions (i.e., G10.15$-$0.34, G10.46+0.03B, G35.05$-$0.52 and G45.48+0.13) and one as a PN (G10.10+0.73; see upper centre panel of Fig.\,\ref{fig:irac_images_contaminating_sources} for an image of this object). Eight of the remaining twelve \citet{wood1989b} sources are unresolved and are probably a mixture of extragalactic background sources and more distant \hii\ regions that fall below the sensitivity of either CORNISH or ATLASGAL.  Using the CORNISH database, we find that G50.23+0.33, G42.90+0.57A and G42.90+0.57B have been classified as extragalactic radio lobes. Two other sources (G44.26+0.10 and G33.50+0.20) are included in the CORNISH catalogue but there is no significant dust emission coincident with the radio position, which would be expected if these were genuine compact or \uchii\ regions.

No radio emission is detected in CORNISH towards three more of the \citet{wood1989b} sample of \uchii\ regions, G54.096$−$0.059, G15.04$−$0.68 and G10.46+0.03A. The first of these is unresolved with a peak flux of 1.8\,mJy\,beam$^{-1}$ and so is likely to fall below the CORNISH catalogue's 7$\sigma$ threshold ($\sim2.8$\,mJy\,beam$^{-1}$). The second is located towards the M17 star forming complex, which has significant bright, extended radio emission, causing locally increased, artefact-dominated noise (\rms\ $\sim$6\,mJy\,beam$^{-1}$) in the associated CORNISH tiles; hence this object was not recovered. The third source is detected in the CORNISH data, but below the 7$\sigma$ threshold. This source was found to have a peak flux of 43.1 mJy\,beam$^{-1}$ by \citet{wood1989b} and so may be variable. 

\begin{figure}
\begin{center}
\includegraphics[width=0.49\textwidth, trim= 0 0 0 0]{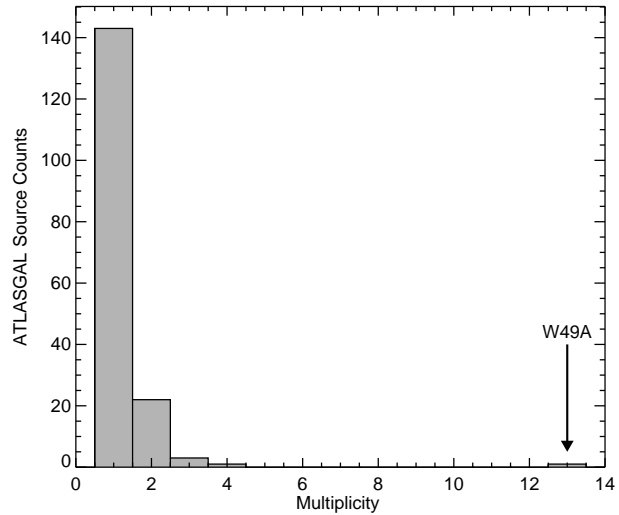}

\caption{\label{fig:cornish_atlas_multiplicity} \uchii\ region density as a function of ATLASGAL clumps. The bin size is 1.} 

\end{center}
\end{figure}

\citet{kurtz1994} identified an additional 29 \uchii\ regions in the ATLASGAL-CORNISH area and we find 28 of these in our matched sample. However, one radio source (i.e., G35.580+0.066) is coincident with the edge of what appears to be a more evolved \hii\ region, and so has been excluded from our final sample (see bottom right panel of Fig.\,\ref{fig:irac_images_contaminating_sources}). The missing radio source (i.e., G53.605+0.046) is unresolved in the \citet{kurtz1994} data, with a 3.6\,cm peak and integrated continuum flux density of 1.1\,mJy\,beam$^{-1}$ and 1.5\,mJy, respectively. Assuming a flat spectral index ($\alpha\sim-0.1$ for thermal free-free emission) we expect a similar flux density in the CORNISH data, which would place it below the 7$\sigma$ threshold.

Of the 89 \uchii\ regions identified by \citet{wood1989b}
and \citet{kurtz1994} in the ATLASGAL-CORNISH region, we have matched 76  with dust emission, though, we classified six as either part of the shell of a more evolved \hii\ region, or a PN. The remaining 13 radio sources not matched to any dust emission and are therefore likely to be part of the extragalactic background population.

Another aspect of the reliability of our sample that we need to consider is the possibility of missing genuine \hii\ regions because their associated dust envelope falls below the ATLASGAL sensitivity. As discussed in Sect.\,2.1 the ATLASGAL survey is sensitive to clumps with masses above 1,000\,\msun\ across the Galaxy ($\sim$20\,kpc). If we assume the \citet{kroupa2001} initial mass function (IMF) and a constant star-formation efficiency of between 10 and 30\,per\,cent (e.g., \citealt{,johnston2009,lada2003}) a minimum mass of $\sim$1-$3\times10^3$\,\msun\ is required to form at least one $\sim$10\,\msun\ star (e.g., \citealt{zinnecker2007}), which is approximately equivalent to a star with a spectral type of B1. Given that the CORNISH survey is sensitive enough to detect an optically thin unresolved \uchii\ region around a star of spectral type B0 or earlier across the Galaxy (\citealt{purcell2013}), it is unlikely we have missed many \uchii\ regions in this way.

Combining the ATLASGAL and CORNISH data sets has provided a simple and reliable method to identify a large sample of molecular clumps associated with \hii\ regions with relatively uniform noise characteristics for both the radio continuum and submillimetre data. Moreover, this constitutes an unbiased and relatively complete sample of \hii\ regions over a large enough region of the Galaxy to provide statistically robust results. 

\begin{figure*}
\begin{center}
\includegraphics[width=0.98\textwidth, trim= 0 0 0 0]{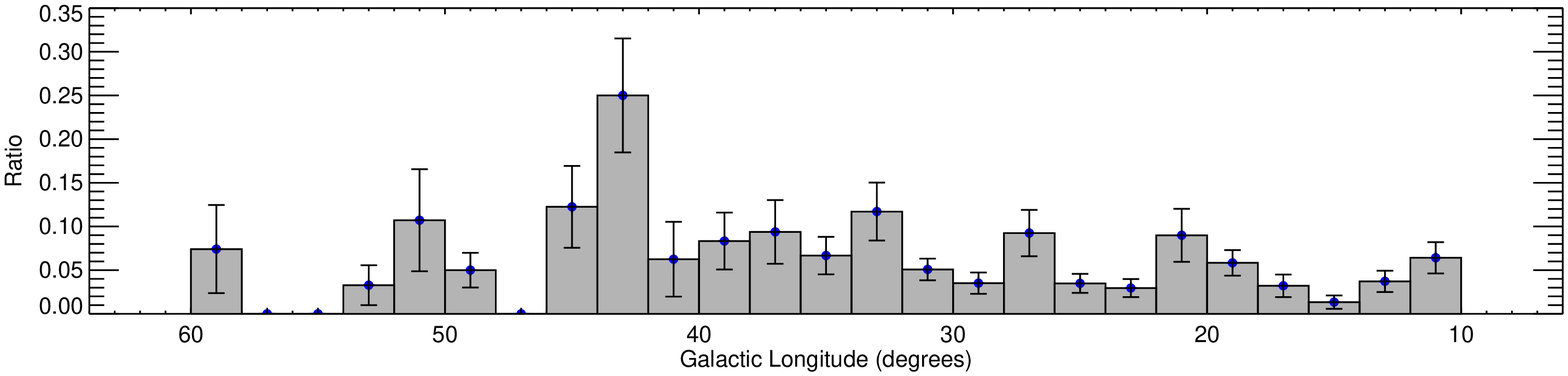}\\
\includegraphics[width=0.98\textwidth, trim= 0 0 0 0]{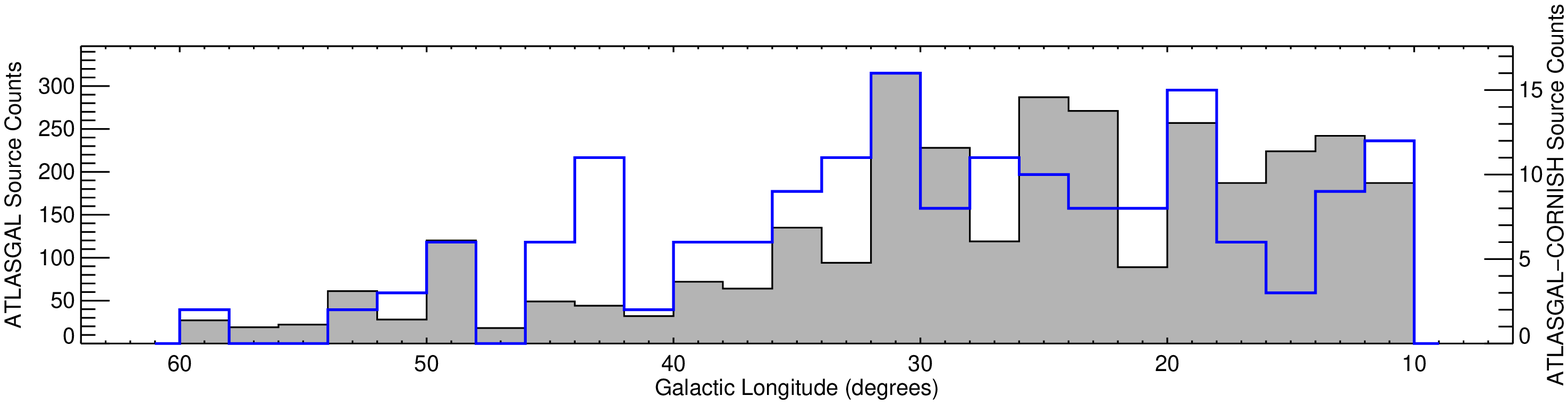}\\

\caption{\label{fig:gal_long} Galactic longitude distribution of ATLASGAL sources (grey filled histogram) and ATLASGAL-CORNISH associated clumps (blue histogram) are presented in the lower panel, while in the upper panel we present the ratio of the two histograms presented in the lower panel. The errors plotted in the upper panel are derived using Poisson statistics. The bin size used is 2\degr.} 

\end{center}
\end{figure*}

\subsection{Multiplicity}

In total, 170 ATLASGAL sources are found to be associated with one or more \hii\ regions. Fig.\,\ref{fig:cornish_atlas_multiplicity} shows the number of \hii\ regions associated with each ATLASGAL source. In the majority of cases (147 or 86\,per\,cent) the ATLASGAL source is associated with a single embedded \hii\ region, and only 23 are associated with two or more \uchii\ regions. Only in five cases are ATLASGAL sources associated with more than two \hii\ regions. We have been particularly careful in our analysis to exclude emission that appears likely to arise from larger \hii\ regions. Furthermore, in each of these cases the radio emission is found to be a bright compact source and, in nearly all cases, is coincident with a discrete mid-infrared point source. We are therefore confident that these represent separate \hii\ regions and not a problem with over resolved emission, which can ofter be an issue with snapshot surveys.

The one object that stands out is AGAL043.166+00.011, which is associated with 13 compact radio sources. This ATLASGAL detection is part of the W49A star-forming complex that also includes two other ATLASGAL objects, i.e., AGAL043.148+00.014 and AGAL043.164$-$00.029, which are themselves associated with five more \hii\ regions. In total, this complex is associated with 18 compact and \uchii\ regions, giving it the highest surface density of \hii\ regions in the part of the Galaxy covered by CORNISH and ATLASGAL.

Three other intense star forming regions (W31, G34.3+0.1 and AGAL020.081$-$00.136) contribute another ten \hii\ regions to the total and so, combined with W49A, these represent a significant fraction of the total number of \hii\ regions identified ($\sim$13\,per\,cent) but only constitute a few per\,cent of the total number of associated clumps. We therefore need to be careful that these rather intense star forming regions do not affect the overall statistical results unduly.

\subsection{Galactic longitude and latitude distribution}
\label{sect:lb_distribution}

\begin{figure}
\begin{center}
\includegraphics[width=0.49\textwidth, trim= 0 0 0 0]{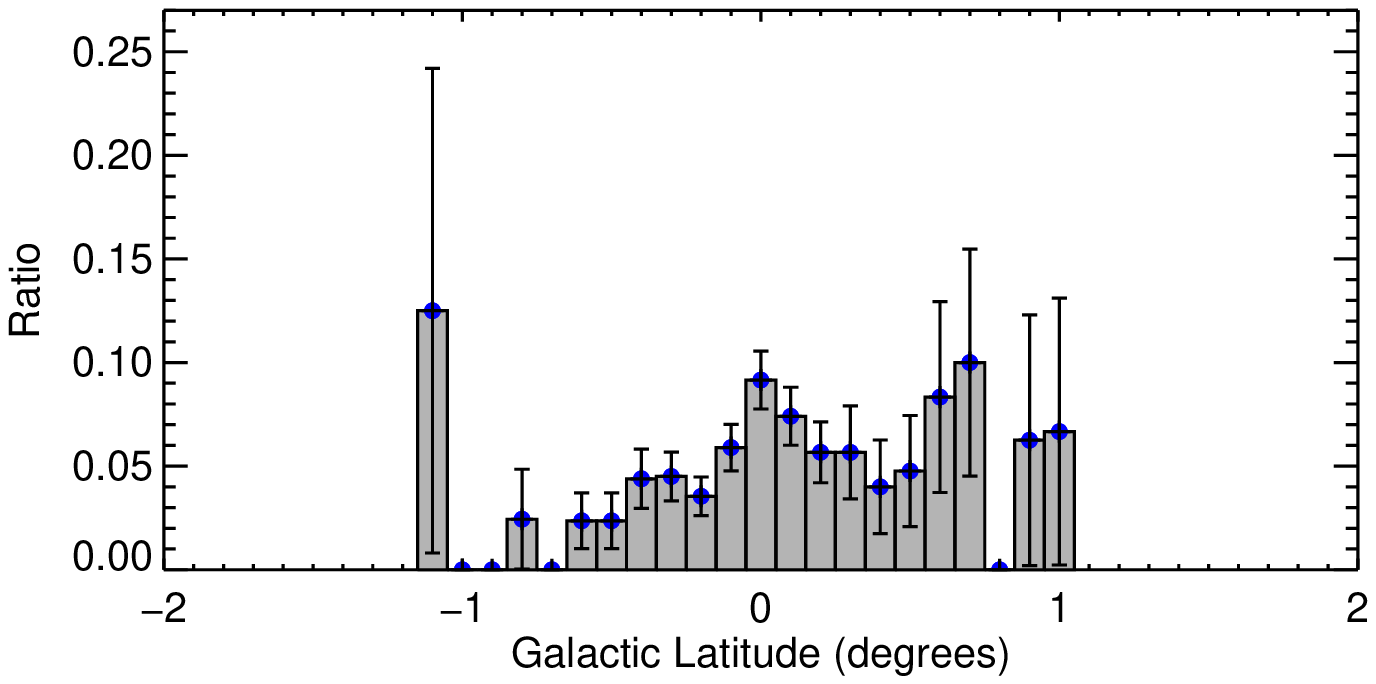}\\
\includegraphics[width=0.49\textwidth, trim= 0 0 0 0]{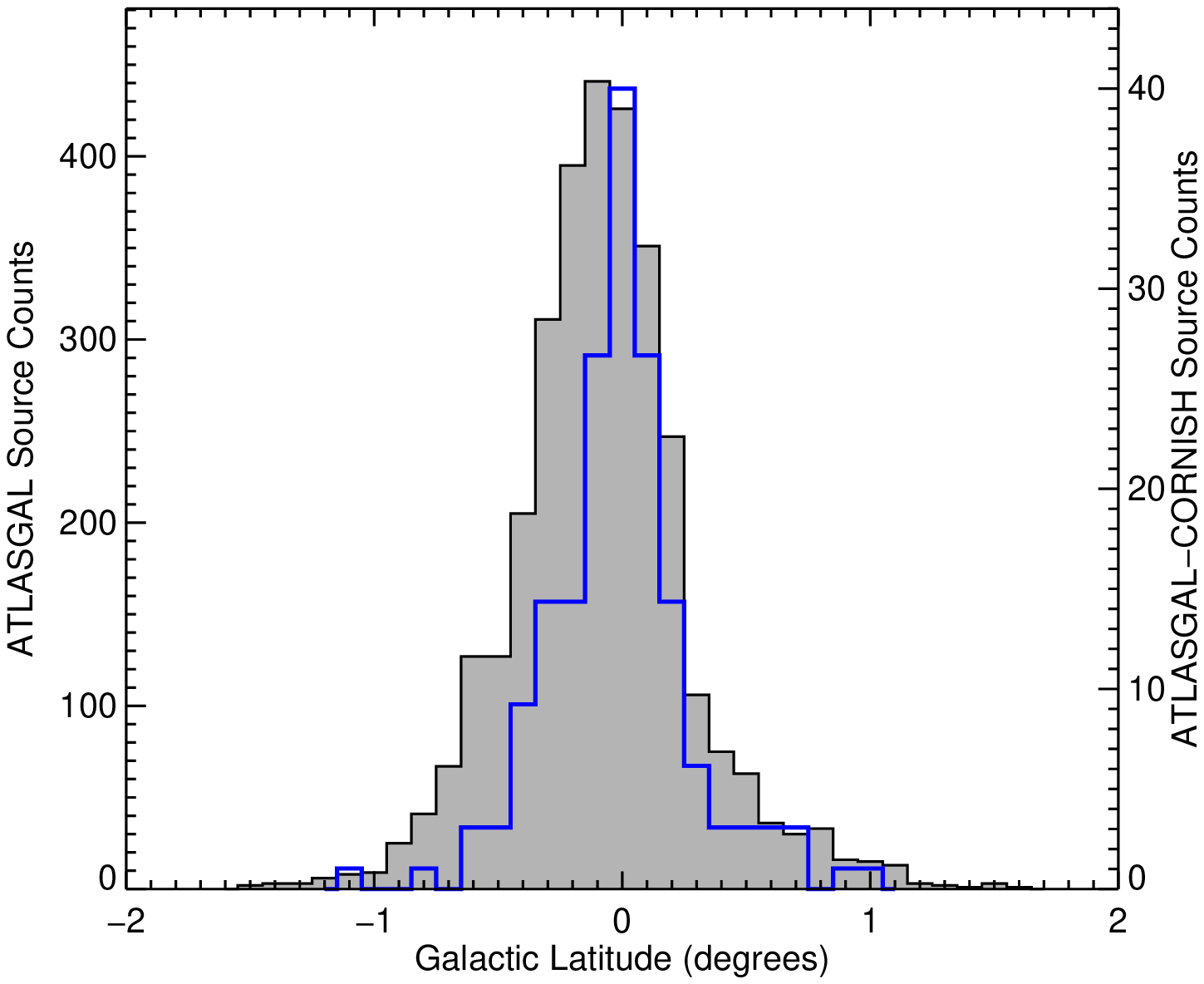}\\

\caption{\label{fig:gal_lat} Galactic latitude distribution of ATLASGAL sources (grey filled histogram) and ATLASGAL clumps associated with one or more \hii\ regions (blue histogram) are presented in the lower panel, while in the upper panel we present the ratio of the two histograms presented in the lower panel. The errors plotted in the upper panel are derived using Poisson statistics. The bin size used is 0.1\degr.} 

\end{center}
\end{figure}

In the lower panels of Figs \ref{fig:gal_long} and \ref{fig:gal_lat} we show the Galactic longitude and latitude distributions of the whole ATLASGAL compact source catalogue (filled grey histogram) and of the associated \hii\ regions identified from the CORNISH catalogue (blue histogram). In the upper panels of these figures we present the ratio of the two distributions, in each case. The ratio of the population of \hii-region associated clumps with the total ATLASGAL source counts can be interpreted as an analogue of, or approximation to, the high-mass star-formation efficiency of dense cores, evaluated on medium to large scales (0.1-1\,kpc) and integrated over the timescale associated with the \uchii\ region phase. Since the latter is relatively short (a few times 10$^5$\,yr; \citealt{davies2011,mottram2011b}), we can assume that any differences in the duration of a phase of star formation from region to region are not a significant influence on this ratio.

For the most part, the ratio is relatively flat and low across the longitude range of the sample. The mean value is 0.06 with sample standard deviation of 0.05, standard error on the mean of 0.01 and the median is 0.05. Almost all the values for each 2-degree longitude bin are within 2$\sigma$ of the mean and most are within 1 sigma. The obvious exception comes in the $\ell=42$-44\degr\ bin in which the ratio is $0.25\pm0.065$, more than 3$\sigma$ above the mean. This bin contains the ``mini-starburst'' star-forming region W49A (see Sect.\,3.3). While several massive star-forming regions have been put forward as Galactic-scale analogues of extragalactic starburst systems, W49A may be genuinely different. Its embedded massive young stellar objects (MYSOs) or proto-clusters have an integrated bolometric luminosity of $6.9 \times 10^6$\,\lsun, dominated by two sources in the infrared-selected Red MSX Source (RMS; \citealt{urquhart2008}) catalogue (G043.1679$-$00.0095 and G043.1650$-$00.0285) with $L_{\rm{bol}} > 10^6$\,\lsun. The infrared-traced star-formation efficiency in W49A is very high ($L_{\rm{bol}}$/$M_{\rm{CO}} = 32\pm6$\,\lsun\,\msun$^{-1}$) and its luminosity function may be flatter than normal (\citealt{moore2012}). Gas temperatures and densities are high at 50-100\,K and $\sim10^6$\,cm$^{−3}$, respectively (\citealt{roberts2011,nagy2012}). Also W49A is a source of very high-energy gamma rays (\citealt{brun2011}) which may be the cause of the additional gas heating characteristic of starbursts (\citealt{papadopoulos2011}).

Several other major star-forming complexes are included in the sample, notably W43 in the Scutum arm tangent at $\ell\sim30$\degr, W51 in the Sagittarius arm tangent at $\ell\sim49$\degr\ and W31 at \mbox{$\ell=10$\degr}, and at least the first two of these have also been labelled as mini-starburst regions. On the scale of individual GMCs, the SFR in these regions may be very high but, on the scales traced by these data, they have little effect on the average SFE (e.g., \citealt{eden2012}). W49A is exceptional in this regard (see also \citealt{moore2012}).

As noted in previous studies of the ATLASGAL data (i.e., \citealt{schuller2009,beuther2012,contreras2013}) the peak in the latitude distribution of the general population of compact ATLASGAL sources is skewed slightly to a negative value of $b$. This is consistent with the Sun's location above the Galactic mid-plane (\citealt{humphreys1995}). However, we note that the peak in the source counts of ATLASGAL clumps associated with \hii\ regions is peaks at $b=0$. This difference in latitude distribution is most likely a distance effect, since the \hii\ region sample has a significantly larger mean distance than the unmatched sources (see also Sect.\,\ref{sect:gal_long_vel}).

\subsection{Correlation between \hii\ regions and their natal clumps}

\subsubsection{Clump-\hii\ region angular correlation}

\citet{thompson2006} reported a significant difference in the angular offsets of embedded \uchii\ regions and of methanol masers, relative to the peak \submm\ flux of their host clumps. They used separations between methanol masers and \submm\ clumps drawn from targeted SCUBA observations reported by \citet{walsh2003}, and compared positions from 850-$\umu$m SCUBA observations to those of \uchii\ regions taken from the literature (i.e., \citealt{wood1989b,kurtz1994, giveon2005a}). Comparing these two samples, \citet{thompson2006} found that the \uchii\ regions had, on average, significantly larger offsets from the \submm\ peaks than the methanol masers and discussed several explanations for this.

\begin{figure*}
\begin{center}
\includegraphics[width=0.49\textwidth, trim= 0 0 0 0]{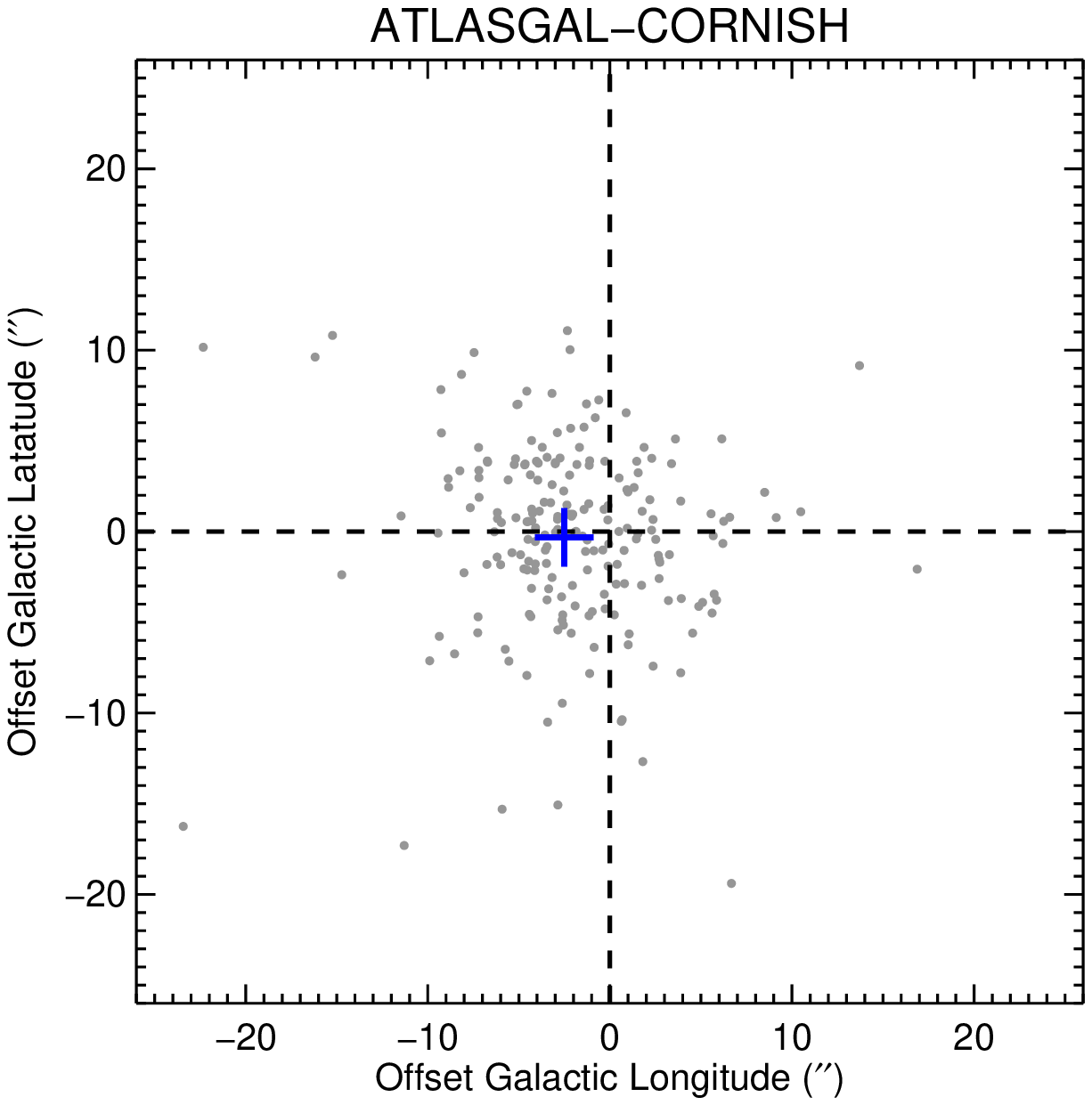}
\includegraphics[width=0.49\textwidth, trim= 0 0 0 0]{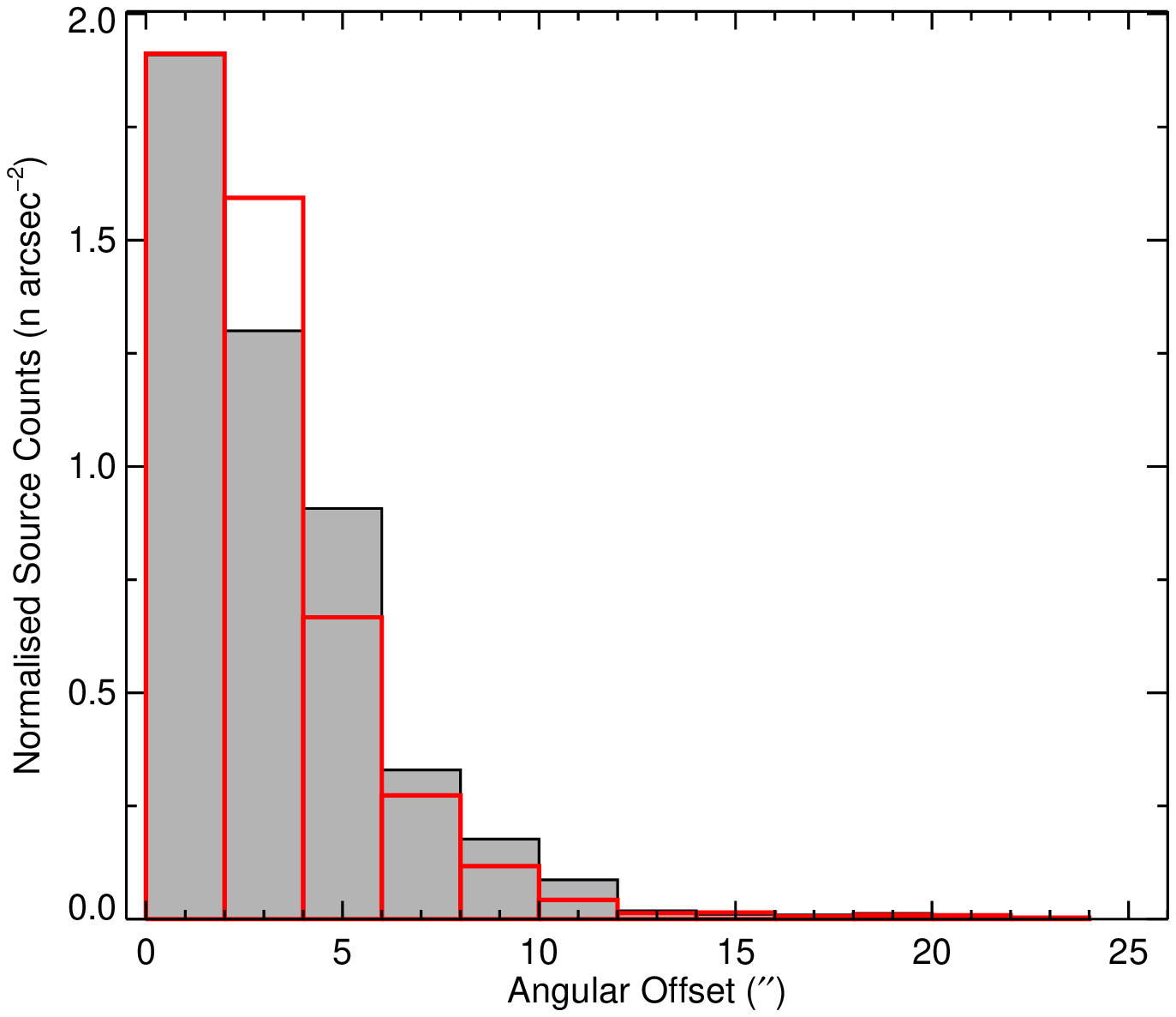}

\caption{\label{fig:cornish_atlas_offset} Left panel: Two-dimensional distribution of the angular offsets between the peak of the ATLASGAL dust emission and the matched CORNISH sources. The dashed vertical and horizontal lines indicate the $x$ and $y=0$ axes, respectively. The blue cross indicates the mean offset in longitude and latitude of the whole sample revealing a systematic offset from the 0,0 position ($\Delta\ell=-1.95\pm0.28$ and $\Delta b= -0.24\pm0.29$). Right panel: Normalised source counts for the \hii\ regions (grey filled histogram), and methanol maser distribution from Paper\,I (red histogram), are shown as a function of separation between their position and that of the peak of the \submm\ emission of their associated ATLASGAL source. The peak of the methanol maser distribution has been normalised to that of the \hii\ region distribution to facilitate comparison of the two samples. A correction has been made to compensate for the systematic offset seen in the left panel of this figure. We have truncated the $x$-axis of this plot at 26\arcsec\ as there are only eight \hii\ regions that have larger separations and the angular surface density effectively falls to zero. The bin size is 2\arcsec.} 

\end{center}
\end{figure*}

In the left panel of Fig.\,\ref{fig:cornish_atlas_offset} we present the 2d distribution of angular offsets for the ATLASGAL-CORNISH sample of \hii\ regions. This plot reveals the presence of a systematic offset of $\sim$2\arcsec\ in Galactic longitude. This offset is approximately six times larger than the standard error on the measurement and is therefore significant. Since the CORNISH data have been referenced to calibrators with positional accuracies much better than 1\arcsec\ this positional discrepancy is likely to be associated with the ATLASGAL data. This positional offset is approximately 10\,per\,cent of the ATLASGAL survey's resolution (19.2\arcsec) and approximately half of the \rms\ pointing error for the ATLASGAL survey ($\sim$4\arcsec; \citealt{schuller2009}) and so is not significant for this work.

The right panel of Fig.\,\ref{fig:cornish_atlas_offset} shows the normalised source counts (i.e., source counts divided by the area of the bin annulus) distribution for this sample, corrected for the systematic offset mentioned in the previous paragraph (grey filled histogram). In addition we have plotted the corresponding distribution for the sample of ATLASGAL clumps matched with methanol masers presented in Paper\,I (red histogram). The two distributions appear very similar and indeed a Kolmogorov-Smirnov (KS) test gives a $p$-value of 0.03 and is therefore unable to reject the null hypothesis that these two distributions are drawn from the same parent population (to reject the null hypothesis with $\ge3\sigma$ confidence the $p$-value must be $ < 0.003$). The positions of the embedded massive stars traced by the methanol masers and \hii\ regions are therefore closely correlated with the peak of the \submm\ emission, which is itself typically found towards the geometric centres of clumps (mean offset is $0.68$\arcsec\ with standard error and deviation of 0.48\arcsec\ and 6.25\arcsec, respectively). %{\color{green} Moore: the last point seems to be an assertion in the context of this paper - is it demonstrated in paper I?  or elsewhere?}

This is in contrast with the results reported by \citet{thompson2006}. Since the angular distribution for the methanol masers and \submm\ clumps is broadly in agreement with that determined by \citet{walsh2003}, we conclude that the difference is associated with the \uchii\ region positions. We tested the astrometry of the ATLASGAL sources and their associated methanol masers in Paper\,I and found it to be in excellent agreement. \citet{thompson2006} were careful to check the astrometry of their \submm\ observations and so the issue is likely to reside in the positional accuracy of the various radio catalogues used by  \citet{thompson2006}. We note that 80\,per\,cent of the sample of \uchii\ regions discussed by \citet{thompson2006} were drawn from the \citet{giveon2005a} catalogue, which was colour selected from a larger sample of 5\,GHz Galactic radio sources presented by \citet{white2005}.  However, a comparison of the positions given for the same sources in these catalogues (i.e., those of \citealt{white2005} and \citealt{giveon2005a}) reveals the presence of a systematic offset of 5\arcsec\ in Declination. It is unclear where this positional error in \citet{giveon2005a} has come from, however, it has led to the larger offset between the radio sources and the peak \submm\ emission of their host clumps reported by \citet{thompson2006}.

\subsubsection{Clump structure and morphology}
\label{sect:size}

\begin{figure}
\begin{center}
\includegraphics[width=0.48\textwidth, trim= 0 0 0 0]{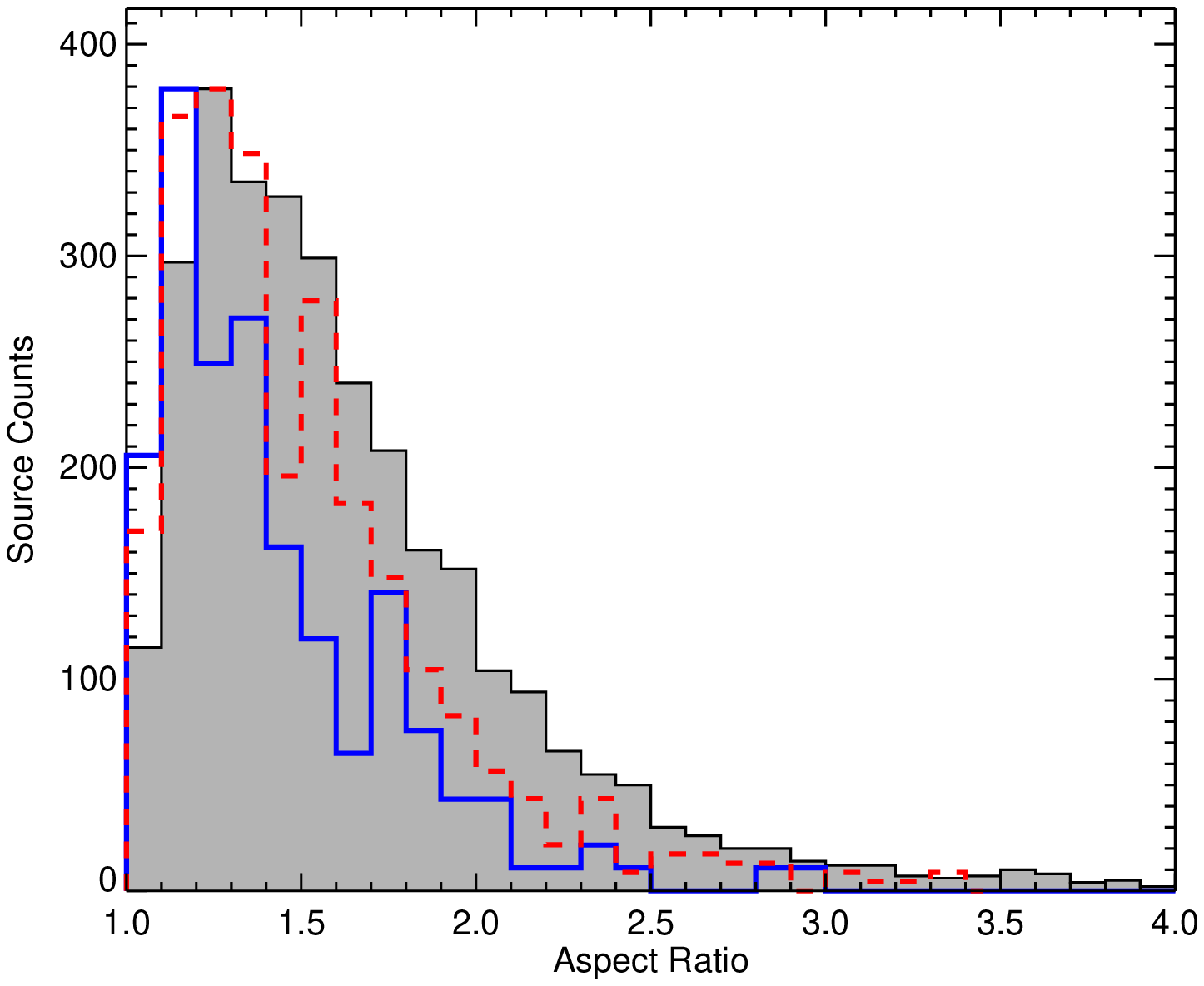}\\
\includegraphics[width=0.48\textwidth, trim= 0 0 0 0]{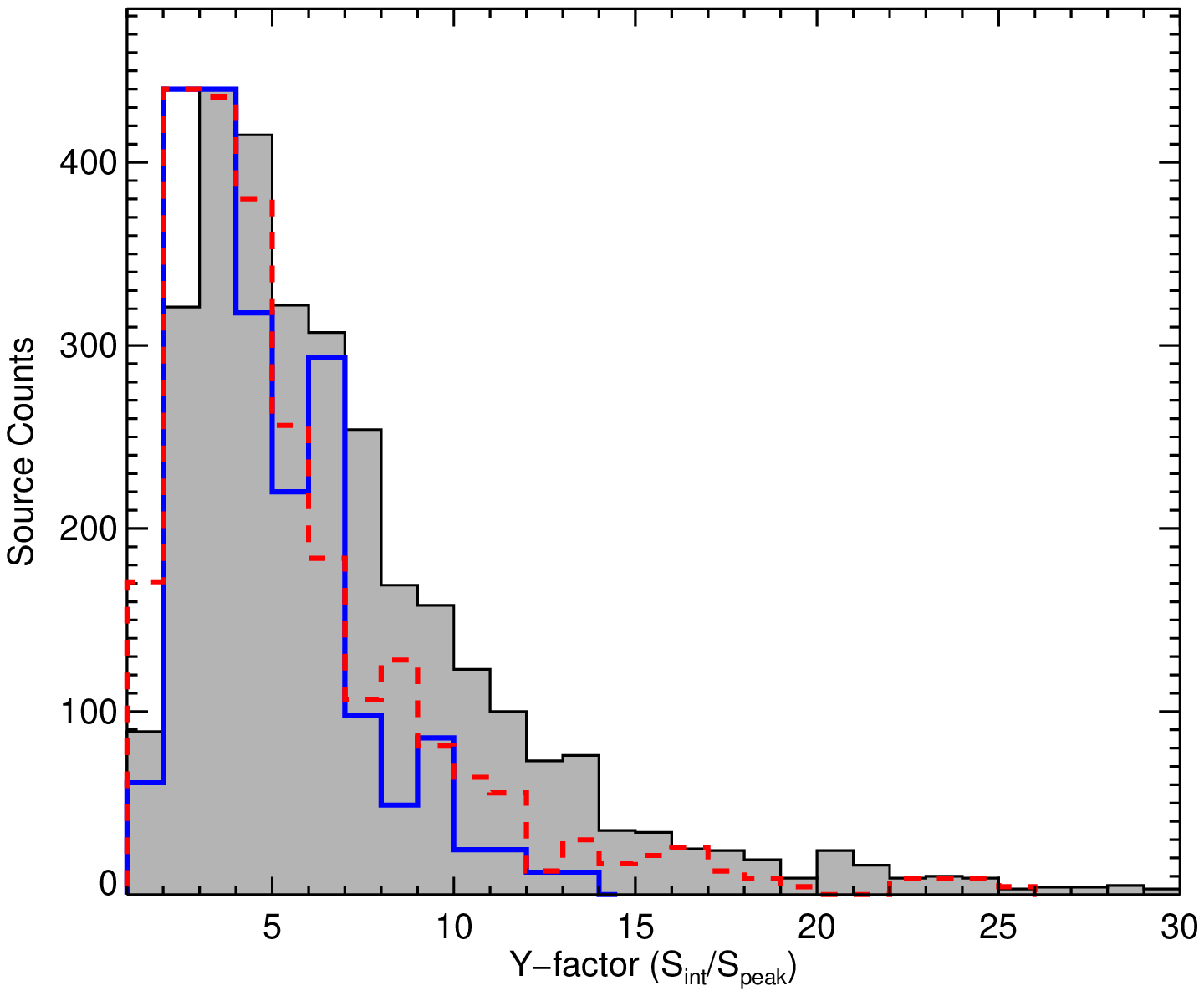}

\caption{\label{fig:clump_structure} The aspect ratio and $Y$-factor distributions for the ATLASGAL-CORNISH  \hii\ region associated clumps are shown in the upper and lower panels, respectively (blue histogram). The distribution of the whole sample of ATLASGAL compact sources located between $\ell = 10$-60\degr\ and $|b|<1$ is shown by the grey filled histogram while the ATLASGAL-MMB associated clumps are shown in red. The peaks of the \hii\ and methanol maser distributions have been scaled to the peak of the ATLASGAL compact source distribution. The bin sizes used for the aspect ratio and $Y$-factor distributions are 0.1 and 1, respectively.} 

\end{center}
\end{figure}

Fig.\,\ref{fig:clump_structure} shows the distributions of the aspect ratio and the $Y$-factor (ratio of integrated to peak \submm\ fluxes) for both the ATLASGAL-CORNISH \hii\ regions and methanol-maser associated clumps discussed in Paper\,I (blue and red histograms, respectively). These have been overlaid on the distributions of the same parameters for the whole ATLASGAL compact source catalogue (grey filled histogram). Examination of these distributions reveals  significant differences between the aspect ratios and $Y$-factors of the \hii\ region-associated clumps and those of the ATLASGAL compact source catalogue, with the \hii\ region clumps having a more spherical structure within which the massive star formation is taking place.

The \hii\ region and methanol-maser samples have very similar aspect-ratio and $Y$-factor distributions. The median aspect ratio and $Y$-factor for the \hii\ region and methanol-maser associated clumps are 1.33 and 1.4, and 4.41 and 4.6, respectively.  For both parameters, a KS test comparing the two samples is unable to reject the null hypothesis that they are drawn from the same parent distribution ($p$-values are 0.03 and 0.05 for the aspect ratio and $Y$-factor, respectively). 

The fact that there is no significant difference in clump structure between the various embedded stages of massive star formation was noted in Paper\,I, as well as by previous studies of high-mass protostellar cores (\citealt{williams2004}) and \uchii\ regions (\citealt{thompson2006}). This has led to the conclusion that the structure of the clump \emph{envelope} does not evolve significantly during the early embedded stages in the massive star-formation process, and this is consistent with our present findings.

\begin{figure}
\begin{center}
\includegraphics[width=0.49\textwidth, trim= 0 0 0 0]{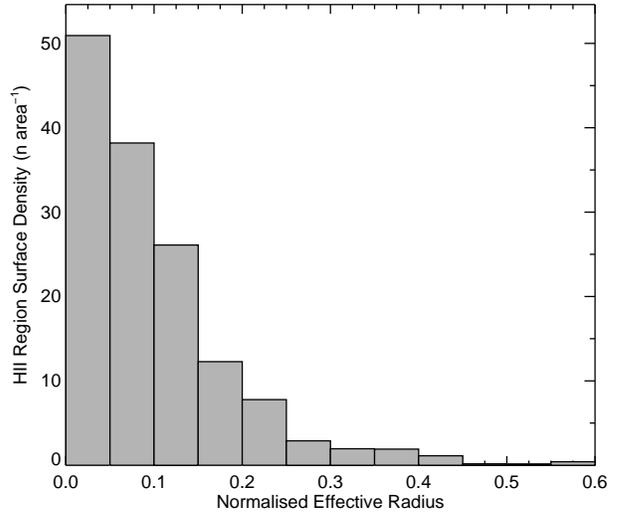}\\

\caption{\label{fig:normalised_radius_offsets} Compact and \uchii\ region surface density as a function of normalised clump effective radius. We have truncated the $x$-axis at a normalised effective radius of 0.6 as there are only seven sources with radii larger than this and the distribution effectively falls to zero beyond this point. The bin size is in units of 0.05 times the effective radius.} 

\end{center}
\end{figure}

The right panel of Fig.\,\ref{fig:cornish_atlas_offset} presents the angular separation between the \hii\ regions and the peak of the \submm\ emission and reveals a strong correlation between the two. However, given the large angular size distribution of the ATLASGAL clumps this plot does not provide a reliable indication of where in the clumps the star formation is taking place. In Fig.\,\ref{fig:normalised_radius_offsets} we present the surface density as a function of the normalised clump effective radii, which shows the distribution of \hii\ regions with respect to the clump radius.\footnote{The effective radii is as defined in \citet{rosolowsky2010}.} There is a very strong correlation between the positions of the \hii\ regions and the centre of the clump's where the highest column densities are found, with the majority being located within the inner 10-20\,per\,cent of their clumps radius. Only 7 sources are found to have normalised clump effective radii larger than 0.6, which corresponds to $\sim$4\,per\,cent of the sample.

The aspect ratio of the clumps and the tight angular correlation between the \hii\ regions  and dense gas found towards the centre of the clumps suggests that the majority of these massive stars are forming in the inner most parts of centrally condensed, spherical structures. Although the mean offset is almost negligible we note that for many cometary \hii\ regions the peaks of the radio and \submm\ emission are measurably offset (e.g., G019.0767$-$00.2882, which is offset by 4\arcsec\ from the peak in the dust emission --- lower right panel of Fig.\,1). However,  these offsets are at the limit of the resolution of these data.

\begin{table*}

\begin{center}\caption{Summary of derived parameters.}
\label{tbl:derived_para}
\begin{minipage}{\linewidth}
\scriptsize
\begin{tabular}{lc.......}
\hline \hline
  \multicolumn{1}{l}{Parameter}&  \multicolumn{1}{c}{Number}&	\multicolumn{1}{c}{Mean}  &	\multicolumn{1}{c}{Standard Error} &\multicolumn{1}{c}{Standard Deviation} &	\multicolumn{1}{c}{Median} & \multicolumn{1}{c}{Min}& \multicolumn{1}{c}{Max}\\
\hline
Angular Offset (\arcsec) &          213&       7.32 &      0.58 &        8.42 & 
      4.93 &      0.32 &       61.28\\
Aspect Ratio &        170&       1.44 &     0.03  &       0.44
 &       1.32  &       1.01  &       4.89 \\
$Y$-factor &          170&       4.87&      0.18 &        2.34
 &       4.36 &       1.52 &       13.82\\
FWHM Line Width (km\,s$^{-1}$) &          126&       2.77 &      0.11 & 
       1.20  &       2.55  &      0.86  &       11.0\\
Kinetic Temperature (K) &          115&       24.65&      0.55 & 
       5.89  &       23.33  &       14.68  &       60.76 \\
Heliocentric Distance (kpc) &          169&       9.09 &      0.32  & 
       4.17 &       9.78  &       1.87  &       18.60 \\
Radius (pc) &          154&       1.52 &     0.07  &       0.88
 &       1.38 &      0.19 &       6.01 \\
Log[Clump Mass (\msun)] &          169&       3.96 &      0.34 &        4.38  &       3.69  &       1.72  &       5.46\\
Log[Viral Mass (\msun)] &           123&       3.93 &      0.37 &        4.07  &       3.74 &       2.08 &       4.89\\
Viral Ratio &          123&       2.47 &      0.17 &        1.88 &       2.08 &      0.25  &       14.90\\

\hline
Log[Lyman Continuum Flux (s$^{-1}$)] &          212&       47.66&
     0.06 &       0.85 &       47.73 &       45.49 & 
      49.69\\
\uchii\ Region Observed Diameter (\arcsec) &          212&       4.44&
      0.25 &        3.61 &       2.88 &       1.5 & 
      23.42\\
\uchii\ Region Diameter (pc) &          212&      0.16 &     0.01  & 
      0.17 &      0.11 &    0.01 &      0.89 \\
\hline\\
\end{tabular}\\

\end{minipage}

\end{center}
\end{table*}

\section{Kinematic properties of the clumps}

\subsection{Molecular line data} 
\label{sect:vlsr}

In order to obtain velocity information for our sample of massive, star-forming clumps, we have searched the literature to find complementary molecular line data. Where available, we have given preference to molecular transitions with high critical densities. The majority of the source velocities are taken from one of three recently reported sets of targeted observations of the lower excitation states of the ammonia inversion transition (NH$_3$ (1,1), (2,2) and (3,3); \citealt{urquhart2011b,dunham2011,wienen2012}). These studies used the Green Bank and Effelsberg 100-m telescopes, with a resolution of $\sim$30 and $\sim$40\arcsec, respectively, and so are reasonably well matched to the submm clump sizes. Between them, these programmes observed 126 of our \hii-region associated sample of clumps, providing unambiguous velocities and line widths. The NH$_3$ (2,2) transition was detected towards 115 of these sources allowing their kinetic temperatures to be estimated.

A further 24 clumps have been observed by \citet{schlingman2011} using HCO$^{+}$ (3-2) and N$_{2}$H$^{+}$ (3-2) as part of follow-up observations of sources detected in the Bolocam Galactic Plane Survey (BGPS; \citealt{aguirre2011}). The velocities for another 13 clumps were drawn from targeted observations made by \citet{bronfman1996}, \citet{blum2000}, \citet{urquhart2009}, \citet{anderson2009a} and \citet{walsh2011}. In total we have been able to obtain velocities from the literature for 163 of the 170 clumps in our sample.

Of the remaining seven sources, three have been observed as part of an ATLASGAL follow-up programme using the IRAM 30-m telescope. Analysis is still ongoing for these data and their results will be discussed in detail in a future publication (Csengeri et al. in prep.). However, we have been able to extract the source velocities from the preliminary data reduction and these should be reliable enough to estimate the kinematic distance to these sources. For the remaining four sources we have obtained a $^{13}$CO (3-2) spectrum from the JCMT HARP survey of the Galactic plane (Toby Moore, private comm.) for two sources, and obtained Director's time on the Mopra radio telescope to observe the other two sources.\footnote{The Mopra radio telescope is part of the Australia Telescope National Facility which is funded by the Commonwealth of Australia for operation as a National Facility managed by CSIRO.} %Plots for this ad-hoc collection of spectra and the fits to the data are presented in Appendix A. {\color{green} Moore: do the spectra need to be published?}

Using a combination of literature values and the ad-hoc observations described in the previous paragraph, we have been able to obtain a radial velocity for every object in our sample. In the following two subsections, we will use the observed molecular properties to investigate the Galactic longitude-velocity distribution, velocity dispersion and temperature distribution of these massive, star-forming clumps.

\subsection{Galactic longitude-velocity distribution} 
\label{sect:gal_long_vel}

\begin{figure*}
\begin{center}
\includegraphics[width=0.98\textwidth, trim= 0 0 0 0]{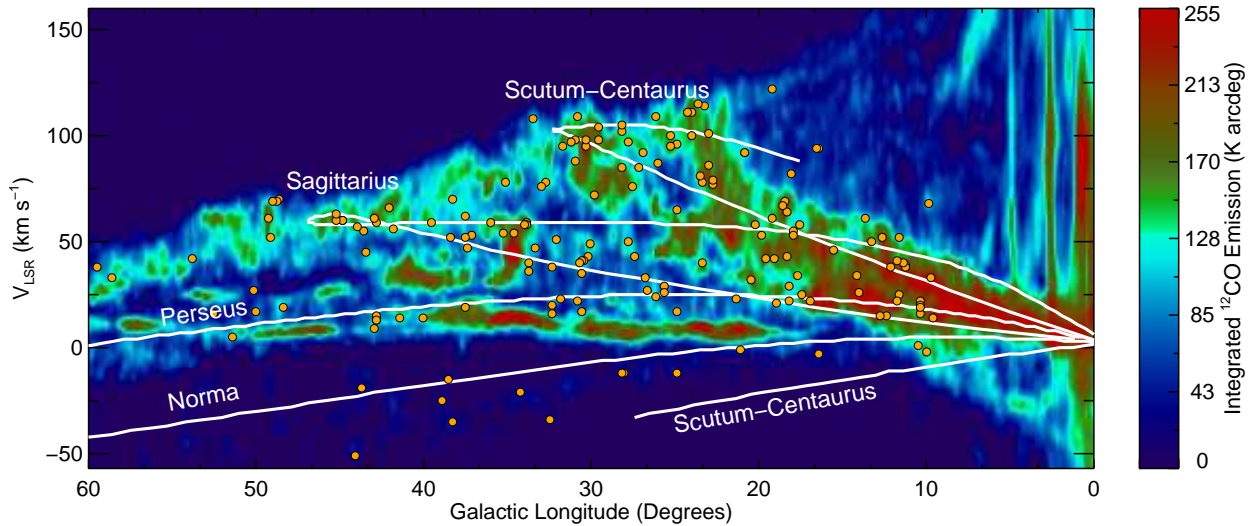}

\caption{\label{fig:lv_distribution}  Galactic longitude-velocity distribution of \hii -region associated clumps. The background image shows the distribution of molecular gas as traced by the integrated $^{12}$CO J=1--0 emission (\citealt{dame2001}); the colour bar on the right shows the relative intensity of the emission. The orange circles mark the positions of the \uchii\ regions and the white lines indicate the location of the spiral arms taken from the model by \citet{tayor1993} and updated by \citet{cordes2004}.}

\end{center}
\end{figure*}

In Fig.\,\ref{fig:lv_distribution} we present the longitude-velocity distribution of our sample with respect to that of the molecular gas as traced by the $^{12}$CO (1-0) emission mapped by \citet{dame2001} and the predicted position of the spiral arms from \citet{tayor1993} and updated by \citet{cordes2004}. The distribution of the compact and \uchii\ regions agrees well with the location of the model spiral arms, in general. However, the \hii-region and molecular-gas distributions correlate significantly less well. %{\color{green} Moore: can this be quantified?}. 

This is in stark contrast to the longitude-velocity distribution of a large sample of ATLASGAL sources observed in ammonia by \citet[][i.e., from the general ATLASGAL catalogue]{wienen2012}. These authors found a much stronger correlation between the distribution of ATLASGAL sources and the molecular gas (see their Fig.\,3). One possible explanation is that the velocities have been incorrectly assigned to our sample, but the majority of these have been drawn from targeted observations using high density tracers (e.g., NH$_3$, CS and HCO$^{+}$ data discussed in the previous subsection). These data all show a single, unambiguous velocity component in the direction of each source and are therefore unlikely to be incorrect.

Inspection of the various distributions shown in Fig.\,\ref{fig:lv_distribution} shows that the CO emission associated with the more distant Norma arm is relatively weak and occurs in small, rather isolated sources but that the HII regions here are quite well correlated with the CO. In  Section\,5 we derive distances for the whole sample of clumps and find that the majority are located at heliocentric distances greater than $\sim$10\,kpc. From these two points we conclude that the apparent anti-correlation between molecular gas and \hii\ region distributions is because the CO emission is dominated by bright, relatively nearby clouds, with emission from the more distant clouds reduced by beam dilution. The higher correlation observed by \citet{wienen2012} suggests that the general ATLASGAL catalogue is also dominated by a large population of nearby sources (see also the discussion in Sect.\,3.4.3).

\subsection{Line widths and temperatures}
\label{sect:linewidths_temps}

Given the variety of molecular tracers used to obtain velocity information, it is not straightforward to directly compare the line widths and kinetic temperatures measured from these different tracers. In this subsection, therefore, we will concentrate on the properties derived from the ammonia observations, and although these are not available for the complete sample, they do include a sufficiently large fraction of the sample ($\sim$70\,per\,cent) such that the statistical results will be reliable.

\begin{figure*}
\begin{center}
\includegraphics[width=0.49\textwidth, trim= 0 0 0 0]{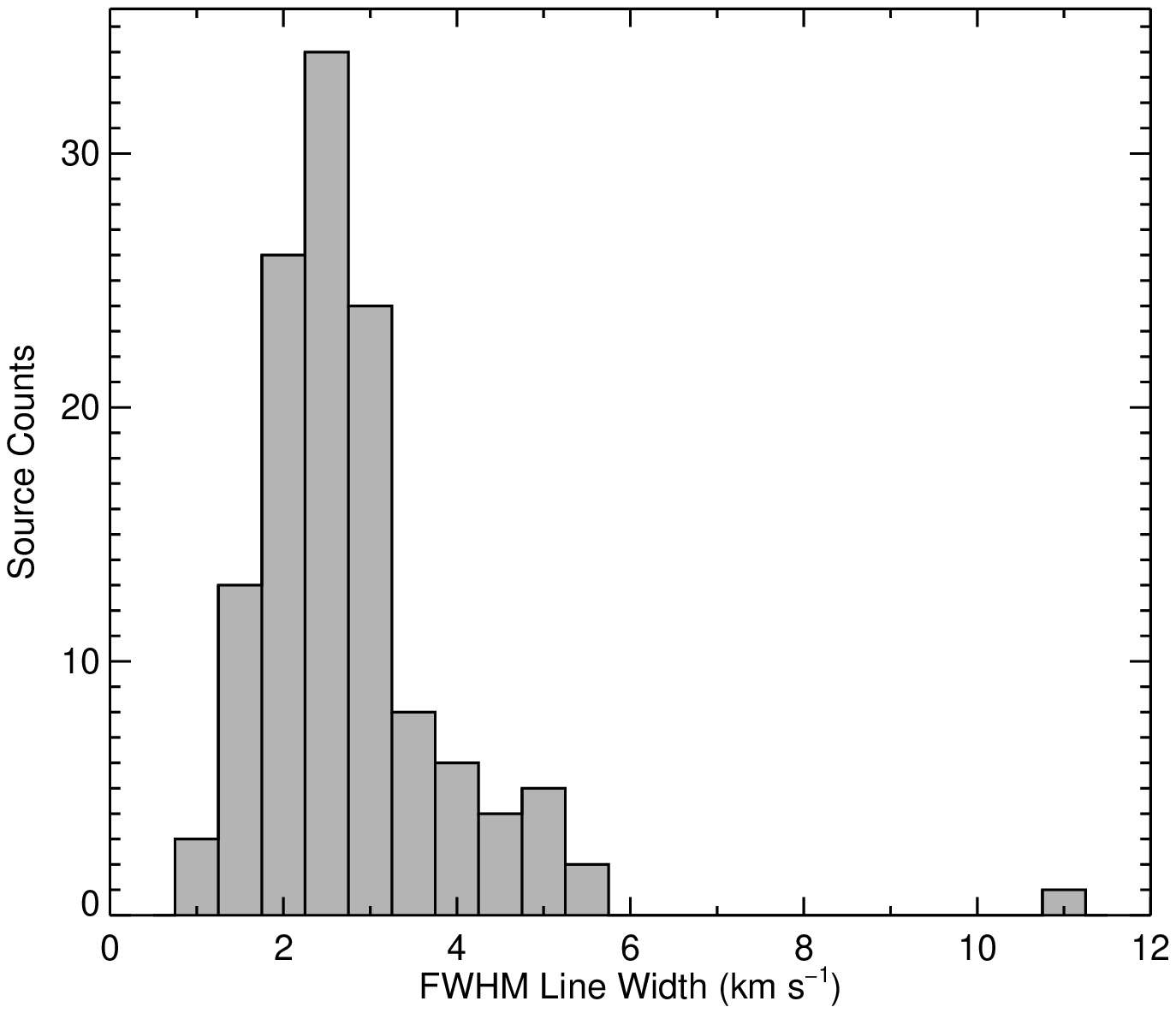}
\includegraphics[width=0.49\textwidth, trim= 0 0 0 0]{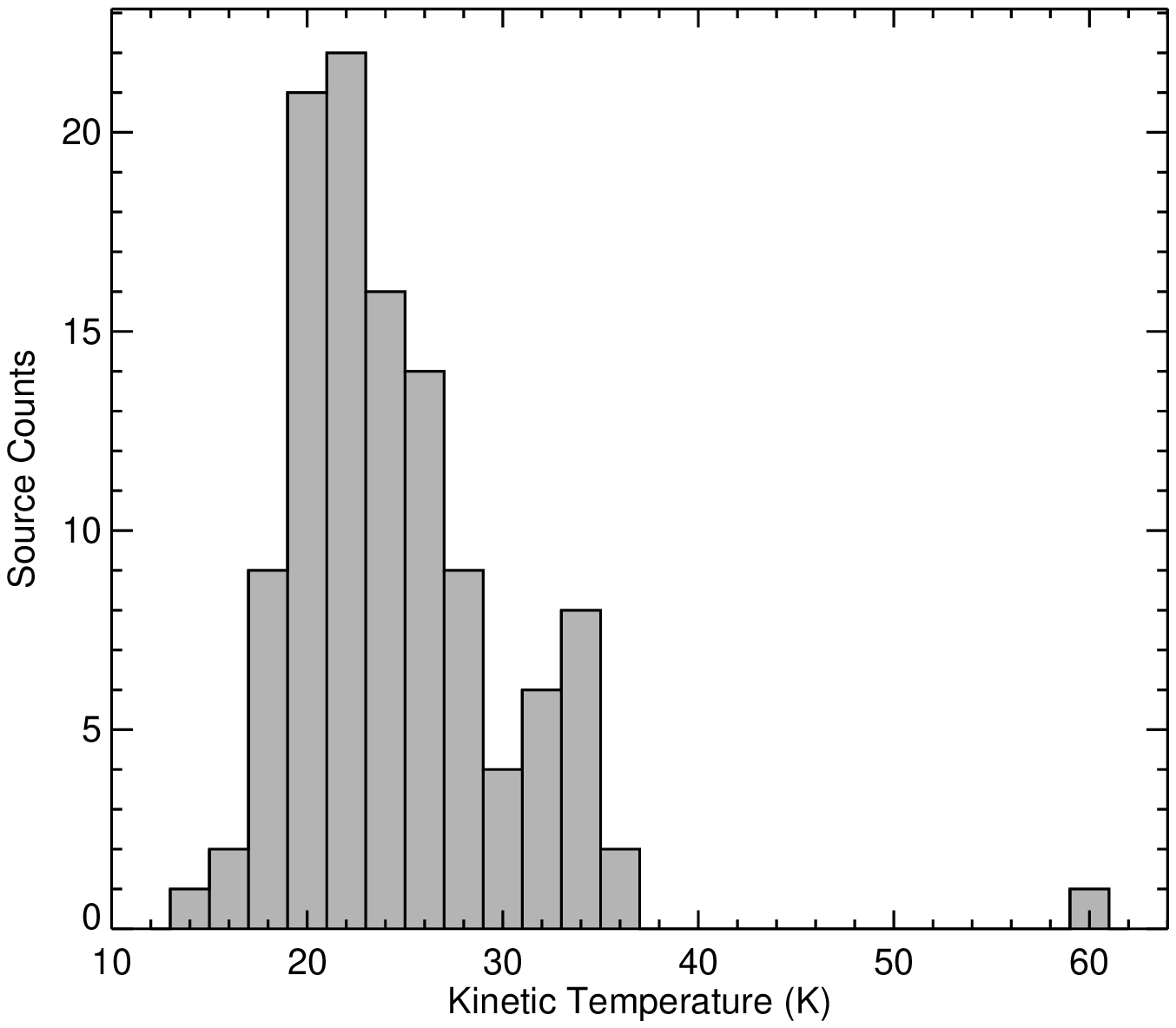}

\caption{\label{fig:line_width_temp}  Left panel: FWHM line width distribution of the NH$_3$ (1,1) inversion transition determined from the 126 clumps observed by \citet{urquhart2011b, dunham2011, wienen2012}. Right panel: distribution of kinetic gas temperatures determined for the 115 clumps towards which both the NH$_3$ (1,1) and (2,2) transitions are detected.} 

\end{center}
\end{figure*}

In Fig.\,\ref{fig:line_width_temp} we present histograms showing the FWHM line width of the ammonia (1,1) inversion transition (left panel) and the kinetic temperature derived from the ratio of the NH$_3$ (1,1) and (2,2) transitions (see \citealt{urquhart2011b} for more details). The statistical values for these two clump properties are tabulated in Table\,\ref{tbl:derived_para}. The ammonia line widths have been determined from a simultaneous fit to the main and satellite features of the NH$_3$ (1,1) line and so these have been corrected for the effects of line broadening due to high optical depths. We note that determination of the kinematic temperature requires the detection of the NH$_3$ (2,2) transition and therefore the distribution shown in the right panel of Fig.\,\ref{fig:line_width_temp} may contain a selection bias to higher temperatures.

The mean line width ($\sim$2.8\,\kms) is typical of those found towards other young massive star formation regions identified by methanol masers (e.g., \citealt{wienen2012, pandian2009}) and MYSOs (\citealt{urquhart2011b}), but significantly larger than for the general population of BGPS clumps reported by \citet{dunham2011} and the subsample of infrared dark ATLASGAL sources reported by \citet{wienen2012} ($\sim$1.9\,km\,s$^{-1}$ for both). The difference is approximately 7 times the standard error of the mean line width. It is unclear at this point whether the increase in line width seen towards the \uchii\ region sample is a result of increased feedback from the embedded star formation or simply reflects that the clumps associated with \hii\ regions are typically found at larger distances and consequently the physical scale probed by the telescope beam is larger; in accordance with the Larson size-line width relation (\citealt{larson1981}). 

We find a mean kinetic temperature of $\sim$25\,K, which is again similar to those of the methanol-maser and MYSO samples. However, it is significantly higher than found for starless clumps that tend to have kinetic temperatures of $\sim$15\,K (e.g., \citealt{pillai2006,dunham2011,wienen2012}). This suggests that young embedded stars are internally heating their host clump on scales large enough to be detected by the present data. It is important to note that all of these ammonia observations have been directed towards the peak of the \submm\ emission or the infrared emission associated with the embedded \hii\ regions. These observations are probing the hottest and most active parts of each clump and therefore the line width and kinetic temperatures are effectively upper limits. It is likely that the kinetic temperature of the gas in the more extended envelope, which contributes the majority of the flux, will be lower. We note that the kinetic temperature of the gas and dust are likely to be significantly higher closer to the \hii\ regions, but the single dish telescopes used for these ammonia observations are unable to probe these size scales. However, the proportion of the total clump mass at higher temperature is very small and unlikely to significantly affect the estimated clump masses. 

\section{Distances}

In the previous section we identified a large sample of compact and \uchii\ regions and their host clumps in which they are still deeply embedded. Before we can begin to investigate the properties of these massive star forming environments (e.g., masses, densities and physical sizes) we need to determine their distances. Many of these \hii\ regions were previously known and so distances are available in the literature for a significant fraction of the sample. We have therefore conducted a comprehensive review of the literature and have identified 24 clumps that have measured maser parallax or spectroscopic distances (e.g., \citealt{reid2009} and \citealt{moises2011}, respectively), and kinematic distances for $\sim$100 additional clumps.

Although maser parallax and spectroscopic distances are the most reliable method for determining distances, these types are only available for a relatively small fraction of our sample. It is therefore necessary to resort to kinematic distances; these will be discussed in more detail in the following subsection.

\subsection{Kinematic distances}
\label{sect:distance}

\begin{figure}
\begin{center}

\includegraphics[width=0.49\textwidth, trim= 0 0 0 0]{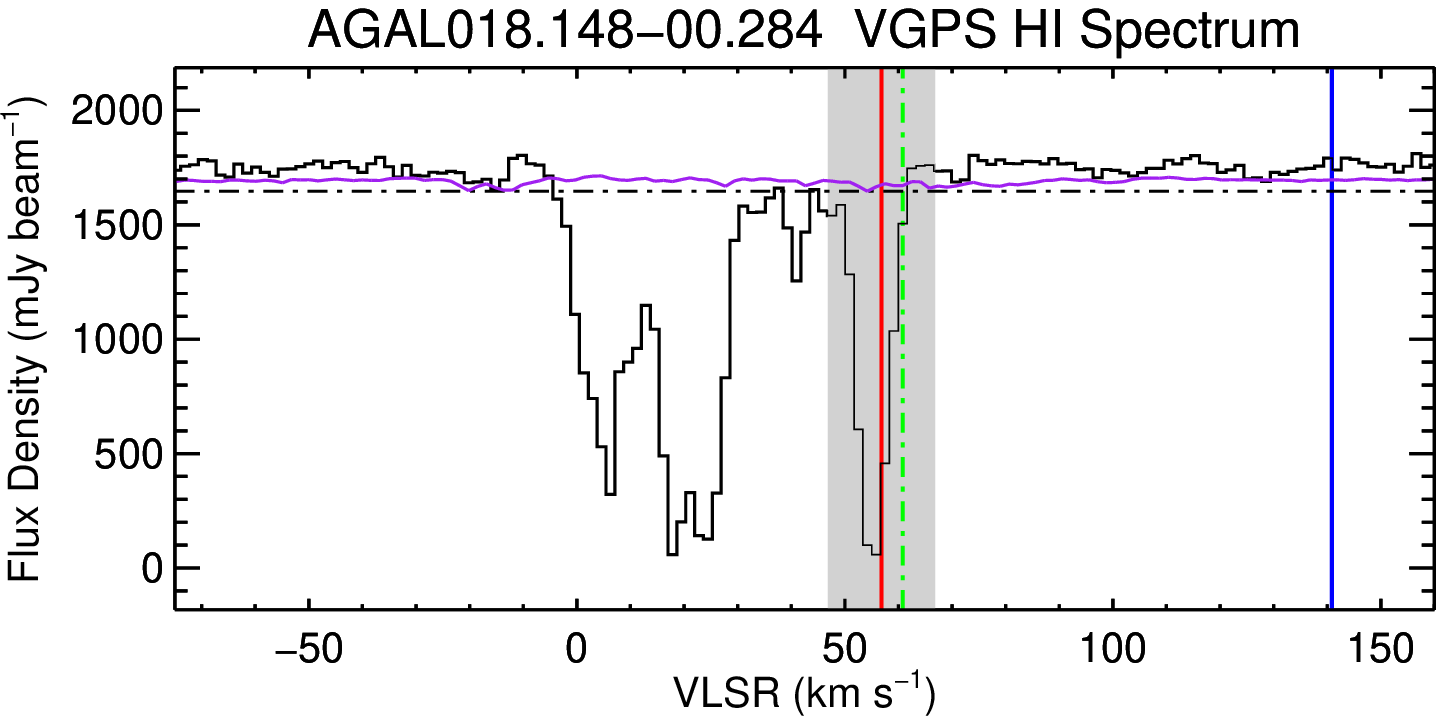}
\includegraphics[width=0.49\textwidth, trim= 0 0 0 0]{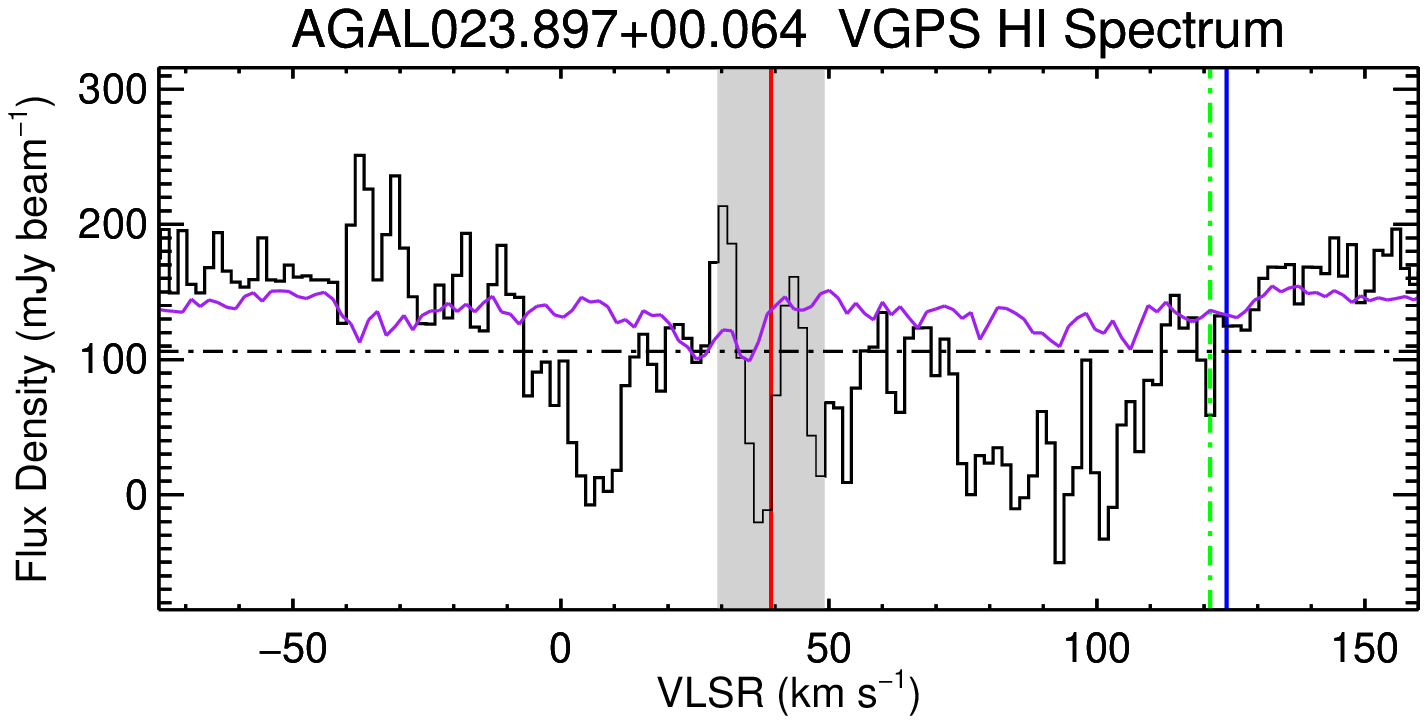}

\caption{\label{fig:uchii_spectral_data2} Source-averaged  
  continuum-included \hi\ spectra extracted from the SGPS and VGPS archives towards  two \hii\,regions for which a distance is not available in the literature. In the upper and lower panels we provide an example of a source located at the near and far distances, respectively. The source velocity ($v_{\rm{s}}$), the velocity of the tangent point ($v_{\rm{t}}$) and the position of the first absorption minimum ($v_{\rm{a}}$) are shown by the red, blue and green vertical lines, respectively. The grey vertical band covers the velocity region 10\,\kms\ either side of the source velocity and is provided to give an indication of the uncertainty associated with it due to streaming motions. The dotted horizontal line shows the 4$\sigma_{\rm{r.m.s}}$ noise level determined from absorption free parts of the spectra. The emission profiles of sources located at the far distance will show absorption dips at all velocities up to the tangent velocity.}

\end{center}
\end{figure}

We use the radial velocities discussed in Sect.\,\ref{sect:vlsr} in conjunction with the Galactic rotation model of \citet[][$\theta_0= 220$\,km\,s$^{-1}$ and $R_0 = 8.5$\,kpc, where $\theta_0$ is the circular rotation speed and $R_0$ is the distance to the Galactic centre]{brand1993} to estimate each source's kinematic distance. For sources located within the Solar circle (i.e., $<$ 8.5\,kpc from the Galactic centre) there are two possible distances that are equally spaced on either side of the tangent position. These distances are known as the \emph{near} and \emph{far} distances and the degeneracy is commonly referred to as the kinematic distance ambiguity (KDA). Resolutions of this problem are discussed below.  To estimate the associated distance uncertainties we have assumed an uncertainty of $\pm$10\,km\,s$^{-1}$ for the radial velocities (i.e., streaming motions; \citealt{reid2009}), which corresponds to an error in the kinematic distances of $\sim$0.6-1\,kpc. These distances are given in Table\,\ref{tbl:derived_clump_para} for individual clumps and their overall statistics are summarised in Table\,\ref{tbl:derived_para}.

Of the 146 clumps for which we were unable to assign a maser parallax or spectroscopic distance we find: twelve have negative velocities and are therefore located outside the Solar circle and so are not affected by KDA; thirteen are located within 10\,km\,s$^{-1}$ of the tangent velocity and so have been placed at that distance; one was placed at the near distance as a far distance allocation would place it at an unrealistically large height above the Galactic mid-plane for a massive star formation region (i.e., $z$ $>$ 120\,pc, which is four times the scale height for massive stars; \citealt{urquhart2011a,reed2000}). From the literature review we were able to identify a further 73 clumps included in previous studies that have already resolved their distance ambiguities.

From the literature and the automatic assignment of sources to the near, far and tangent distances, described at the beginning of the previous paragraph, we have obtained distances for 124 clumps ($\sim$75\,per\,cent of the sample). For the remaining 46 clumps we have extracted \hi\ spectra from the Southern Galactic Plane Survey (SGPS; \citealt{mcclure2005}) and the VLA Galactic Plane Survey (VGPS; \citealt{stil2006}). There are two \hi-related methods that can be employed to resolve KDAs; these are known as the absorption against a continuum  (\citealt{fish2003, kolpak2003,anderson2009a,roman2009,urquhart2012}) and the \hi\ self-absorption technique (\citealt{jackson2003,roman2009,green2011b}).

\subsubsection{\hi\ absorption against a continuum source}
\label{sect:HIEA}
	
This method is the most reliable and works by identifying \hi\ absorption features towards strong continuum sources such as \hii\ regions (see \citealt{anderson2009a} for details and comparison with the \hi\ self-absorption method). The technique works on the principle that \hi\ gas associated with the molecular clouds that contain an \hii\ region will produce an absorption dip at the same velocity as the cloud. There is a high likelihood that numerous \hi\ clouds will lie along the line of sight towards a distant continuum source and so \hi\ spectra towards sources at far distances will show absorption features at all intervening velocities, including at or near the tangent velocity. On the other hand, \hi\ spectra towards \hii\ regions at near distances will show only absorption features at velocities between the velocity of the \hii\ region and the observer.

We identified 22 sources for which this method could be reliably employed and followed the method defined by \citet{kolpak2003} and the procedures described in \citet{urquhart2012}. We provide the continuum images and 21-cm emission profiles in Fig.\,11 and give the assigned distance in Table\,\ref{tbl:derived_clump_para}.

\subsubsection{\hi\ self-absorption}

This technique is similar to the absorption against a continuum method in that objects located at the near distance are associated with extended envelopes of cool \hi\ that absorb the background emission from warmer \hi\ in the interstellar medium at the same radial velocity as the target. However, for sources located at the far distance there is effectively no background \hi\ to absorb and no absorption signature is seen. This method can be used for the \hii\ regions that do not produce sufficient continuum emission to be detected in the VGPS/SGPS data but is inherently less secure as there are frequently multiple absorption components, most of which are extremely weak against the \hi\ background ($\sim$80\,per\,cent reliability; \citealt{busfield2006}, see also discussion in \citealt{anderson2009a}). However, in the absence of more accurate distance measurements and strong continuum emission it is necessary to use this method.

A comprehensive description of this method can be found in \citealt{jackson2003} and \citet{roman2009}. We provide the \hi\ profiles in Fig.\,12 and give the assigned distance in Table\,\ref{tbl:derived_clump_para}. Using this method we are able to resolve the distance ambiguities for 23 of the remaining 24 sources. The only source we have been unable to resolve the distance to is AGAL017.986+00.126.

\begin{figure}
\begin{center}

\includegraphics[width=0.49\textwidth, trim= 0 0 0 0]{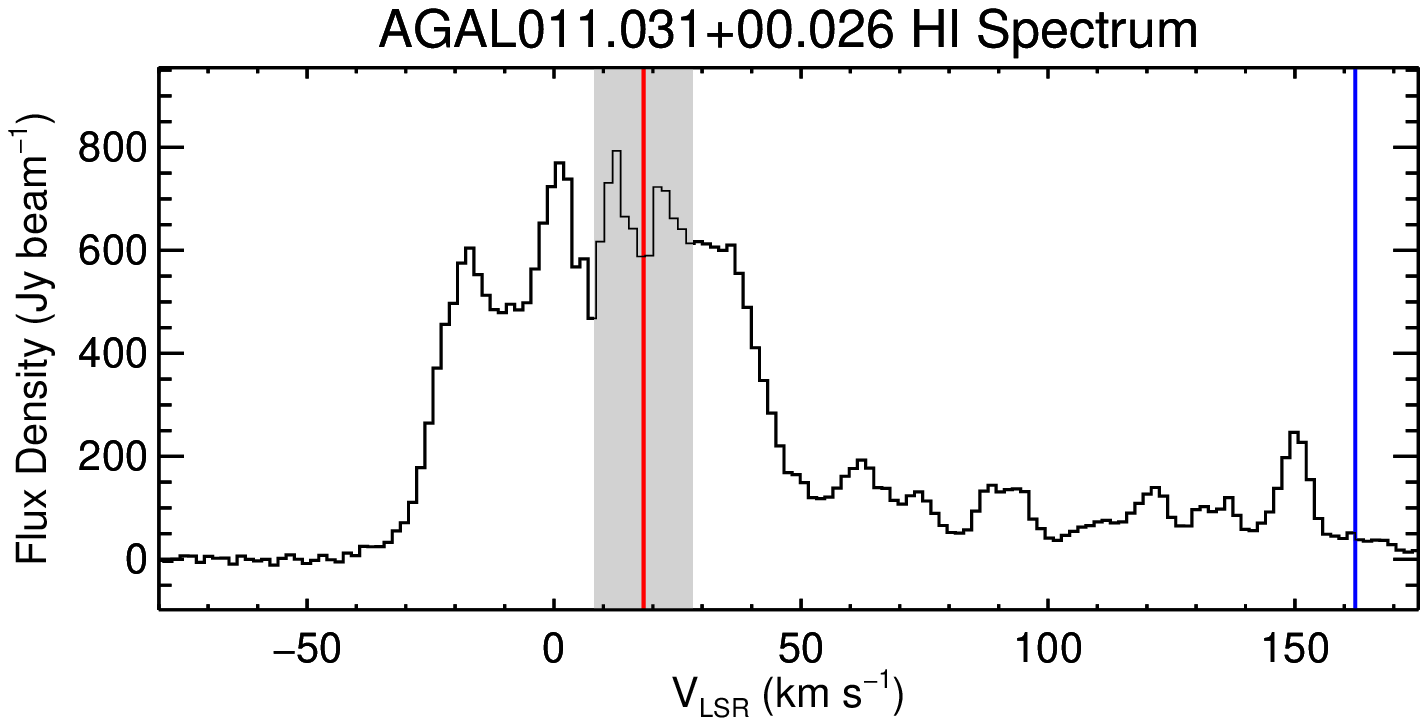}
\includegraphics[width=0.49\textwidth, trim= 0 0 0 0]{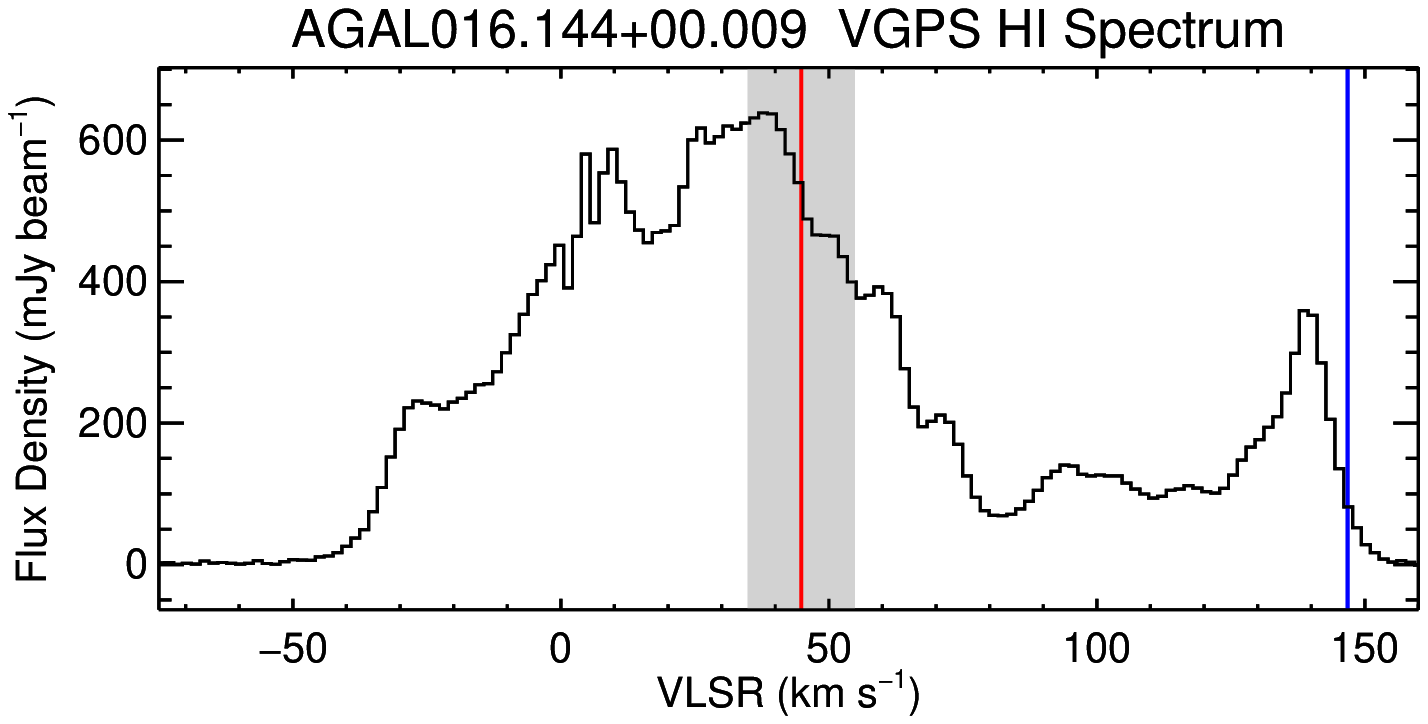}

\caption{\label{fig:hisa_spectral_data} Example \hi\ spectra extracted from the SGPS and VGPS archives. In the upper and lower panel we show and example of a source located at the near and far distance, respectively. The vertical red and blue lines are as described in Fig.\,11. }

\end{center}
\end{figure}

\setcounter{figure}{12}

\subsection{Distance distribution}
\label{sect:distance_distribution}

Using a combination of maser parallax, spectroscopic and kinematic distances drawn from the literature and discussed in the previous two subsections we have compiled distances for 169 of the 170 clumps identified in Sect.\,\ref{sect:matching_stats}.

\begin{figure}
\begin{center}
\includegraphics[width=0.49\textwidth, trim= 0 0 0 0]{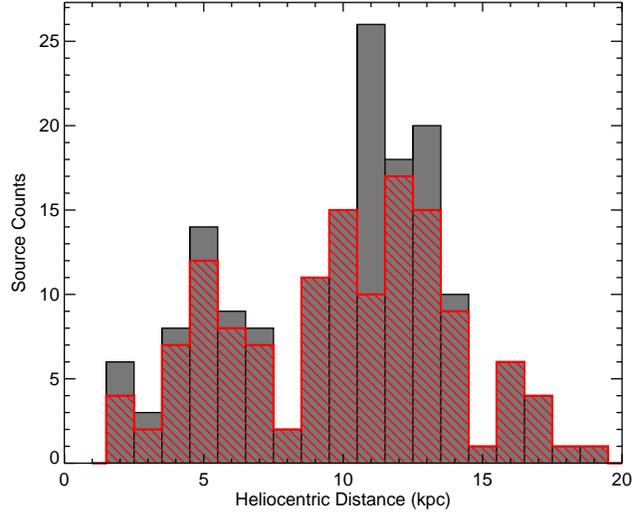}

\caption{\label{fig:atlas_radio_distance_hist} Heliocentric distance distribution for the 212 \hii\ regions and the 169 associated clumps are shown by the filled grey and red hatched histograms, respectively. The strong peak seen in the \hii\ region distribution at $\sim$11\,kpc is due to the presence of the W49A star forming complex in this bin. The bin size used is 1\,kpc.} 

\end{center}
\end{figure}

Fig.\,\ref{fig:atlas_radio_distance_hist} shows the distribution of \hii\ regions (grey filled histogram) and their host clumps (red hatched histogram) as a function of heliocentric distance. The two samples show a similar distribution with five peaks located at 2, 5, 10, 12 and 16\,kpc.  These roughly coincide with the distances of, respectively, the near side of the Sagittarius arm, the end of the bar and Scutum-Centaurus arm, the far sides of the Sagittarius and Perseus arms (these two structures are somewhat blended), and the Norma spiral arm. The two distributions differ in source counts at $\sim$11\,kpc; this is the result of multiple \hii\ regions associated with a single clump where more intense star formation is taking place. The large difference in number of \hii\ regions and clumps seen in this bin is almost entirely due to the W49A star forming complex. 

The distribution has a mean and median distance of 9.2$\pm$0.3\,kpc and 9.8\,kpc, respectively, and therefore the majority of the sample is located on the far side of the tangent position. This is consistent with the peak in the latitude source counts being slightly positive as discussed in Sect.\,\ref{sect:lb_distribution}, and their apparent anti-correlation with regions of intense CO emission discussed in Sect.\,\ref{sect:gal_long_vel}.

\setlength{\tabcolsep}{3pt}

%\begin{landscape}
\begin{table*}
%Select replace(t1.atlas_name,'-','$-$'),'\&',if(t1.complex_name is null,'\\multicolumn{1}{c}{$\\cdots$}',t1.complex_name),'\&',count(t2.atlas_name),'\&', round(t1.elongation,1),'\&',round(t1.submm_flux_peak,2),'\&',round(t1.submm_flux_int,2),'\&',round(t1.submm_flux_int/t1.submm_flux_peak,2),'\&',round(t1.vlsr,1),'\&', replace(vlsr_reference_flag,' ',''),'\&',if(t1.bb_distance is null,'\\multicolumn{1}{c}{$\\cdots$}',round(t1.bb_distance,1)),'\&',if(t1.bb_distance is null,'\\multicolumn{1}{c}{$\\cdots$}',distance_reference_flag),'\&',round(if(rgc_paper is null, rgc,rgc_paper),1) as rgc,'\&',if(t1.bb_distance is null,'\\multicolumn{1}{c}{$\\cdots$}',if(radius_sign like '<', concat('\\multicolumn{1}{c}{<',radius_paper,'}'), radius_paper)) as radius,'\&',round(log10(t1.dust_col_den_jochen),2) as col_den,'\&',if(t1.bb_distance is null,'\\multicolumn{1}{c}{$\\cdots$}',round(log10(dust_mass),2)) as dust_mass,'\&',if(t1.virial_mass is null,'\\multicolumn{1}{c}{$\\cdots$}',round(log10(t1.virial_mass),2)) as virial_mass,'\\\\' from cornish.atlas_cornish_matches_sex_distances as t1 left join cornish.atlas_cornish_matches_sex as t2 on t1.atlas_name=t2.atlas_name and uchii_flag is not null group by t1.atlas_name limit 30

\begin{center}\caption{Derived clump parameters.}
\label{tbl:derived_clump_para}
\begin{minipage}{\linewidth}
\scriptsize
\begin{tabular}{lcc......c.c....}
\hline \hline
  \multicolumn{1}{c}{ATLASGAL}&  \multicolumn{1}{c}{Complex}&	\multicolumn{1}{c}{\hii}  &\multicolumn{1}{c}{Aspect}  &\multicolumn{1}{c}{Peak Flux}  &\multicolumn{1}{c}{Int. Flux}  &\multicolumn{1}{c}{$Y$-factor}  &\multicolumn{2}{c}{V$_{\rm{LSR}}$}   &	\multicolumn{2}{c}{Distance} &\multicolumn{1}{c}{R$_{\rm{GC}}$}&\multicolumn{1}{c}{Radius}&\multicolumn{1}{c}{Log(N(H$_2$))} &	\multicolumn{1}{c}{Log($M_{\rm{clump}}$)} &	\multicolumn{1}{c}{Log($M_{\rm{vir}}$)} \\
  
    \multicolumn{1}{c}{Name }&  \multicolumn{1}{c}{Name }&  \multicolumn{1}{c}{Density} & \multicolumn{1}{c}{Ratio}  &\multicolumn{1}{c}{(Jy beam$^{-1}$)}  &\multicolumn{1}{c}{(Jy)}  &\multicolumn{1}{c}{ }  &	\multicolumn{1}{c}{(km\,s$^{-1}$)}&\multicolumn{1}{c}{Ref. }&	\multicolumn{1}{c}{(kpc)} &\multicolumn{1}{c}{Ref. }&\multicolumn{1}{c}{(kpc)}&\multicolumn{1}{c}{(pc)}&\multicolumn{1}{c}{(cm$^{-2}$)} &	\multicolumn{1}{c}{(\msun)} &	\multicolumn{1}{c}{(\msun)} \\
\hline
AGAL010.299$-$00.147	&	W31-North	&	1	&	1.9	&	7.67	&	54.18	&	7.06	&	12.8	&	1	&	2.4	&	1	&	6.2	&	0.79	&	23.18	&	3.23	&	3.44	\\
AGAL010.321$-$00.257	&	W31-South	&	1	&	2.8	&	2.39	&	14.93	&	6.24	&	32.2	&	1	&	3.5	&	1	&	5.1	&	1.22	&	22.87	&	3.02	&	3.28	\\
AGAL010.472+00.027	&	\multicolumn{1}{c}{$\cdots$}	&	2	&	1.5	&	35.01	&	88.12	&	2.52	&	67.0	&	1,2,3	&	11.0	&	2,3	&	3.1	&	2.99	&	23.94	&	4.77	&	4.30	\\
AGAL010.624$-$00.384	&	W31-North	&	4	&	1.4	&	33.10	&	116.87	&	3.53	&	-3.2	&	1,2,3	&	2.4	&	1	&	6.2	&	0.76	&	23.81	&	3.57	&	3.61	\\
AGAL010.957+00.022	&	\multicolumn{1}{c}{$\cdots$}	&	1	&	1.8	&	3.03	&	13.55	&	4.48	&	21.4	&	2,3	&	13.7	&	2,3	&	5.6	&	2.82	&	22.81	&	4.15	&	3.77	\\
AGAL010.964+00.011	&	\multicolumn{1}{c}{$\cdots$}	&	1	&	1.4	&	1.33	&	5.40	&	4.07	&	19.2	&	2,3	&	2.7	&	4	&	5.8	&	0.2	&	22.53	&	2.35	&	2.85	\\
AGAL011.031+00.026	&	\multicolumn{1}{c}{$\cdots$}	&	1	&	1.1	&	0.47	&	1.39	&	2.95	&	18.1	&	4	&	2.6	&	4	&	5.9	&	\multicolumn{1}{c}{$<$0.29}	&	22.22	&	1.72	&	\multicolumn{1}{c}{$\cdots$}	\\
AGAL011.034+00.061	&	\multicolumn{1}{c}{$\cdots$}	&	1	&	1.5	&	1.77	&	6.05	&	3.41	&	15.4	&	4	&	14.7	&	5	&	6.6	&	1.7	&	22.66	&	3.86	&	\multicolumn{1}{c}{$\cdots$}	\\
AGAL011.109$-$00.397	&	\multicolumn{1}{c}{$\cdots$}	&	1	&	1.9	&	3.53	&	42.04	&	11.92	&	0.1	&	2,3	&	16.8	&	4	&	8.6	&	6.01	&	23.04	&	4.81	&	4.13	\\
AGAL011.902$-$00.141	&	\multicolumn{1}{c}{$\cdots$}	&	2	&	1.2	&	3.17	&	20.03	&	6.32	&	38.4	&	1	&	4.1	&	6	&	4.6	&	0.99	&	23.12	&	3.27	&	3.44	\\
AGAL011.936$-$00.616	&	\multicolumn{1}{c}{$\cdots$}	&	1	&	1.2	&	7.22	&	45.64	&	6.32	&	37.0	&	1	&	4.0	&	2,3,6	&	4.6	&	1.04	&	23.33	&	3.61	&	3.27	\\
AGAL011.946$-$00.036	&	\multicolumn{1}{c}{$\cdots$}	&	1	&	1.8	&	1.75	&	7.62	&	4.36	&	39.1	&	3,4	&	12.6	&	4	&	4.7	&	1.98	&	22.65	&	3.83	&	\multicolumn{1}{c}{$\cdots$}	\\
AGAL012.198$-$00.034	&	\multicolumn{1}{c}{$\cdots$}	&	1	&	1.1	&	2.92	&	7.42	&	2.54	&	50.8	&	1,2,3	&	11.9	&	6	&	4.0	&	1.32	&	22.97	&	3.76	&	3.37	\\
AGAL012.208$-$00.102	&	\multicolumn{1}{c}{$\cdots$}	&	1	&	1.2	&	11.58	&	37.39	&	3.23	&	24.5	&	1	&	13.6	&	5	&	5.6	&	2.78	&	23.38	&	4.58	&	4.26	\\
AGAL012.431$-$00.049	&	\multicolumn{1}{c}{$\cdots$}	&	1	&	2.4	&	0.72	&	5.71	&	7.95	&	21.1	&	1	&	13.9	&	7	&	5.9	&	1.41	&	22.39	&	3.78	&	3.07	\\
AGAL012.431$-$01.114	&	\multicolumn{1}{c}{$\cdots$}	&	1	&	1.6	&	4.28	&	14.73	&	3.44	&	39.7	&	1,2,3	&	4.1	&	4	&	4.6	&	0.62	&	23.00	&	3.14	&	2.85	\\
AGAL012.804$-$00.199	&	W33	&	2	&	1.9	&	36.48	&	339.23	&	9.30	&	36.5	&	4	&	2.1	&	8	&	6.5	&	1.04	&	23.97	&	3.92	&	\multicolumn{1}{c}{$\cdots$}	\\
AGAL012.998$-$00.357	&	\multicolumn{1}{c}{$\cdots$}	&	1	&	1.3	&	2.29	&	9.86	&	4.32	&	14.4	&	1	&	1.9	&	4	&	6.6	&	0.3	&	22.87	&	2.32	&	2.85	\\
AGAL013.209$-$00.144	&	\multicolumn{1}{c}{$\cdots$}	&	1	&	1.3	&	3.09	&	30.64	&	9.90	&	51.1	&	1,3	&	4.6	&	5	&	4.2	&	1.2	&	23.05	&	3.55	&	3.08	\\
AGAL013.384+00.064	&	\multicolumn{1}{c}{$\cdots$}	&	1	&	1.4	&	1.11	&	5.45	&	4.92	&	14.1	&	4	&	1.9	&	4	&	6.7	&	0.26	&	22.45	&	2.02	&	\multicolumn{1}{c}{$\cdots$}	\\
AGAL013.872+00.281	&	\multicolumn{1}{c}{$\cdots$}	&	1	&	1.3	&	5.52	&	29.72	&	5.39	&	48.5	&	1	&	4.4	&	2	&	4.4	&	1.1	&	23.13	&	3.50	&	3.35	\\
AGAL014.246$-$00.071	&	\multicolumn{1}{c}{$\cdots$}	&	1	&	1.5	&	2.98	&	11.43	&	3.84	&	60.2	&	1	&	11.6	&	4	&	4.0	&	2.05	&	23.06	&	3.93	&	3.59	\\
AGAL014.607+00.012	&	\multicolumn{1}{c}{$\cdots$}	&	1	&	1.6	&	2.74	&	16.62	&	6.06	&	25.5	&	1,2,3	&	2.8	&	4	&	5.8	&	0.58	&	22.87	&	2.86	&	2.95	\\
AGAL014.777$-$00.334	&	\multicolumn{1}{c}{$\cdots$}	&	1	&	1.1	&	0.95	&	4.02	&	4.24	&	33.2	&	4	&	13.1	&	4	&	5.3	&	1.16	&	22.39	&	3.58	&	\multicolumn{1}{c}{$\cdots$}	\\
AGAL016.144+00.009	&	\multicolumn{1}{c}{$\cdots$}	&	1	&	1.1	&	0.99	&	2.51	&	2.53	&	44.8	&	2,3	&	12.4	&	4	&	4.9	&	1.08	&	22.62	&	3.33	&	3.04	\\
AGAL016.942$-$00.072	&	\multicolumn{1}{c}{$\cdots$}	&	1	&	1.1	&	1.43	&	6.94	&	4.86	&	-4.2	&	4	&	17.0	&	4	&	9.2	&	2.04	&	22.56	&	4.05	&	\multicolumn{1}{c}{$\cdots$}	\\
AGAL017.029$-$00.071	&	\multicolumn{1}{c}{$\cdots$}	&	1	&	1.2	&	1.27	&	3.42	&	2.69	&	93.2	&	1	&	10.4	&	6	&	3.4	&	0.93	&	22.63	&	3.31	&	3.01	\\
AGAL017.112$-$00.114	&	\multicolumn{1}{c}{$\cdots$}	&	1	&	1.8	&	0.62	&	1.82	&	2.93	&	93.1	&	2,3	&	10.4	&	6	&	3.4	&	\multicolumn{1}{c}{$<$1.17}	&	22.17	&	3.04	&	\multicolumn{1}{c}{$\cdots$}	\\
AGAL017.554+00.167	&	\multicolumn{1}{c}{$\cdots$}	&	1	&	1.2	&	0.52	&	2.38	&	4.60	&	20.6	&	4	&	14.1	&	4	&	6.5	&	1.22	&	22.12	&	3.42	&	\multicolumn{1}{c}{$\cdots$}	\\
AGAL017.986+00.126	&	\multicolumn{1}{c}{$\cdots$}	&	1	&	1.6	&	0.42	&	1.27	&	3.04	&	24.3	&	5	&	\multicolumn{1}{c}{$\cdots$}	&	\multicolumn{1}{c}{$\cdots$}	&	6.3	&	\multicolumn{1}{c}{$\cdots$}	&	22.17	&	\multicolumn{1}{c}{$\cdots$}	&	\multicolumn{1}{c}{$\cdots$}	\\
\hline\\
\end{tabular}\\
\normalsize
References --- Velocity: (1) \citet{wienen2012}, (2) \citet{urquhart2011b}, (3) \citet{urquhart_13co_north}, (4) \citet{schlingman2011}, (5) this paper, (6) \citet{dunham2011b}, (7) Csengeri et al. in prep., (8) \citet{kolpak2003}, (9) \citet{blum2000}, (10) Radio recombination line velocity from HOPS (\citealt{walsh2011}; priv. comm. A. Walsh), (11) \citet{bronfman1996}, (12) \citet{anderson2009a}  

References --- Distance: (1) \citet{moises2011}, (2) \citet{sewilo2003}, (3) \citet{pandian2009}, (4) this paper, (5) \citet{pandian2008}, (6) \citet{green2011b}, (7) \citet{wood1989b}, (8) \citet{Immer2012}, (9) \citet{roman2009}, (10) \citet{anderson2009a}, (11) \citet{kolpak2003}, (12) \citet{dunham2011}, (13) \citet{fish2003}, (14) \citet{russeil2003}, (15) \citet{downes1980}, (16) \citet{xu2011}, (17) \citet{araya2002}, (18) \citet{watson2003}, (19) \citet{stead2010}, (20) \citet{nagayama2011}, (21) \citet{sato2010b}, (22) \citet{urquhart2012}, (23) Zhang et al. (2013) %\citet{zhang2013}

Notes: Only a small portion of the data is provided here, the full table is available in electronic form at the CDS via anonymous ftp to cdsarc.u-strasbg.fr (130.79.125.5) or via http://cdsweb.u-strasbg.fr/cgi-bin/qcat?J/MNRAS/.
\end{minipage}

\end{center}
\end{table*}
%\end{landscape}
\setlength{\tabcolsep}{6pt}

\section{Clump properties}

In this section we will concentrate on the physical properties of the molecular clumps within which the compact and \uchii\ regions are embedded.  We presented a detailed description of the derivation of many of these parameters and the associated caveats in Paper\,I  and so we only provide a brief outline here. The parameter values for individual clumps are presented in Table\,\ref{tbl:derived_clump_para}, while in Table\,\ref{tbl:derived_para} we summarise the global properties of the sample.

\begin{figure}
\begin{center}
\includegraphics[width=0.49\textwidth, trim= 0 0 0 0]{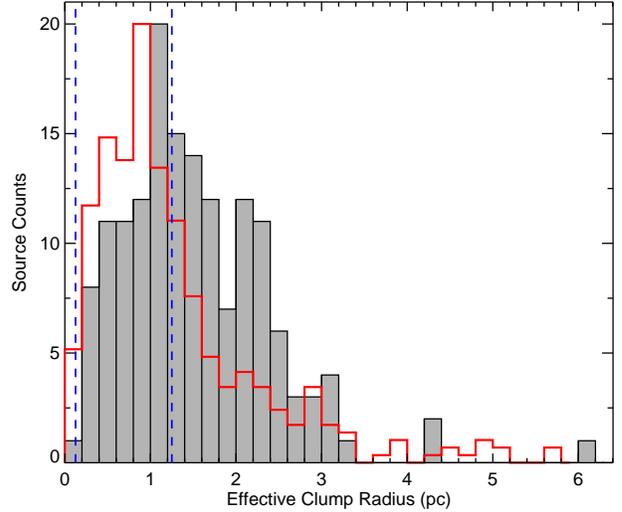}

\caption{\label{fig:physical_size} Distribution of the effective radii of the \hii -region associated clumps (filled grey histogram) and that of methanol-maser associated clumps (red). The methanol maser distribution has been scaled to the peak of the \hii-region associated clump distribution. The dashed vertical lines indicate the radii separating cores from clumps (0.125\,pc), and clumps from clouds (1.25\,pc), respectively. The bin size is 0.2\,pc.} 

\end{center}
\end{figure}

\subsection{Clump radius}
\label{sect:size}

We have estimated the physical radius of the each clump using the heliocentric distances derived in the previous section and their effective radius. In Fig.\,\ref{fig:physical_size} we show the distribution of effective radii for the \hii\ region associated clumps (grey histogram) and the methanol maser associated clumps discussed in Paper\,I (blue histogram). The dashed vertical lines mark radii of 0.15 and 1.25\,pc, indicating the nominal boundary between cores and clumps, and clumps and clouds, respectively (e.g., \citealt{bergin2007}). However, as also noted in Paper\,I, neither sample of clumps shows a sharp break in its distribution at these sizes and so these definitions are probably somewhat arbitrary.  The clumps in this sample appear to be rather spherical in structure, are centrally condensed and the vast majority are associated with a single \hii\ region. It is likely these clumps are in the process of forming single clusters of stars and we will therefore continue to refer to them as clumps.

In general the radii of the clumps hosting \hii\ regions are significantly larger than those associated with methanol masers (median vales of $\sim$1.38\,pc and 0.97\,pc, respectively). The \hii\ region and methanol maser samples of massive star forming clumps are largely drawn from different Quadrants of the Galaxy (First and Fourth Quadrants, respectively). It is likely that at least some of the difference in physical size is simply a result of the different spiral structures dominating these Quadrants. To test this we performed a KS test on a distance limited sample of methanol and \hii\ region associated clumps located between 4 and 6\,kpc and find we are unable to reject the null hypothesis that these two samples are drawn from the same parent population ($p$-value = 0.4).

The distributions of both samples peak at radii of $\sim$1\,pc and it is therefore likely that the embedded \hii\ regions and methanol masers are tracing the most massive member of a young proto-cluster. The derived clump properties, bolometric luminosities and Lyman continuum fluxes are therefore likely to be related to the embedded cluster rather than a single star and we need to bear this in mind when interpreting these results.

\subsection{Isothermal clump masses and column densities}
\label{sect:mass}

\begin{figure*}
\begin{center}

\includegraphics[width=0.49\textwidth, trim= 0 0 0 0]{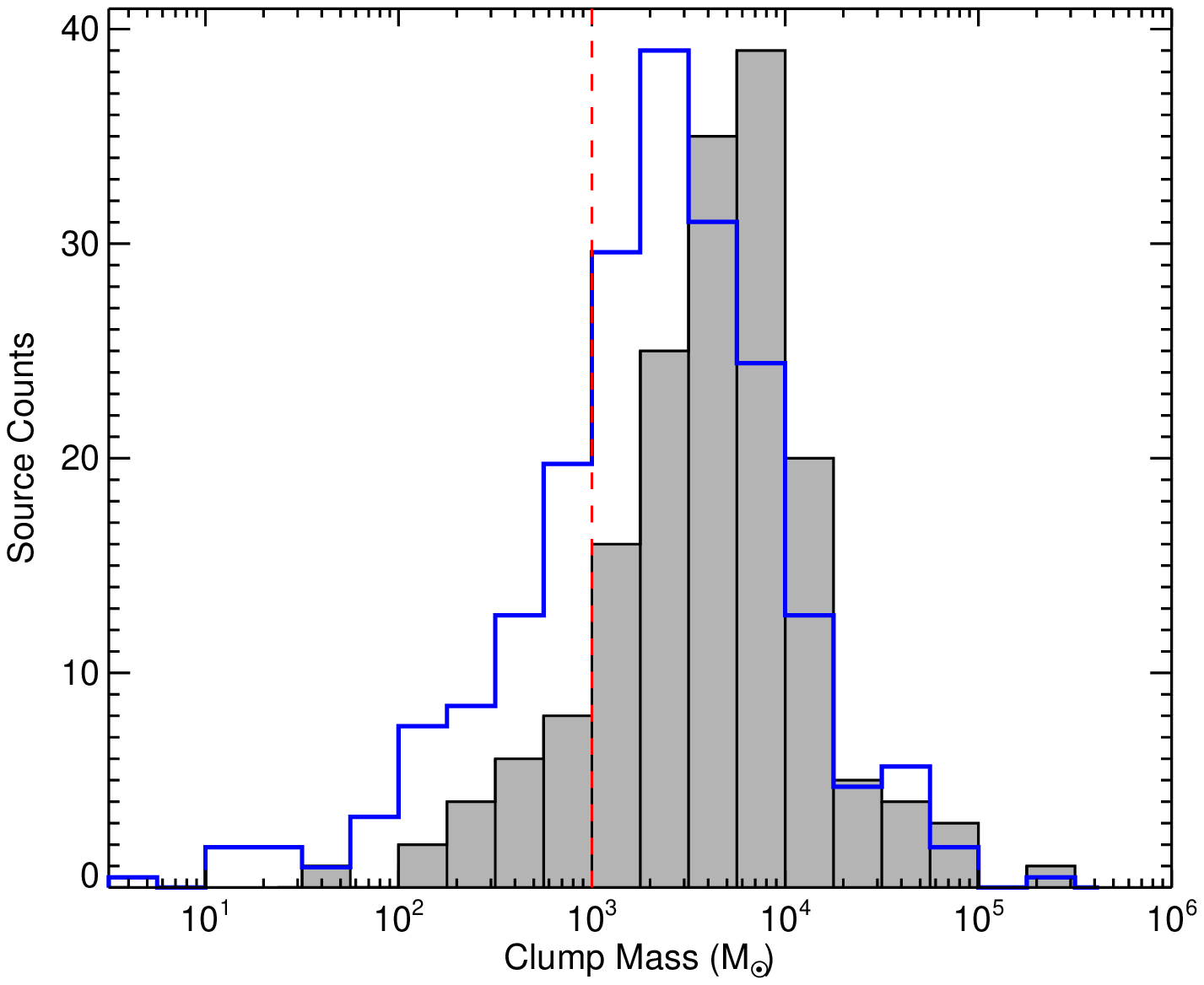}
\includegraphics[width=0.49\textwidth, trim= 0 0 0 0]{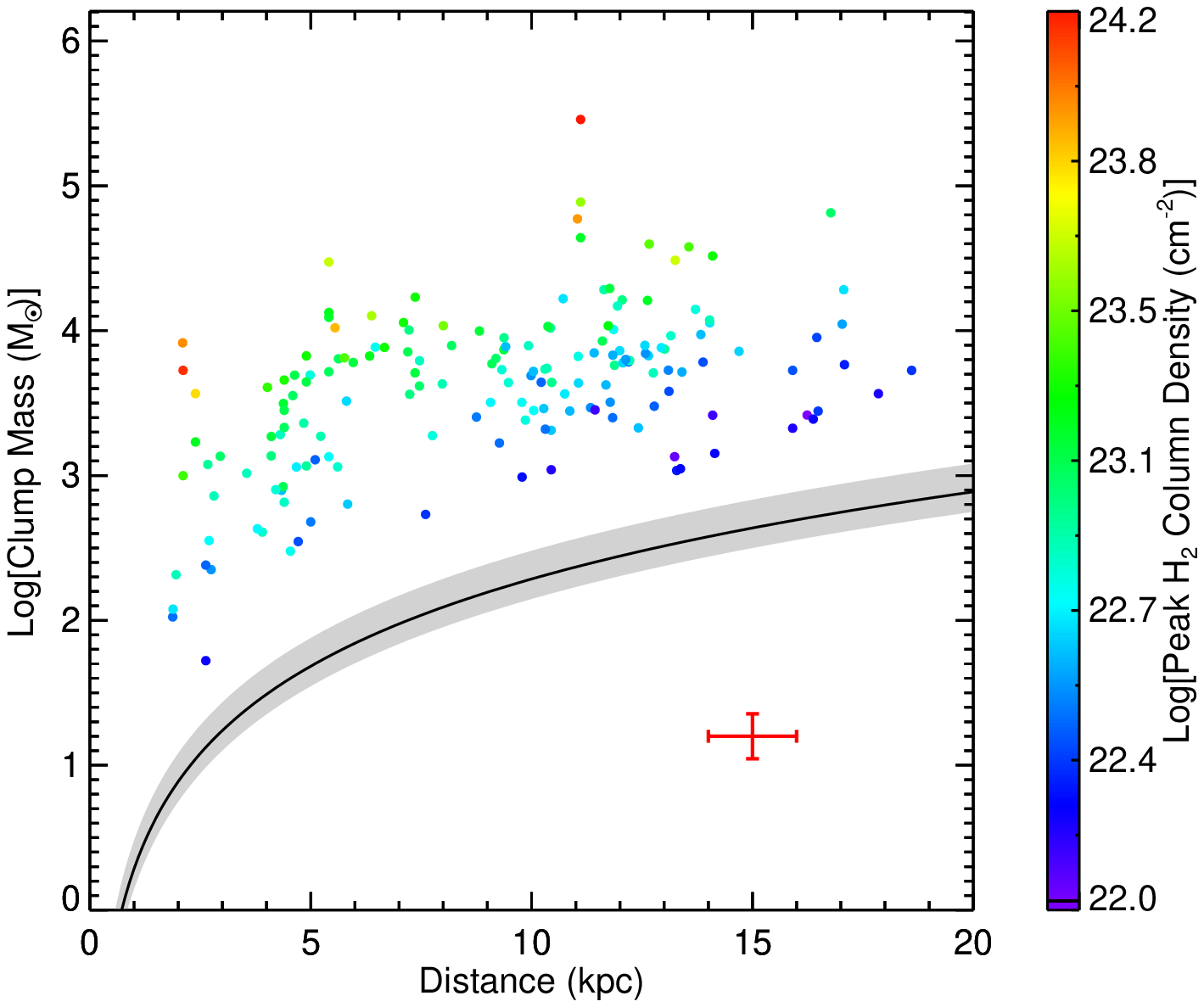}

\caption{\label{fig:clump_mass} Left panel: The isothermal clump-mass distribution of the \hii -region and the methanol-maser associated clumps, shown as filled grey and blue histograms, respectively. A dust temperature of 20\,K has been assumed for both samples. The vertical red line indicates the completeness limit (see text for details). The bin size is 0.5 dex. Right panel: Mass distribution of \hii -region associated clumps as a function of heliocentric distance. The colours indicate the peak column density of each source (see colour bar on the right for values) and the solid black line and the grey filled region indicated the mass sensitivity limit of the survey and its associated uncertainty assuming a dust temperature of 20$\pm5$\,K. Characteristic error bars for these parameters are shown in the lower right corner of the plot.}

\end{center}
\end{figure*}

We estimate isothermal clump masses following the method of \citet{hildebrand1983} and assuming that the total clump mass is proportional to the total flux density integrated over the source such that:

\begin{equation}
\left(\frac{M_{\rm{clump}}}{\rm{M}_\odot}\right) \, = \, \left(\frac{D}{\rm{kpc}}\right)^2 \, \left(\frac{S_\nu}{\rm{mJy}}\right) \, \frac{R}{B_\nu(T_{\rm{dust}}) \, \kappa_\nu},
\end{equation}

%$R$ {\color{green} Moore: R is also used for radii}

\noindent where $S_\nu$ is the integrated 870\,$\umu$m flux, $D$ is the heliocentric distance to the source, $R$ is the gas-to-dust mass ratio (assumed to be 100), $B_\nu$ is the Planck function for a dust temperature $T_{\rm{dust}}$, and $\kappa_\nu$ is the dust absorption coefficient taken as 1.85\,cm$^2$\,g$^{-1}$, which was derived by \citet{schuller2009} by interpolating to 870\,$\umu$m from Table\,1, Col.\,5 of \citet{ossenkopf1994}.

For consistency with the masses determined in Paper\,I, we assume a dust temperature of 20\,K. Although we have seen in Sect.\,\ref{sect:linewidths_temps} the gas temperatures range between 20 and 30\,K, the ammonia observations were made towards the peak \submm\ emission, and given that the majority of the dust emission comes from the cooler extended envelope (i.e., mean $Y$-factor $\sim$5; see Sect.\,\ref{sect:size} for details) a lower kinetic temperature is considered more appropriate. Moreover, the choice of 20\,K facilitates comparison of the derived masses with similar studies that have used the same temperature (e.g., \citealt{motte2007,hill2005}).

We estimate column densities from the peak flux density of the clumps using the following equation:

\begin{equation}
\left(\frac{N_{\rm{H}_2}}{\rm{cm}^{-2} }\right) \, = \, \left(\frac{S_\nu}{\rm{mJy}}\right) \, \frac{R}{B_\nu(T_{\rm{dust}}) \, \Omega \, \kappa_\nu \, \mu
\, m_{\rm{H}}},
\end{equation}

\noindent where $\Omega$ is the beam solid angle, $\umu$ is the mean molecular weight of the gas, which we assume to be equal to 2.8 (\citealt{kauffmann2008}), $m_{\rm{H}}$ is the mass of the hydrogen atom, and $\kappa_\nu$ and $R$ are as previously defined. We again assume a dust temperature of 20\,K.

The uncertainties in the derived clump mass and column density are both $\sim$70\,per\,cent. When estimating this error we assume an uncertainty of $\pm$5\,K for the gas temperature and a $\sim$10\,per\,cent error in the distance, which we add in quadrature with the absolute flux measurement uncertainty (15\,per\,cent; \citealt{schuller2009}). Although quite large, these errors are unlikely to have a significant impact on the overall distribution or the statistical analysis of these parameters.
 
In the left panel of Fig.\,\ref{fig:clump_mass} we present the isothermal gas mass distribution for both the \hii\ region and methanol maser associated clumps, while in the right panel we show the mass distribution for only the \hii\ region associated clumps as a function of heliocentric distance. It is clear from the latter that we are sensitive to all of the \hii\ region associated clumps with masses above 1,000\,\msun\ across the Galaxy within the observed region, and our statistics should be complete above this level; this was also the case for the methanol maser associated clumps. (This completeness limit is indicated on the left panel of Fig.\,\ref{fig:clump_mass} by the vertical dashed red line.) 

There are a small number of \hii\ regions that are associated with clumps below the mass expected for the formation of a massive star (i.e., $<$1,000\,\msun), which may indicate that the distances to these sources have been incorrectly assigned. However, checking the distance assignments for these lower mass sources ($<$300\,\msun) we find them to be correct given the quality of the data at hand (mostly VGPS data with a angular resolution of $\sim$1\arcmin). We note that these lower mass clumps are also the most compact  ($R_{\rm{eff}} < 0.2$\,pc) and therefore may be forming individual or small groups of stars where the IMF is not expected to apply.

The compact and \uchii\ associated clump mass distribution has a mean and median value of $\sim$10,000\,\msun\ and $\sim$5,000\,\msun, respectively, which are both significantly larger than the completeness limit (see left panel of Fig.\,\ref{fig:clump_mass}) and so the drop off in source counts between the completeness limit and the peak is very likely to be a genuine feature of the distribution. We also note that there are significant differences between the mass distributions of the methanol maser and \hii\ region associated clumps. The masses of the methanol maser associated clumps peak at a few thousand \msun, which is also above the completeness limit, but a factor of two lower than found for the \hii-region associated clumps. However, as discussed in the previous subsection these two samples are drawn from different regions of the Galaxy and have different distance distributions as a result.

To mitigate this possible distance bias we again perform a KS test on a distance limited sample of \hii-region and methanol-maser associated clumps (between 4-6\,kpc); this reveals no significant difference in the mass distribution of the two samples of clumps. We therefore conclude that the differences in the mass and radii distributions is likely to be a result of the different Galactic structure present in the First and Fourth Quadrants from which these samples are primarily drawn (e.g., \citealt{georgelin1976}).

\subsection{Virial mass}

To assess the gravitational stability of these clumps we compare thermal masses and virial masses for the 123 resolved clumps towards which ammonia (1,1) emission is been detected using:

\begin{equation}
\left (\frac{M_{\rm{vir}}}{\rm{M}_\odot} \right) \; = \; \frac{5}{8ln2} \frac{1}{a_1 a_2 G} \left(\frac{R_{\rm{eff}}}{\rm{pc}}\right) \left(\frac{\Delta v_{\rm{avg}}}{\rm{km\,s}^{-1}}\right)^{2}
\end{equation}
\noindent where $R_{\rm{eff}}$ is the effective radius of the clump, $\Delta v_{\rm{avg}}$ is the velocity dispersion in the gas calculated from the measured line width via Eqn.\,5, $G$ is the gravitational constant and $a_1$ and $a_2$ are corrections for the assumptions of uniform density and spherical geometry, respectively ({\citealt{bertoldi1992}). For aspect ratios less than 2 the value $a_2 \sim 1$, which is suitable for this sample of clumps, and the value of $a_1$ is given by:

\begin{equation}
a_1 \; = \; \frac{1-p/3}{1-2p/5} \; {\rm{for}} \; p < 2.5
\end{equation}

\noindent where $p$ is the power-law index of the density profile where [$n(r) \propto r^{-p}$]. We have adopted a value for $p$ of 1.6 (i.e., $a_1=1.3$) as determined for a similar sample of massive star forming clumps by \citet{beuther2002}.

The measured ammonia line width needs to be corrected to estimate the average velocity dispersion of the total column of gas. Following \citealt{fuller1992}:

\begin{equation}
\Delta v^{2}_{\rm{avg}} \; = \; \Delta v^{2}_{\rm{corr}}+8ln2\times \frac{k_{\rm{b}}T_{\rm{kin}}}{m_{\rm{H}}}\left(\frac{1}{\umu_{\rm{p}}}-\frac{1}{\umu_{\rm{NH_3}}}\right)
\end{equation}

\noindent where $\Delta v_{\rm{corr}}$ is the observed NH$_3$ line width corrected for the resolution of the spectrometer (i.e., $\Delta v^2_{\rm{corr}}=\Delta v^{2}_{\rm{obs}}-\Delta v^{2}_{\rm{channel\,width}}$; 0.4\,\kms\ and 0.7\,\kms\ for the GBT and Effelsberg observations, respectively), $k_{\rm{b}}$ is the Boltzmann constant, $T_{\rm{kin}}$ is the kinetic temperature of the gas (again taken to be 20\,K) and $\umu_{\rm{p}}$ and $\umu_{\rm{NH_3}}$  are the mean molecular mass of molecular hydrogen and ammonia, respectively; these are taken as 2.33 (i.e., \citealt{fuller1992}) and 17, respectively. The error in the virial mass is of order 20\,per\,cent allowing for a $\sim$10\,per\,cent error in the distance, the fitted line width, and spectrometer channel width, and the error in the measured source size.

In calculating the virial mass we are making the assumption that the observed NH$_3$ line width is representative of the whole clump, however, \citet{zinchenko1997} found that the line width decreases towards the edges of the clumps hosting embedded protostellar objects. This may lead to the virial mass being overestimated (see \citealt{dunham2011} for a more comprehensive discussion).

\begin{figure}
\begin{center}

\includegraphics[width=0.49\textwidth, trim= 0 0 0 0]{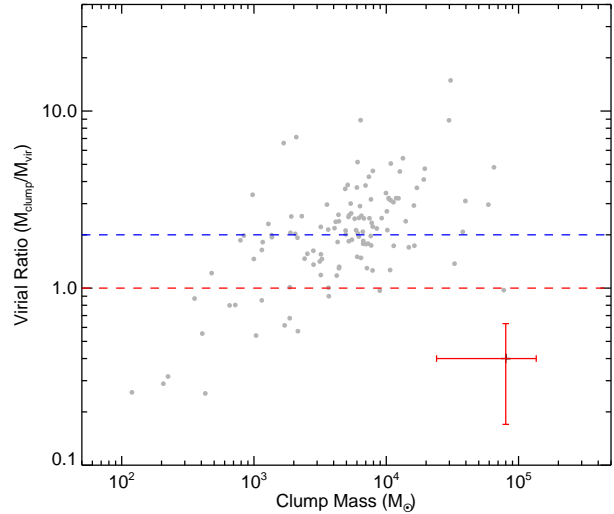}

\caption{\label{fig:virial_mass} Virial ratio ($M_{\rm{clump}}/M_{\rm{vir}}$) as a function of clump mass ($M_{\rm{clump}}$) for the subsample of 121 clumps detected in ammonia (see Sect.\,\ref{sect:vlsr} for details). The red horizontal lines indicate the locus of gravitational equilibrium for thermal and kinematic energies, while the blue dashed lines indicates the locus of gravitational equilibrium assuming an equipartition of kinetic and magnetic energy. Characteristic error bars for these parameters are shown in the lower right corner of the plot.} 

\end{center}
\end{figure}

In Table\,\ref{tbl:derived_clump_para} we present the estimated virial mass for each source and in Fig.\,\ref{fig:virial_mass} we plot the virial ratio ($M_{\rm{clump}}/M_{\rm{vir}}$) versus the clump mass for the subsample of sources with ammonia data. The dashed red horizontal line marks the line of gravitational stability.  Clumps located below this line are likely to be unbound while those above this should may be unstable against gravity. The clumps have a mean and median virial ratio values of 3.45$\pm0.30$ and 3.35, respectively, which are both significantly above unity.

However, the equation used to estimate the virial mass does not take the magnetic field strength into account, which can provide a significant amount of magnetic support to counteract gravity. The magnetic field has not been measured, however, if we assume an equipartition of energy between the kinetic and magnetic energy we find that $M_{\rm{clump}} = 2\times M_{\rm{vir}}$ (e.g., \citealt{bertoldi1992}). The dashed blue horizontal line shown on Fig.\,\ref{fig:virial_mass} indicates the virial ratio for gravitational stability when magnetic support is included. We might therefore expect clumps to be unbound if their virial ratio were $\lesssim$2, however, even allowing for an equal amount of magnetic support we find the mean and median values for the virial ratio are significantly larger than 2, which would suggest that the majority of clumps are likely to be unstable against gravity and collapsing.

\begin{figure*}
\begin{center}

\includegraphics[width=0.49\textwidth, trim= 0 0 0 0]{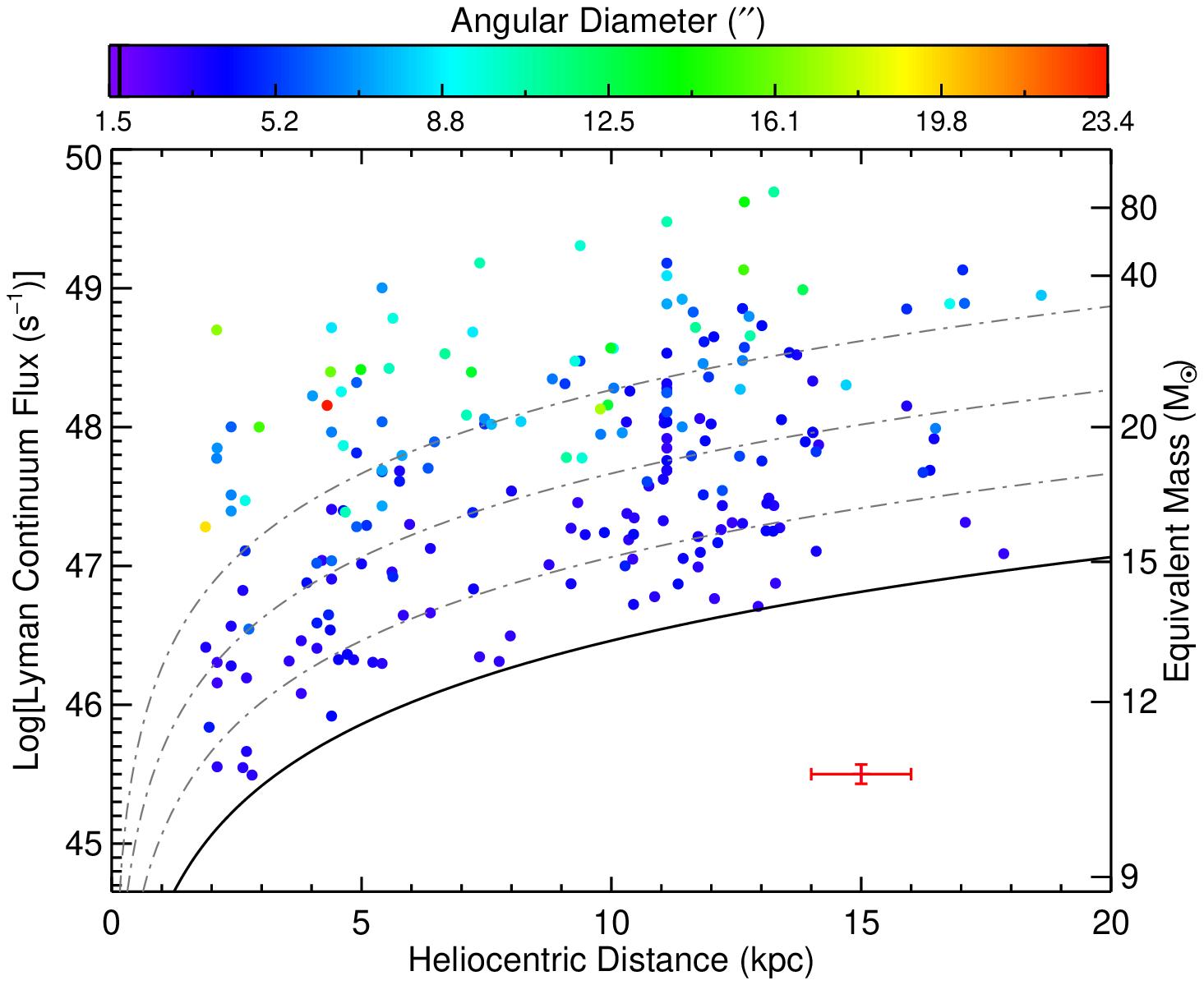}
\includegraphics[width=0.49\textwidth, trim= 0 0 0 0]{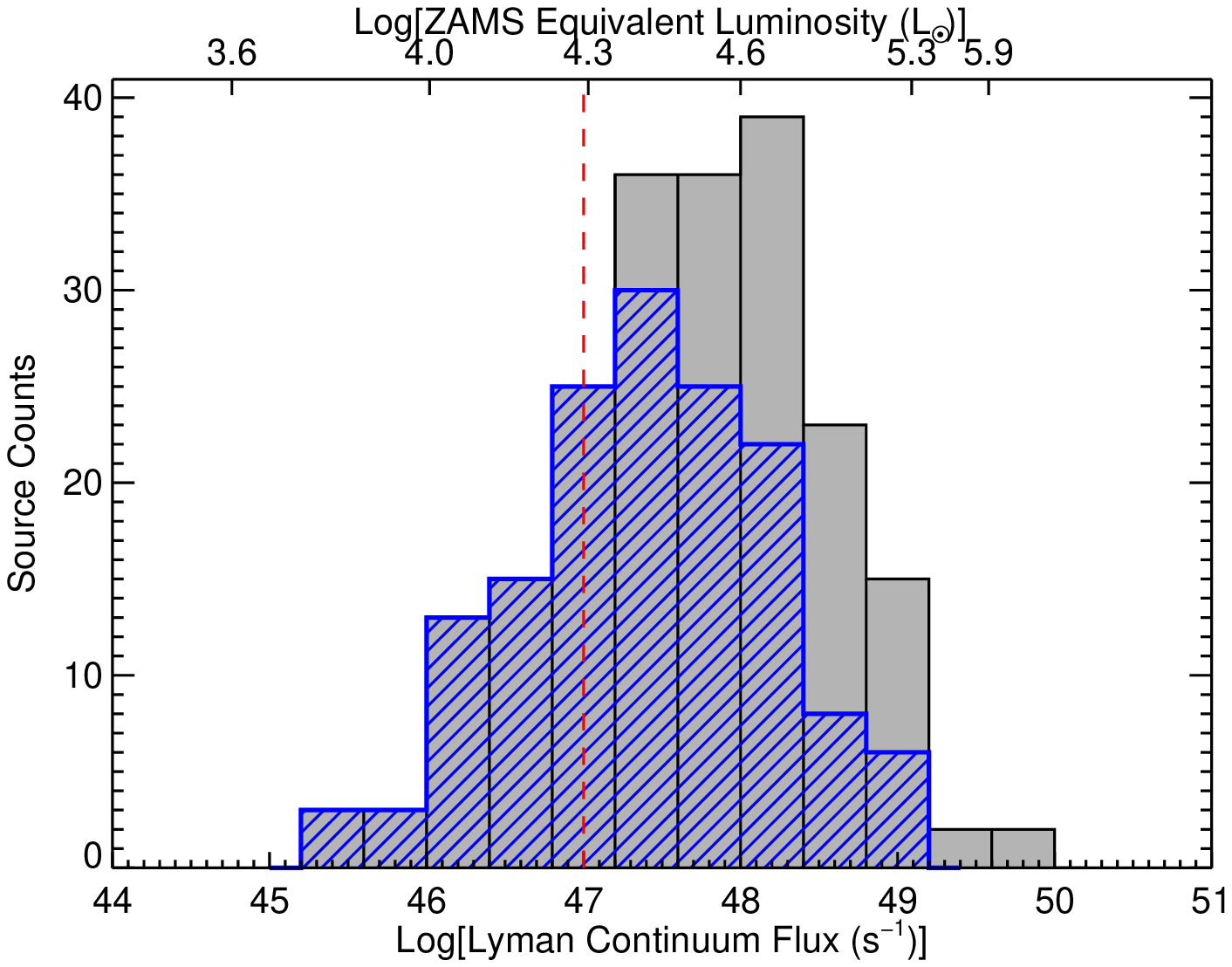}

\caption{\label{fig:ni_distribution} Left panel: Lyman continuum flux as a function of heliocentric distance. The colours correspond to the angular sizes of the \hii\ regions (see colour bar for approximate values) and the solid and dash-dotted curved lines indicate the CORNISH sensitivity to different angular size scales; these correspond to FWHM of 1.5, 3, 6 and 12\arcsec, respectively. Characteristic error bars for these parameters are shown in the lower right corner of the plot. Right panel: (grey histogram) The ionising photon flux distribution of the whole ATLASGAL-CORNISH matched sample of embedded \hii\ regions; (blue histogram) the same, but for a subset of sources with sizes $< 5''$. The dashed red vertical line indicates the CORNISH completeness level. The bin size is 0.4 dex.} 

\end{center}
\end{figure*}

\section{\hii\ region properties}

\setlength{\tabcolsep}{6pt}

\begin{table*}
%select replace(t1.cornish_name,'-','$-$'),'&', replace(t1.atlas_name,'-','$-$'),'&', round(t1.offset,1),'&',round(radio_flux_p,2),'&',round(radio_flux_i,2),'&', round(radio_angular_size_deconv,2),'&',if(radio_size_pc is null,'\\multicolumn{1}{c}{$\\cdots$}',round(radio_size_pc,2)),'&',round(ni+log10(power(t3.bb_distance,2)),2),'&',if(msx_name is null,'\\multicolumn{1}{c}{$\\cdots$}',round(log10(bol_lum),2)),'\\\\' from atlas_cornish_matches_sex_distances as t3,atlas_cornish_matches_sex as t1 left join atlas_cornish_rms_matches_final as t2 on (t1.cornish_name=t2.cornish_name and t2.reject_flag is null) where t3.atlas_name=t1.atlas_name and uchii_flag is not null order by t1.cornish_name limit 30

\begin{center}\caption{Derived \hii\ region parameters.}
\label{tbl:derived_hii_para}
\begin{minipage}{\linewidth}
\scriptsize
\begin{tabular}{ll.......}
\hline \hline
  \multicolumn{1}{c}{CORNISH} & \multicolumn{1}{c}{ATLASGAL} &  	\multicolumn{1}{c}{Offset} & \multicolumn{1}{c}{Peak Flux}  & \multicolumn{1}{c}{Integrated Flux}  &\multicolumn{2}{c}{Diameter}  & \multicolumn{1}{c}{Log($N_{\rm{Lyman}}$)}&	\multicolumn{1}{c}{Log($L_{\rm{bol}}$) } \\
\cline{6-7} 
    \multicolumn{1}{c}{Name }&  \multicolumn{1}{c}{Name }&  \multicolumn{1}{c}{(\arcsec)}  &\multicolumn{1}{c}{(mJy beam$^{-1}$)}  &\multicolumn{1}{c}{(mJy)}  &\multicolumn{1}{c}{(\arcsec)}  &	\multicolumn{1}{c}{(pc)} &\multicolumn{1}{c}{(photon\,s$^{-1}$)}&	\multicolumn{1}{c}{(\lsun)} \\
\hline

G010.3009$-$00.1477	&	AGAL010.299$-$00.147	&	6.2	&	56.37	&	631.39	&	5.24	&	0.06	&	47.51	&	\multicolumn{1}{c}{$\cdots$}	\\
G010.3204$-$00.2586	&	AGAL010.321$-$00.257	&	4.1	&	14.61	&	18.20	&	0.74	&	0.01	&	46.31	&	3.11	\\
G010.4724+00.0275	&	AGAL010.472+00.027	&	0.5	&	22.34	&	38.43	&	1.67	&	0.09	&	47.62	&	4.82	\\
G010.4736+00.0274	&	AGAL010.472+00.027	&	4.1	&	12.30	&	19.30	&	1.36	&	0.07	&	47.33	&	\multicolumn{1}{c}{$\cdots$}	\\
G010.6218$-$00.3848	&	AGAL010.624$-$00.384	&	8.8	&	16.81	&	37.06	&	1.73	&	0.02	&	46.28	&	\multicolumn{1}{c}{$\cdots$}	\\
G010.6223$-$00.3788	&	AGAL010.624$-$00.384	&	20.5	&	97.90	&	483.33	&	5.56	&	0.06	&	47.40	&	\multicolumn{1}{c}{$\cdots$}	\\
G010.6234$-$00.3837	&	AGAL010.624$-$00.384	&	3.2	&	305.58	&	1952.22	&	4.39	&	0.05	&	48.00	&	3.90	\\
G010.6240$-$00.3813	&	AGAL010.624$-$00.384	&	10.4	&	38.23	&	71.65	&	1.45	&	0.02	&	46.57	&	\multicolumn{1}{c}{$\cdots$}	\\
G010.9584+00.0221	&	AGAL010.957+00.022	&	3.6	&	109.11	&	195.97	&	1.61	&	0.11	&	48.52	&	5.36	\\
G010.9656+00.0089	&	AGAL010.964+00.011	&	8.7	&	7.86	&	51.75	&	4.65	&	0.06	&	46.55	&	3.55	\\
G011.0328+00.0274	&	AGAL011.031+00.026	&	9.2	&	3.58	&	5.69	&	1.15	&	0.01	&	45.55	&	\multicolumn{1}{c}{$\cdots$}	\\
G011.0339+00.0616	&	AGAL011.034+00.061	&	3.0	&	21.01	&	103.36	&	6.93	&	0.49	&	48.30	&	5.14	\\
G011.1104$-$00.3985	&	AGAL011.109$-$00.397	&	6.0	&	51.55	&	305.37	&	8.22	&	0.67	&	48.89	&	5.68	\\
G011.9032$-$00.1407	&	AGAL011.902$-$00.141	&	3.6	&	5.59	&	25.57	&	2.84	&	0.06	&	46.59	&	\multicolumn{1}{c}{$\cdots$}	\\
G011.9039$-$00.1411	&	AGAL011.902$-$00.141	&	6.3	&	11.11	&	16.81	&	1.07	&	0.02	&	46.41	&	\multicolumn{1}{c}{$\cdots$}	\\
G011.9368$-$00.6158	&	AGAL011.936$-$00.616	&	4.5	&	163.93	&	1155.90	&	5.69	&	0.11	&	48.23	&	4.77	\\
G011.9446$-$00.0369	&	AGAL011.946$-$00.036	&	5.1	&	61.58	&	943.58	&	14.56	&	0.89	&	49.13	&	5.50	\\
G012.1988$-$00.0345	&	AGAL012.198$-$00.034	&	4.0	&	19.68	&	62.71	&	2.22	&	0.13	&	47.90	&	4.64	\\
G012.2081$-$00.1019	&	AGAL012.208$-$00.102	&	2.5	&	72.18	&	207.87	&	2.41	&	0.16	&	48.54	&	\multicolumn{1}{c}{$\cdots$}	\\
G012.4294$-$00.0479	&	AGAL012.431$-$00.049	&	7.2	&	13.78	&	45.17	&	2.26	&	0.15	&	47.89	&	\multicolumn{1}{c}{$\cdots$}	\\
G012.4317$-$01.1112	&	AGAL012.431$-$01.114	&	9.8	&	9.67	&	69.01	&	4.13	&	0.08	&	47.02	&	\multicolumn{1}{c}{$\cdots$}	\\
G012.8050$-$00.2007	&	AGAL012.804$-$00.199	&	6.2	&	287.90	&	12616.40	&	16.16	&	0.16	&	48.70	&	\multicolumn{1}{c}{$\cdots$}	\\
G012.8131$-$00.1976	&	AGAL012.804$-$00.199	&	32.5	&	131.04	&	1500.39	&	5.22	&	0.05	&	47.77	&	\multicolumn{1}{c}{$\cdots$}	\\
G012.9995$-$00.3583	&	AGAL012.998$-$00.357	&	7.8	&	4.76	&	20.14	&	2.70	&	0.03	&	45.84	&	\multicolumn{1}{c}{$\cdots$}	\\
G013.2099$-$00.1428	&	AGAL013.209$-$00.144	&	5.6	&	40.72	&	946.76	&	8.21	&	0.18	&	48.25	&	4.52	\\
G013.3850+00.0684	&	AGAL013.384+00.064	&	15.3	&	10.49	&	603.94	&	19.12	&	0.17	&	47.28	&	3.53	\\
G013.8726+00.2818	&	AGAL013.872+00.281	&	3.5	&	24.71	&	1447.55	&	15.36	&	0.33	&	48.40	&	\multicolumn{1}{c}{$\cdots$}	\\
G014.2460$-$00.0728	&	AGAL014.246$-$00.071	&	7.3	&	11.40	&	51.26	&	3.64	&	0.20	&	47.79	&	\multicolumn{1}{c}{$\cdots$}	\\
G014.5988+00.0198	&	AGAL014.607+00.012	&	40.6	&	2.89	&	4.39	&	1.08	&	0.01	&	45.49	&	3.28	\\
G014.7785$-$00.3328	&	AGAL014.777$-$00.334	&	6.4	&	11.02	&	18.25	&	1.75	&	0.11	&	47.45	&	4.43	\\
\hline\\
\end{tabular}\\
\normalsize
Notes: Only a small portion of the data is provided here, the full table is available in electronic form at the CDS via anonymous ftp to cdsarc.u-strasbg.fr (130.79.125.5) or via http://cdsweb.u-strasbg.fr/cgi-bin/qcat?J/MNRAS/.
\end{minipage}

\end{center}
\end{table*}
\setlength{\tabcolsep}{6pt}

In this paper we will only investigate the relationships between the ionisation rates and bolometric luminosities of the embedded \hii\ regions and the masses of the surrounding clumps, plus the overall Galactic distribution of the sample. The morphologies, infrared and radio properties of the \hii\ regions themselves will be presented in two future publications (i.e., Purcell et al., and Hoare et al., both in prep.). The physical properties of the individual \hii\ regions are presented in Table\,\ref{tbl:derived_hii_para} and a summary of their overall statistics is presented in Table\,\ref{tbl:derived_para}.

\subsection{Lyman continuum flux}
\label{sect:lyman_continuum_flux}

The Lyman continuum output rate from a massive star, $N_{i}$, can be estimated using Eqn.\,7 from \citet{carpenter1990} relating $N_{i}$ to the measured radio continuum flux density, i.e.,

\begin{equation}
\left(\frac{N_{i}}{\rm{photon\,s^{-1}}}\right)=9\times10^{43}\,\,\left(\frac{S_{\nu}}{\rm{mJy}}\right)\,\, \left(\frac{D^{2}}{\rm{kpc}}\right) \,\, \left(\frac{\nu^{0.1}}{\rm{5\,GHz}}\right)
\label{eqn:ni}
\end{equation}

\noindent where $S_{\nu}$ is the integrated radio flux density measured at frequency  $\nu$ and $D$ is the heliocentric distance to the source. This assumes that the \hii\ regions are optically thin and will significantly underestimate the Lyman flux for more compact \hii\ regions that are optically thick at 5\,GHz. The typical error in the derived Lyman flux is $\sim$20\,per\,cent and takes into account the 10\,per\,cent error in both the distance and integrated flux measurement. Using the CORNISH 7$\sigma$ limit of $\sim$3\,mJy and maximum distance of 20\,kpc we find we are complete to unresolved optically thin \hii\ regions with Lyman fluxes of $\ge$10$^{47}$\,photon\,s$^{-1}$, which corresponds to a star of mass 15\,\msun\ (i.e., a star of spectral type B0 or earlier). In the case of hyper-compact HII regions, where the ionized nebula is likely to be optically thick below $\sim$50\,GHz (e.g., \citealt{kurtz2005a}), the measured flux would be two orders of magnitude lower at 5\,GHz. We are therefore only sensitive to \hchii\ regions associated with stars with masses of $\ge40$\,\msun, equivalent to a spectral type of O6-O5 or earlier at a distance of 20\,kpc. 

In the left panel of Fig.\,\ref{fig:ni_distribution} we present the Lyman photon flux as a function of heliocentric distance. The right panel of this figure shows the distribution for the whole sample with the completeness limit indicated by the red vertical dashed line. A peak in the Lyman flux distribution occurs at approximately 10$^{48}$\,photon\,s$^{-1}$, significantly above the completeness limit for unresolved, optically thin \hii\ regions (i.e., 10$^{47}$\,photon\,s$^{-1}$). However, given a typical IMF, we should expect the peak in the distribution to coincide with the completeness level. This would suggest that we are missing a significant number of \hii\ regions with Lyman fluxes between 10$^{47-48}$\,photon\,s$^{-1}$. 

The colours of the symbols in the left panel of Fig.\,\ref{fig:ni_distribution} indicate the observed angular sizes of each \hii\ region (see the horizontal colour bar for corresponding source diameters). While the solid black curve shows the point-source sensitivity of the CORNISH survey, it is clear that many of the \hii\ regions are resolved. As an \uchii\ region expands, its total flux density is conserved but its surface brightness decreases. We have therefore added the dashed-dotted lines to show the sensitivity to more extended sources and this shows that we are not sensitive to \hii\ regions with $N_{i} < 10^{47}$\,photon\,s$^{-1}$ that are larger than 12\arcsec, 6\arcsec\ and 3\arcsec\ at distances greater than 2, 5 and 12\,kpc, respectively.  For \hii\ regions with $N_{i} < 10^{48}$\,photon\,s$^{-1}$, we are not sensitive to sources larger than 12\arcsec\ and 6\arcsec\ at more than 7 and 15\,kpc, respectively, but are sensitive to \uchii\ regions with sizes of 3\arcsec\ or smaller across the Galaxy.

It is clear that we are less sensitive to more extended \hii\ regions and that this has a bigger impact on the nearby early B type star population of the sample than the generally more distant O type star population. To illustrate this, we overlaid the $N_{i}$ distribution of the more compact \hii\ regions ($<$5\arcsec) on the right panel of Fig.\,\ref{fig:ni_distribution} (blue hatched histogram). This shows that the distribution of the most compact \hii\ regions peaks nearer the completeness limit than that of the full sample. It is this more limited sensitivity to the less compact \hii\ regions with ionizing fluxes below 10$^{48}$\,photon\,s$^{-1}$ that produces the higher than expected turnover for the whole sample seen in the right panel of Fig.\,\ref{fig:ni_distribution}.

Another interesting feature arising from Fig.\,\ref{fig:ni_distribution} is that the proportion of compact to less-compact \hii\ regions decreases steeply as the ionising photon flux increases.

\subsection{Physical sizes}
\label{sect:hii_size}

In Fig.\,\ref{fig:physical_size_uchii} we show the physical size distribution for this sample of \hii\ regions (grey histogram). Additionally we separate this sample into two subsamples using a $N_{i}$ threshold of 10$^{48}$\,photon\,s$^{-1}$. This is the approximately the ionizing photon output rate expected from a $\sim$20\,\msun\ zero age main sequence (ZAMS) star (\citealt{martins2005,mottram2011b,davies2011}), roughly equivalent to an O9 spectral type and thus effectively separates late O and early B type star populations. For simplicity, we will refer to these two subsamples as O and B type \hii\ regions.  The distribution of \hii\ regions above and below this threshold are plotted in Fig.\,\ref{fig:physical_size_uchii} as red and blue hatched histograms, respectively.  

\begin{figure}
\begin{center}

\includegraphics[width=0.49\textwidth, trim= 0 0 0 0]{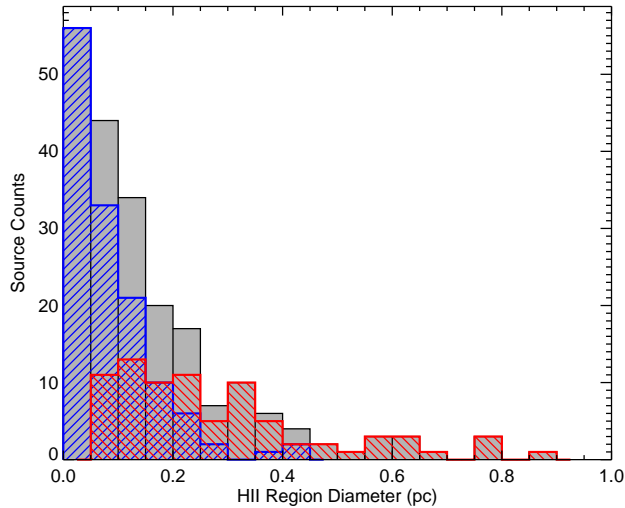}

\caption{\label{fig:physical_size_uchii}  Distribution of \hii\ region sizes. The whole sample is traced by the filled grey histogram, while the red and blue histograms show the \hii\ regions that have Lyman continuum fluxes above and below 10$^{48}$\,photon\,s$^{-1}$, respectively. The bin size used is 0.05\,pc.} 

\end{center}
\end{figure}

 The overall distribution of \hii\ region sizes shows a peak for diameters less than 0.05\,pc, after which it decreases rapidly before flattening into a long tail at $\sim$0.4\,pc; the largest measured size being $\sim$0.8\,pc. The classification as a \uchii\ region is generally reserved for sources with diameters less than 0.1\,pc and applying this to sample we find only 100 \hii\ regions satisfy this requirement (slightly less than 50\,per\,cent). Although the overall distribution looks fairly smooth, in that there are no obvious breaks or evidence of bimodality, the size distributions of O and B type \hii\ regions show some striking differences.

The B type \hii\ region size distribution peaks at the same position as that of the whole sample, but drops off much more quickly, with the largest size measured being $\sim$0.4\,pc. There are 131 \hii\ regions in this subsample with the majority falling into the \uchii\ region classification (89 or $\sim$68\,per\,cent). The lower number of more extended B type \hii\ regions can be understood by considering the relationship between the angular size of the \hii\ region and the surface brightness sensitivity of CORNISH as discussed in the previous subsection.

The distribution of O type \hii\ regions is significantly flatter and more skewed towards larger diameters, with a peak at 0.0-0.05\,pc. This subset provides the majority of the more extended \hii\ regions ($>$0.2\,pc). Noticeably, there are no O type \hii\ regions smaller than 0.05\,pc and only 11 of the 81 \hii\ regions in this sample could be classified as \uchii\ regions ($\sim$14\,per\,cent). Due to the lower sensitivity of CORNISH to more extended \hii\ regions, particularly those associated with B type stars, we are likely to be incomplete to these types of \hii\ regions. However, we should be sensitive to all optically thin \uchii\ regions driven by O stars across the Galaxy and therefore the low number of these in the sample is significant (this is discussed in more detail in Sect.\,\ref{sect:lifetimes}). Additionally, although not explicitly shown in Fig.\,\ref{fig:physical_size_uchii}, we find no compact or \uchii\ regions with $N_{i} > 10^{49}$\,photon\,s$^{-1}$ and physical diameters smaller than 0.2\,pc.

\subsection{Bolometric luminosity}

Many of our sample of  \hii\ regions are also found in the RMS catalogue (\citealt{urquhart2008}) and therefore have estimated bolometric luminosities (i.e., \citealt{mottram2010,mottram2011a}). The RMS survey has characterised the Galactic population of mid-infrared-selected \hii\ regions and MYSOs and is complete to a few 10$^4$\,\lsun. This is well matched to the CORNISH completeness and makes the RMS sample well suited for comparison.

Using a search radius of 10\arcsec, we have matched 135 \hii\ regions with their RMS counterparts and adopted the bolometric luminosities from the RMS database.\footnote{http://rms.leeds.ac.uk/cgi-bin/public/RMS\_DATABASE.cgi.} The fluxes used to estimate these luminosities are effectively clump-average values and are therefore a measure of the total luminosity of the embedded proto-cluster.

\begin{figure}
\begin{center}

\includegraphics[width=0.49\textwidth, trim= 0 0 0 0]{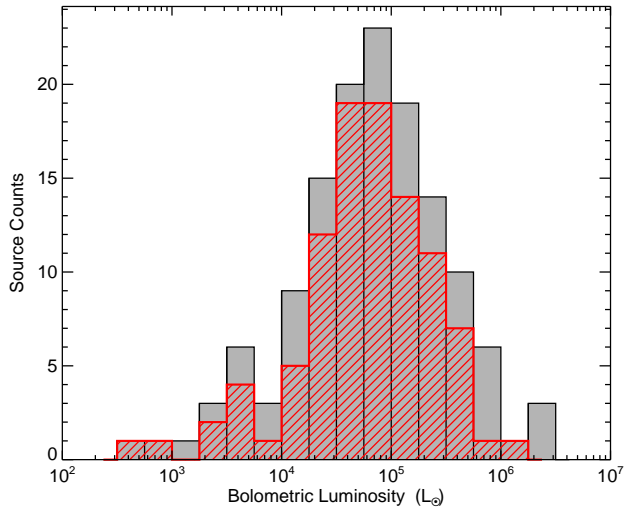}

\caption{\label{fig:bol_lum} Bolometric luminosity distribution of the \hii\ region sample. The bin size is 0.25\,dex.} 

\end{center}
\end{figure}

The bolometric luminosities for \hii\ regions and associated clusters range from $\sim$1000\,\lsun, which is effectively the lower limit required for a B3 star, to 10$^6$\,\lsun, which is approximately equivalent to an O3 star (\citealt{martins2005}). These stellar classes have been estimated assuming that a single star is responsible for the measured luminosity but in reality the luminosity of the most massive star will be a few subclasses later than those stated. \citet{mottram2011a} estimates the typical errors for the derived luminosities to be approximately 34\,per\,cent.

\section{Discussion}

\subsection{Empirical mass-size relationship for massive star formation}
\label{mass-size-relationship}

In Fig.\,\ref{fig:mass_radius_distribution} we reproduce the clump mass-radius plot from Fig.\,15 of Paper\,I, with the addition of the \hii -region data presented in this paper.  The dashed red and blue lines show the least-squares fits to the \hii -region and methanol-maser associated clump samples, respectively. The similarity between their overall distributions and the fit parameters, which agree within the errors, provides strong evidence that the methanol-maser sources and the \hii\ regions form in the same population of dense clumps. The two data sets provide us with a combined sample of over 500 massive, star forming-clumps, forming a fairly continuous distribution over almost four orders of magnitude in mass and two orders of magnitude in radius. The least squares fit to the combined sample is Log($M_{\rm{clump}}) = 3.4\pm0.013 + (1.67\pm0.036)\times {\rm{Log}}(R_{\rm{eff}}$).

\begin{figure}
\begin{center}
\includegraphics[width=0.49\textwidth, trim= 0 0 0 0]{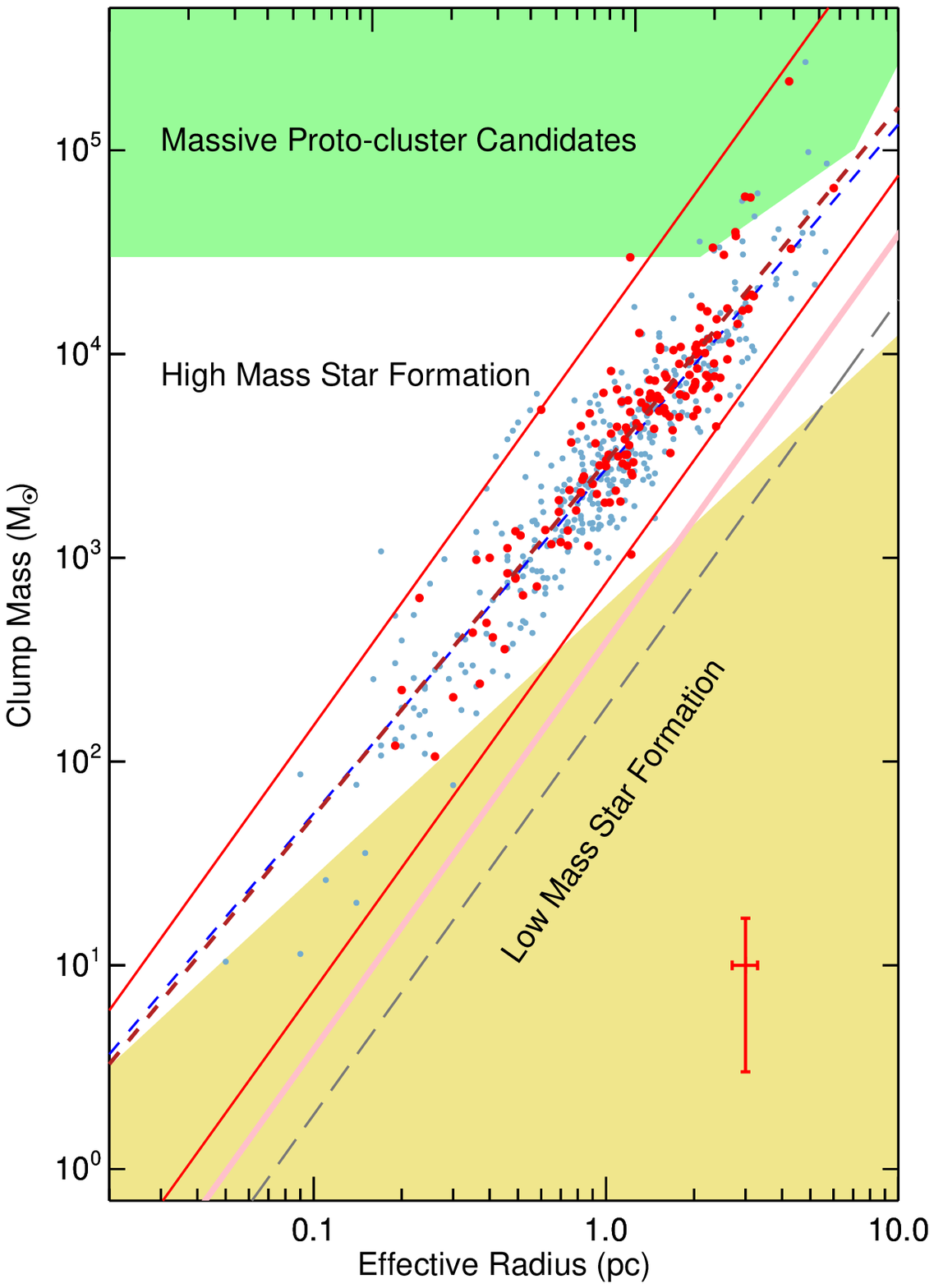}

\caption{\label{fig:mass_radius_distribution}The mass-size relationship of the current \hii -region sample (red dots) and methanol-maser associated clumps from Paper\,I (blue dots). The yellow shaded region shows the part of the parameter space found to be devoid of massive star formation that satisfies the relationship $m(r) \le 580$\,\msun\, $(R_{\rm{eff}}/{\rm{pc}})^{1.33}$ (cf. \citealt{kauffmann2010c}). The green shaded region indicates the region in which young massive cluster progenitors are expected to be found (i.e., \citealt{bressert2012}). The dashed blue and red lines show the result of linear least-squares fits to the methanol-maser and \hii -region associated clumps, respectively. The grey dashed line shows the sensitivity of the ATLASGAL survey and the upper and lower solid red line shows the surface densities of 1\,g\,cm$^{-2}$ and 0.05\,g\,cm$^{-2}$, respectively. The diagonal pink band fills the gas surface density ($\Sigma(\rm{gas})$) parameter space between 116-129\,\msun\,pc$^{-2}$ suggested by \citet{lada2010} and \citet{heiderman2010}, respectively, to be the threshold for ``efficient'' star formation. Characteristic error bars for these parameters are shown in the lower right corner of the plot.}

\end{center}
\end{figure} 

The lower and upper solid diagonal red lines in Fig.\,\ref{fig:mass_radius_distribution} denote surface densities of 0.05 and 1\,g\,cm$^{-2}$, the empirical lower and upper bounds for the clump-averaged surface densities of the methanol-maser associated clumps of Paper\,I. The upper bound is equal to the value of 1\,g\,cm$^{-2}$ predicted to be a lower limit for massive star formation (i.e., \citealt{mckee2003,krumholz2008}). However, the latter was determined for core-sized structures and is clearly not applicable to the larger clumps resolved by ATLASGAL.  These limits also provide reasonable constraints on the \hii -region associated clump distribution and thus on the surface density required for massive star formation.

To facilitate the comparison of our results with other studies we also include the threshold derived by \citet{lada2010} and \citet{heiderman2010} (116 and 129\,\msun\,pc$^{-2}$, respectively, which approximately corresponds to A$_V \simeq8$\,mag; hereafter LH threshold) for ``efficient'' star formation (solid pink line) and the mass-size criterion for massive star formation determined by \citet[][i.e., where $m(r) \ge 580$\,\msun\ $(R_{\rm{eff}}/{\rm{pc}})^{1.33}$; yellow shaded region]{kauffmann2010b}.\footnote{Note that when deriving this relationship \citet{kauffmann2010b} reduced the dust opacities of \citet{ossenkopf1994} by a factor of 1.5. This reduced value for the opacities has not been applied when determining the clump masses presented here and therefore we have rescaled the original coefficient given by \citet{kauffmann2010b} by this factor (i.e., from 870\,\msun\ to 580\,\msun; cf. \citealt{dunham2011}).} The LH threshold was determined from studies of nearby molecular clouds and is therefore not specific to massive star formation. The \citet{kauffmann2010b} criterion was determined from an investigation of the mass-radius relationship of nearby molecular clouds ($<$500\,pc, i.e., Ophiuchus, Perseus, Taurus and the Pipe Nebula; \citealt{kauffmann2010a, kauffmann2010b}) and known samples of high-mass star formation such as those of \citet{beuther2002}, \citet{hill2005}, \citet{motte2007} and \citet{Mueller2002}, \citet{kauffmann2010b}. However, as stated in Paper\,I the lower bound of 0.05\,g\,cm$^{-2}$ is approximately twice the LH-threshold and provides a better constraint than Kauffmann et al.\ for the high-mass end of the distribution (i.e., $R_{\rm{eff}} \gtrsim 0.5$\,pc or $M_{\rm{clump}} \gtrsim $500\,\msun).

So while the LH threshold can be considered a reasonable criterion for ``efficient'' star formation the value of 0.05\,g\,cm$^{-2}$, found to fit the lower envelope of the mass-radius relationship for both methanol maser and \hii\ region associated clumps, can be considered the threshold required for ``efficient'' high mass star formation. 

The area highlighted in green in the upper portion of Fig.\,\ref{fig:mass_radius_distribution} shows the region of parameter space where massive proto-cluster (MPC) candidates are expected to be located (\citealt{bressert2012}; see Paper\,I for details). Six of our sources are located in this region. Three are associated with W49A (AGAL043.148+00.014, AGAL043.164$-$00.029 and AGAL043.166+00.011) and have a combined mass $\sim4\times10^5$\,\msun. One clump is associated with each of the W31 and W51 star-forming regions (AGAL010.472+00.027 and AGAL049.489$-$00.369, respectively). The remaining source is AGAL019.609$-$00.234. All six have masses above 10$^4$\,\msun\ and are therefore possible progenitors of future young massive clusters (YMCs) such as the Arches and Quintuplet clusters (\citealt{portegies_zwart2010}). The MPC candidates in W31, W49A and W51 were previously identified by \citet{ginsburg2012} from an analysis of 1.1\,mm continuum data from the BGPS and so AGAL019.609$-$00.234 is the only new potential MPC identified by this work.\footnote{AGAL019.609$-$00.234 was also identified by \citet{ginsburg2012} as a massive clump but its derived mass was slightly below $3\times10^4$\,\msun\ and so did not make it into their final sample. However, the mass derived by us would suggest this source is a good MPC candidate.}

In Paper\,I we identified 7 MPC candidates, six of which were new, and together with those identified by \citet{longmore2012a} and \citet{ginsburg2012} we estimated the total population of MPC candidates is likely to be $\le20\pm6$. However, comparing this estimate with the number of known YMCs we found that there are far more MPCs than would be expected, assuming a lifetime of 10\,Myr and formation time of 1\,Myr for the YMCs (e.g., \citealt{longmore2012a}}. We suggested a couple of possible explanations for this apparent discrepancy: either there are far more YMCs yet to be discovered or the SFE is significantly lower than the 30\,per\,cent assumed by \citet{bressert2012} when defining their mass-size criterion. The first of these possibilities would require the total number of YMCs in the Galaxy to be an order of magnitude larger than currently known, which seems unlikely. The second seems more plausible given that a SFE of $\sim$10\,per\,cent (e.g., \citealt{johnston2009} and see Sect.\,\ref{sect:mass_lum_comparison} of this paper) would reduce the number of MPC candidates to no more than a handful.  

\subsection{Galactic distribution of compact and \uchii\ regions}
\label{gal_distribution}

\begin{figure}
\begin{center}
\includegraphics[width=.49\textwidth, trim= 0 0 0 0]{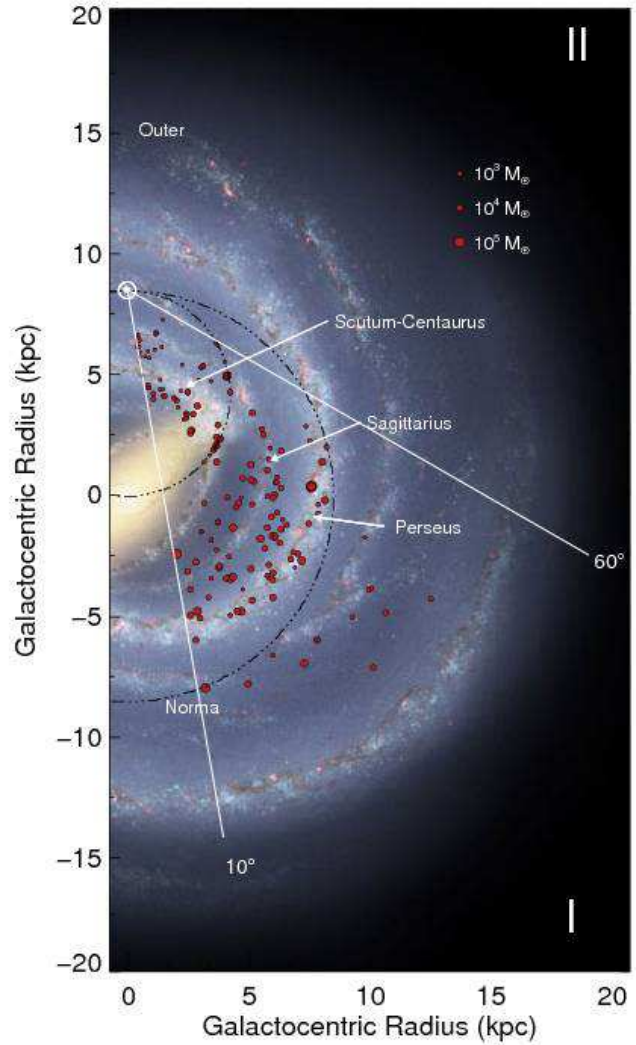}

\caption{\label{fig:galactic_mass_radius_distribution}Galactic distribution of the ATLASGAL-CORNISH associations with known distances and masses above the completeness limit (1,000\,\msun). The Galactic positions of our sample derived from the maser parallax, photometric and kinematic distances and their Galactic longitude are plotted as red circles, the sizes of which give an indication of their mass, as depicted in the upper right corner. We have superimposed the clump distribution over the sketch of the Galaxy produced by Robert Hurt of the Spitzer Science Center in consultation with Robert Benjamin (Courtesy NASA/JPL-Caltech).  The position of the Sun is shown by the small white circle above the Galactic centre. The two white solid lines enclose the region of the Galactic plane overlapped by the ATLASGAL and CORNISH surveys. The black dot-dashed circles represent the locus of tangent points and the Solar circle, respectively.} 

\end{center}
\end{figure}

We have already seen some hints of the large scale structural feature of the Galaxy in the longitude-velocity diagram presented in Sect.\,\ref{sect:gal_long_vel} and the heliocentric distribution presented in Sect.\,\ref{sect:distance_distribution}. In this subsection we will use the distances and Galactic longitude to trace the 2-dimensional distribution of compact \hii\ regions for the first quadrant. As these \hii\ regions are still deeply embedded in their dust cocoons and have lifetimes of at most a few times 10$^5$\,yr (\citealt{davies2011,mottram2011b}) we can assume that they have not traveled far from where they formed. Furthermore, since massive star formation is found to be almost exclusively associated with the spiral arms in nearby spiral galaxies (e.g., \citealt{kennicutt2005}), the Galactic distribution of this sample of \hii\ regions may be able to trace the spiral arms of the Milky Way.

In Fig.\,\ref{fig:galactic_mass_radius_distribution} we present the 2-D Galactic distribution of ATLASGAL-CORNISH associations. The positions of the ATLASGAL-CORNISH associations are in reasonable agreement with the main structural features of the Galaxy, as determined by other tracers (taking into account the kinematic distance uncertainty due to peculiar motions is of order $\pm$1\,kpc; \citealt{reid2009}). We note there is a large amount of scatter in these data, particularly at the far side of the Sagittarius arm. However, the width of the inter-arm region is approximately twice the uncertainty in the distance measurement and therefore it is possible, and even likely, that sources located in this inter-arm region are actually associated with either the Sagittarius or Perseus arms. Two previous studies have performed a similar analysis using methanol masers (Paper\,I) and a sample of infrared-selected massive young stellar objects (MYSOs) and \hii\ regions (\citealt{urquhart2011a}), and with both finding a higher degree of correlation with the spiral arms. However, both studies had larger samples and so sample size may be an issue here.

\begin{figure}
\begin{center}

\includegraphics[width=0.49\textwidth, trim= 0 0 0 0]{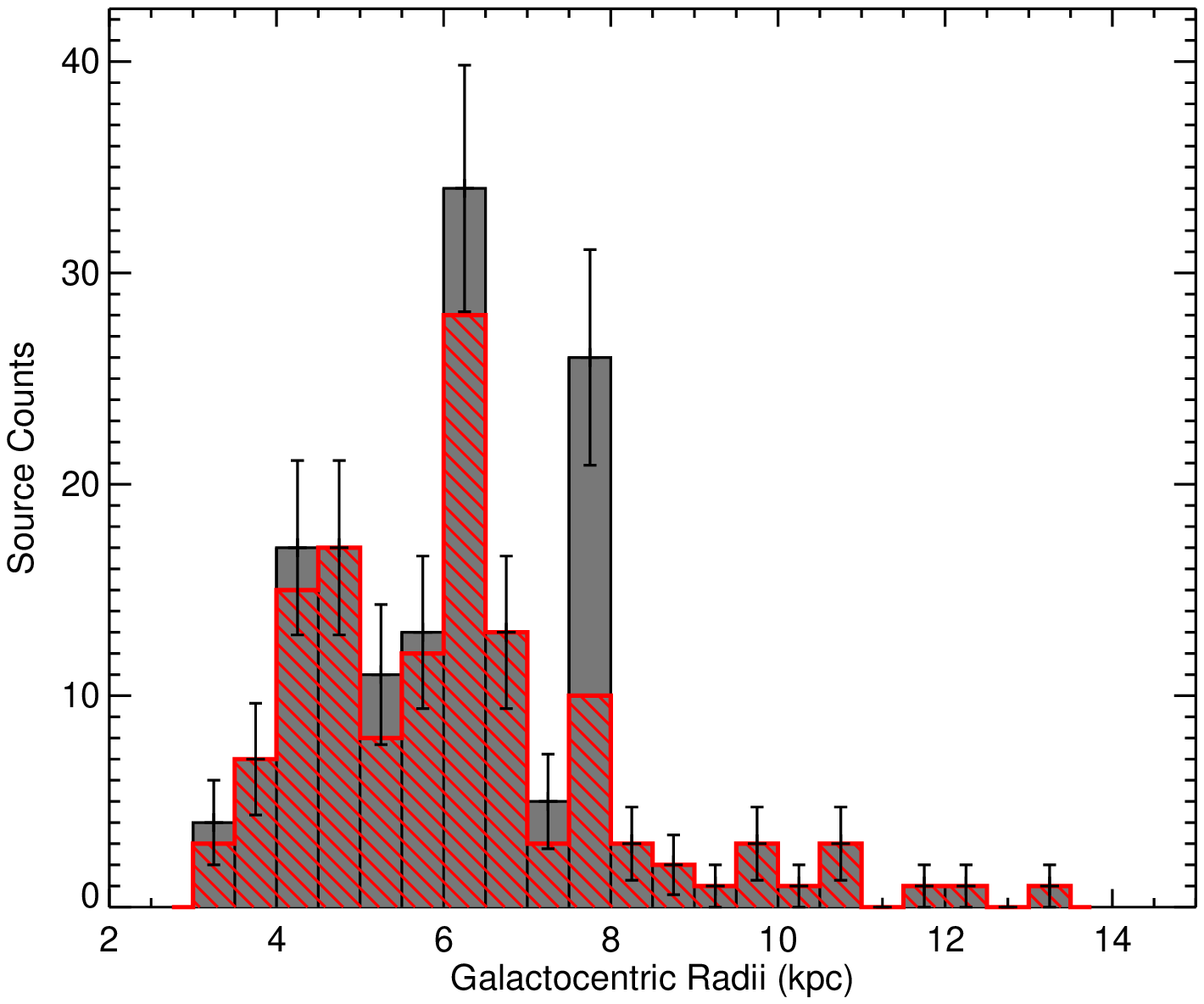}
\includegraphics[width=0.49\textwidth, trim= 0 0 0 0]{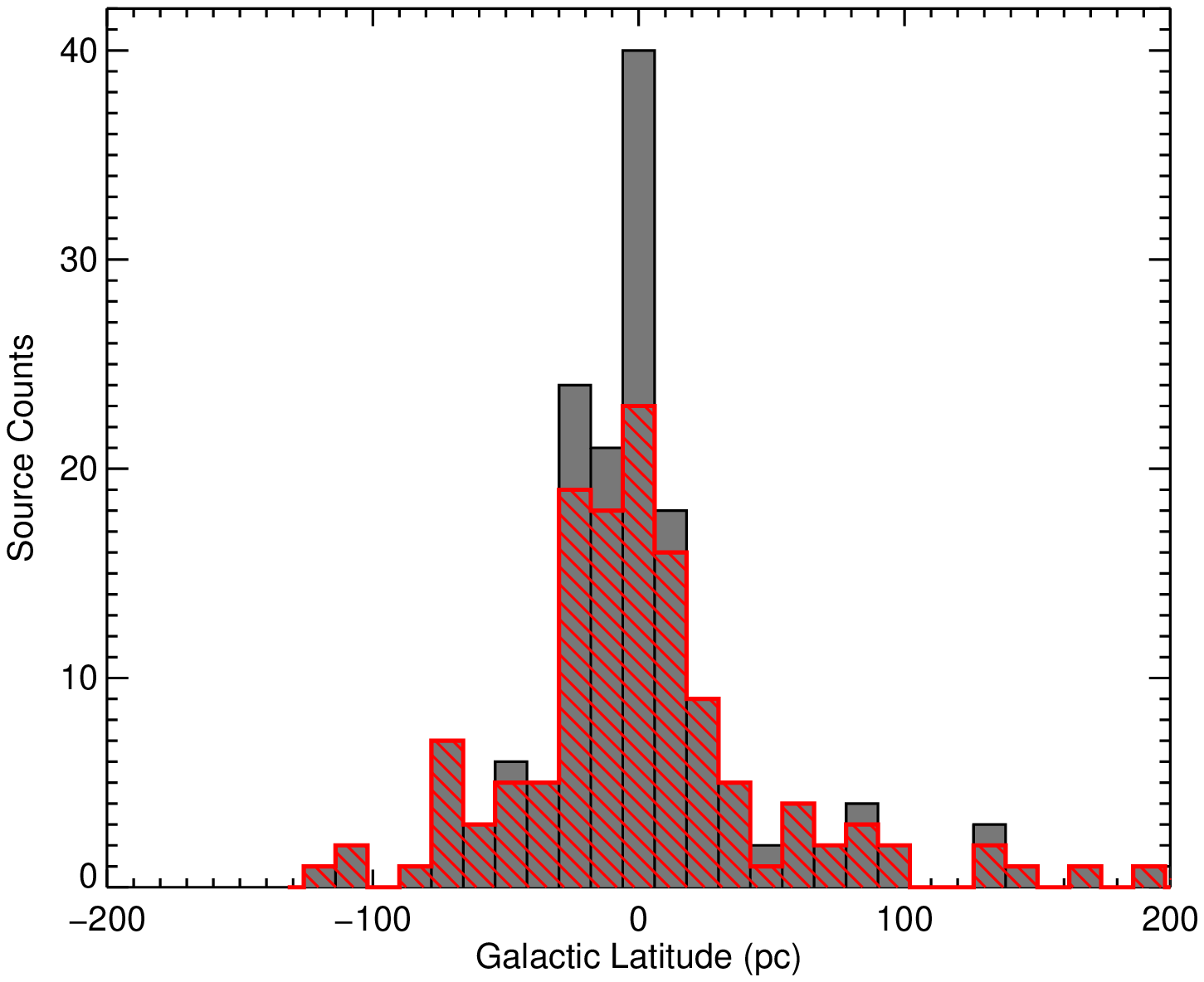}

\caption{\label{fig:rgc_scaleheight} Galactocentric and Galactic latitude distributions of the \hii\ regions (grey) and clumps (red hatching) are shown in the upper and lower panels, respectively. In both of these plots we have only included \uchii\ regions with ionising fluxes and masses above the completeness limit ($Ni > 10^{47}$\,photon\,s$^{-1}$  and $M_{\rm{clump}} > 1,000$\,\msun; 163 \hii\ regions and 132 clumps). The strong peak seen at $\sim$8\,kpc is almost entirely due to the presence of W49A in this bin. The bins sizes are 0.5\,kpc and 12\,pc in the upper and lower panels, respectively.} 

\end{center}
\end{figure}

The Galactocentric and latitude distributions of the ATLASGAL-CORNISH matched sources are presented in the upper and lower panel of Fig.\,\ref{fig:rgc_scaleheight}, respectively. As with the heliocentric distance distribution presented in Fig.\,\ref{fig:atlas_radio_distance_hist}, we plot the distribution of the \hii\ regions and clumps separately (filled grey and red hatched histogram, respectively); this is primarily to evaluate the impact of extreme star formation regions, such as W49A, have on the overall \hii\ region distribution.

Although the association with individual spiral arms is fairly weak in the 2-D distribution shown in Fig.\,\ref{fig:galactic_mass_radius_distribution}, the Galactocentric distribution shows three distinct peaks at $\sim$4, 6 and 8\,kpc. Peaks at similar radii have been previously noted in the northern Galactocentric radial distribution from a sample of IRAS sources observed in CS~(J=2-1) \citep{bronfman2000} and the distribution of the RMS sample of young massive stars and their associated CO emission \citep{urquhart2011a,moore2012}. The first of these corresponds to the position of the Scutum-Centaurus arm tangent and the end of the long bar, which is located at a Galactic longitude of $\sim$25\degr. The peak at $\sim$6\,kpc is largely associated with the Sagittarius arm. The third peak at $\sim$8\,kpc is again almost entirely due to the W49A star forming complex, however, this peak is still significant and is coincident with the expected Galactocentric radius of the far-side of the Perseus arm.

The peaks at 4 and 6\,kpc in the Galactocentric radius distribution have also been reported by \citet{anderson2009a} from a study of compact and diffuse \hii\ regions, however, the relative intensity of the peaks is reversed, presumably due to a larger proportion of nearby sources in their sample. The lack of sources at distances less than 3\,kpc is not significant as it simply results from the minimum longitude range used in this study. 

An interesting feature of this plot is that the vast majority of the massive star formation is taking place within $\sim$8\,kpc of the Galactic centre, with relatively little taking place outside the Solar circle. Similarly, significantly lower surface densities of methanol masers (discussed in Paper\,I), \hii\ regions and massive young stellar objects identified by the RMS survey (\citealt{lumsden2013}) have been reported in the outer Galaxy than found inside the Solar circle. Since we have selected only sources above our completeness limit this is unlikely to be a selection effect and suggests that high mass clumps are more likely to be found inside the Solar circle where the majority of the Galactic molecular material resides. The sharp drop in surface density occurs at approximately the same distance from the Galactic centre as the co-rotation radius ($\sim$8\,kpc), which also coincides with a sharp drop in the metallicity gradients of open clusters (\citealt{lepine2011}). This drop in the metallicity gradients is interpreted by the authors as being due to a gap in the dense gas at the co-rotation radius,  with smaller gas flows and consequently fewer mergers producing massive clumps outside the co-rotation radius than inside (see \citealt{lumsden2013} for a more detailed discussion). While spiral arms are thought to play an important role in forming dense molecular structures within the co-rotation radius (e.g., \citealt{heyer1998} and \citealt{dobbs2006}) it is likely that this process is regulated by a different mechanism in the outer Galaxy, such as supernovae (\citealt{dibs2009}). Alternatively, the lack of massive clumps in the outer Galaxy could simply be due to statistics rather than any physics since there is a power-law mass function for clumps (CMF) and so a smaller population of clumps will have far fewer massive clumps, and none, if the clump population density is below some level (e.g., \citealt{snell2002}).

Peaks in both the \hii\ region and clump latitude distribution are coincident with the Galactic mid-plane ($|z| \sim 0$). The peak of the \hii\ distribution is noticeably higher than found for the clumps, however, again this is primarily the presence of W49A in this bin. The presence of a few extreme star-forming regions close to the mid-plane, not only increases the overall peak in the \hii\ region distribution with respect to the clump distribution, but also has a knock on effect to the measured scale height. The scale height of the \hii\ region and clump samples is 20.7$\pm$1.7\,pc and 29.1$\pm$3.0\,pc, respectively. The scale height derived from the methanol maser distribution reported by \citet{green2011b} is 27$\pm$1\,pc, which is in good agreement with the scale height derived from our clump distribution.

As seen in the heliocentric distance distribution discussed in Sect.\,\ref{sect:distance_distribution} and for the Galactocentric and latitude distributions presented here it is clear that the presence of a small number of extreme star formation regions, such as W31, W33, W42 and W49A, can have a significant impact of the overall distributions of \hii\ regions.

\subsection{Estimating the Galactic population of \hii\ regions and their lifetimes}
\label{sect:lifetimes}

\begin{figure}
\begin{center}

\includegraphics[width=0.49\textwidth, trim= 0 0 0 0]{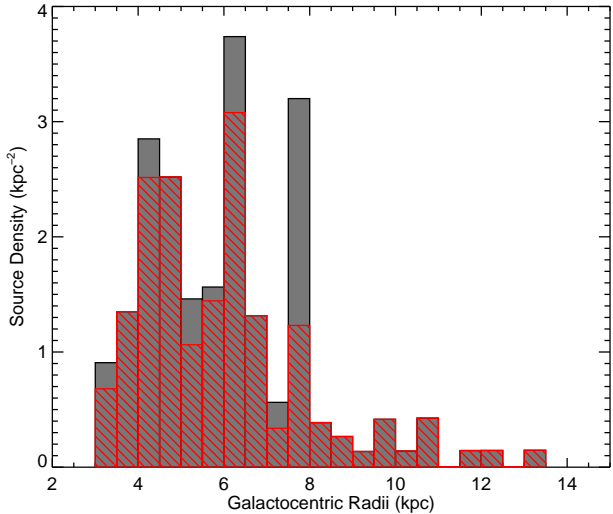}

\caption{\label{fig:rgc_surface_density} Surface density of \hii\ regions and their host clumps. Colours same as for Fig.\,\ref{fig:rgc_scaleheight}. The strong peak seen at $\sim$8\,kpc is almost entirely due to the presence of W49A in this bin. The bin size is 0.5\,kpc.} 

\end{center}
\end{figure}

Our catalogue of \hii\ regions covers a large fraction of the Galaxy and has been compiled from two unbiased surveys.\footnote{More precisely this includes $\sim$36\,per\,cent of the Galacic surface area within the Solar circle and $\sim$25\,per\,cent of the area within 15\,kpc of the Galactic centre.} It should therefore constitute a good representation of the Galactic population as a whole and be able to provide reliable estimates of the number of \hii\ regions. We begin by calculating the surface density of our sample of \hii\ regions as a function of Galactocentric distance (as shown in Fig.\,\ref{fig:rgc_surface_density}) by dividing the source counts given in the upper panel of Fig.\,\ref{fig:rgc_scaleheight} by the surface area of the portion of each annulus included in survey.  Assuming this surface density is representative of the Galactic population as a whole, we can simply multiply each bin by the area of the corresponding complete annulus to estimate the total number of \hii\ regions in each bin, and then sum all of the bins to get the total population. Using this method we estimate that the total number of \hii\ regions around B0 and earlier type stars (the nominal CORNISH sensitivity limit) to be approximately 400. Performing the same procedues using the surface density of O6 stars and earlier would suggest there are only $\sim$45 detectable \hii\ regions of these types in the Galaxy. 

The total number of \hii\ regions is slightly lower than the Galactic population estimated by the RMS survey. However, the RMS sample includes some more extended \hii\ regions that have been identified from their mid-infrared morphology and whose radio continuum emission falls below the CORNISH surface brightness sensitivity. More interesting is the low number of \hii\ regions associated with a star of spectral type O6 and earlier. \citet{wood1989} estimated the number of embedded O6 stars expected from the \hii\ expansion time and the number of field O stars to be of order a few hundred. Comparing this with the number expected from the 1,600 candidate \uchii\ regions identified from their IRAS colours led \citet{wood1989} to conclude that the embedded phase is much longer than would be expected from an uninhibited expansion of the \hii\ region --- this is often referred to as the \emph{lifetime problem}. It is likely that their sample of \hii\ region candidates is significantly contaminated from evolved stars and nearby B stars. Furthermore, it is very likely that given the IRAS resolution (ranging from 0.5\arcmin\ to 5\arcmin\ for the 12-100\,\mum\ bands) that many of the luminosities calculated for their sample of \uchii\ regions have been significantly overestimated. This would reduce the total number of embedded O6 stars, which in turn would reduce the need for an extended embedded \uchii\ region phase.

\citet{wood1989b} estimated the \uchii\ region phase to be $\sim3$-$6\times10^5$\,yr, which is supported by a Galactic population synthesis analysis \citep{davies2011} and estimates of their statistical lifetimes \citep{mottram2011b}, both of which reported similar \hii\ region lifetimes (a few 10$^5$\,yr). However, we find only a small proportion of \hii\ regions associated with late O type stars, and none of those associated with early O type stars, can be classified as \uchii\ regions. This may suggest that their expansion is extremely rapid and that the \uchii\ region lifetime is relatively short for late O type stars, and possibly so short for early O-type stars that their \uchii\ region stage is effectively unobservable. Alternatively, the early O-type stars evolve from late O-type stars via accretion of material through their \uchii\ and compact \hii\ region (\citealt{keto2007}). In this picture the \hii\ region first forms when the MYSO reaches $\sim$30\,\msun\ (approximately equivalent to a ZAMS O7 type star) and contracts down to the main sequence and expands rapidly while continuing to accrete material. By the time the central star has accreted enough mass to be classified as an early O-type star the \hii\ region has expanded beyond the sizes of typical of ultra compact and compact \hii\ regions.

Another possibility is that there is a significant number of O-type stars surrounded by optically thick ionized nebulae intermixed with the sample of B-type \hii\ regions; these extremely compact, optically thick \hii\ regions are classified as hyper-compact (HC) \hii\ regions. The Lyman continuum flux of these \hchii\ regions may be underestimated by up to two orders of magnitude (see Sect.\,\ref{sect:lyman_continuum_flux}). Although it is conceivable that very high accretion rates might impede the expansion of the \hii\ region it can do nothing to suppress the infrared luminosity. We might therefore expect to find a number of radio quiet, embedded objects with bolometric luminosities of 10$^{5}$\,\lsun\ and higher. However, the fact that the RMS survey (\citealt{mottram2011a,mottram2011b}) has not found any embedded MYSOs more luminous than $\sim10^{5}$\,\lsun\ rules this possibility out.

\subsection{Comparison between methanol maser- and \hii\ region-associated clumps}

Comparing the derived parameters for the \hii\ region associated clumps presented in this paper with those for the methanol maser associated clumps presented in Paper\,I we find them to be very similar. The overall clump structure (aspect ratios and $Y$-factors) for the two samples are indistinguishable from each other. This confirms the finding reported in earlier studies of high-mass protostellar objects (HMPO; \citealt{williams2004}) and \uchii\ regions (\citealt{thompson2006}) that the \emph{envelope} structure does not significantly change during the embedded stages of the massive star's or associated cluster's evolution. This would imply that the clumps are already in place before the star formation begins (as also suggested by \citealt{lumsden2013}) and that evolution on clump scales has a much longer timescale than the formation process for massive stars.

We find no significant difference in the location within clumps of massive stars as traced by \hii\ regions and methanol masers. Both tracers are closely associated with the peak of the \submm\ emission which is likely to be the highest column density. This would place the massive stars at the centres of the clumps at the bottom of the gravitational potential, which is broadly in agreement with the predictions of the competitive accretion model of massive star formation (\citealt{bonnell1997,bonnell2001}). However, as mentioned in Paper\,I, higher resolution observations are required to investigate the internal structure and mass distribution of these clumps before any detailed comparison with the models can be performed. Comparing the mass-radius distribution of these two samples of clumps we find them to be again indistinguishable from each other, both covering the same narrow band of parameter space and both satisfying the criteria for massive star formation. 

The similarities between the properties of clumps associated with methanol masers and \hii\ regions, given that only $\sim$20\,per\,cent of the methanol masers are associated with the later stages of formation (i.e., MYSOs and \hii\ regions; Paper\,I) provide strong support for the hypothesis that methanol masers are almost exclusively associated with the earliest stages of massive star formation. Approximately 80\,per\,cent of the detected methanol masers are not bright enough at mid-infrared to have been detected as a discrete source by $MSX$. These could be deeply embedded accreting protostars in a very early evolutionary stage having not yet warmed their dust envelope sufficiently to be observed by \emph{MSX}. However, the majority of the methanol masers, although not bright enough in the mid-infrared to have been detected by \emph{MSX}, are detected in \emph{Spitzer} IRAC images (\citealt{gallaway2013}). These weaker mid-infrared sources may still be very embedded MYSOs but alternatively could also be more evolved intermediate-mass YSOs that have started to clear their surroundings, lowering the line of sight extinction and allowing them to become visible in the mid-infrared.

We also note that the total number of \hii\ regions around stars of stellar type B0 and earlier is similar to the expected number of methanol masers ($\sim$600; Paper\,I), which would suggest these two stages have similar statistical lifetimes.

\subsection{Correlation between \uchii\ regions and their natal clumps}
\label{sect:mass_lum_comparison}

\subsubsection{Monte-Carlo Simulation}

In the previous section we derived the ionizing photon fluxes and bolometric luminosities for a large proportion of our sample of \hii\ regions. These parameters can provide a useful indication of the current level of star formation taking place within each clump. As discussed in Sects.\,\ref{sect:size} and \ref{sect:mass} the clump median radius and mass are $\sim$1.4\,pc and 5,000\,\msun, respectively. It is likely these clumps are host to embedded proto-clusters (e.g., \citealt{lada2003,motte2003}) rather than single massive stars. Furthermore, massive stars are almost exclusively found to be associated with clusters (e.g., \citealt{de-wit2004,gvaramadze2012}) and we therefore want to compare these integrated parameters with the emission expected from an embedded cluster.

To do this, we have followed the procedure described by \citet{walsh2001} using a Monte-Carlo simulation to randomly sample a \citet{kroupa2001} IMF (i.e., $N \propto M^\alpha$ where $\alpha$ = 1.3 for 0.08\,\msun\ $ \le M \le $ 0.5\,\msun\ and $\alpha = 2.3$ for $M >$ 0.5\,\msun). We used a stellar mass range of 0.1 to 120\,\msun\ and have assumed an average massive star formation efficiency (SFE) of 10\,per\,cent (e.g., \citealt{johnston2009}).

\begin{figure}
\begin{center}

\includegraphics[width=0.49\textwidth, trim= 0 0 0 0]{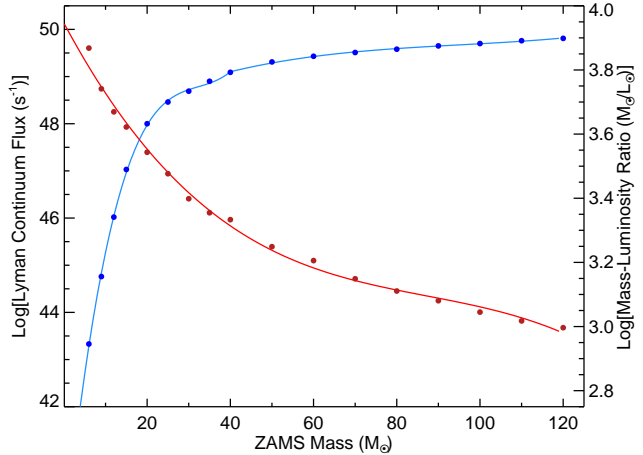}

\caption{\label{fig:mass_lum_model} Mass-Lyman continuum flux and mass-luminosity ratio for  OB stars derived from stellar atmosphere models are shown in blue and red, respectively. The filled circles show the data taken from \citet[][and reference therein]{davies2011}, while the curves show the polynomial fit to these points. } 

\end{center}
\end{figure}

A random number is generated and compared to the probability assigned to each mass interval in the range to determine the mass of star. This process is repeated until the total mass of the stars is approximately equal to the clump mass\,$\times$\,SFE. We have used polynomial fits to the stellar atmosphere models \citep[i.e.,][see Fig.\,\ref{fig:mass_lum_model} for details]{davies2011} to determine each star's contribution to the total cluster luminosity and Lyman continuum flux. This process was repeated 1,000 times for each mass increment for clump masses in the range 100\,\msun\ to $5\times10^5$\,\msun. The results of these simulations will be used in the following subsections.

\subsubsection{Luminosity-Lyman continuum flux relation}

\begin{figure}
\begin{center}

\includegraphics[width=0.49\textwidth, trim= 0 0 0 0]{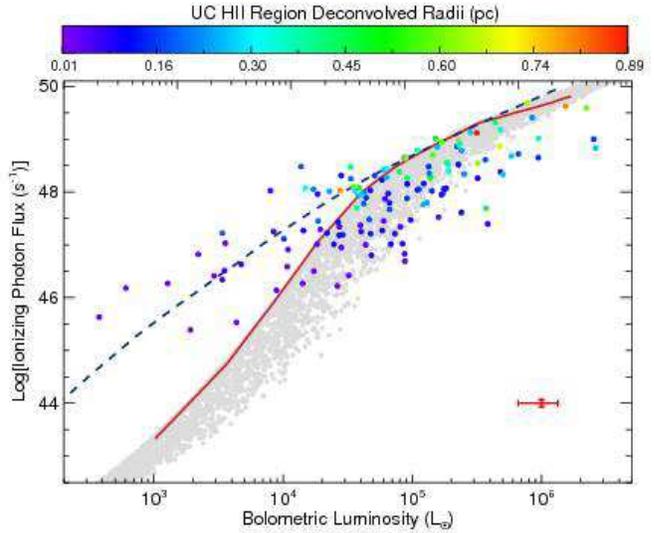}

\caption{\label{fig:lum_ni_comparison} \hii\ region Lyman continuum photon flux as a function of bolometric luminosity of the associated embedded cluster. The coloured circles show the source parameters derived from the data and indicate \hii\ region sizes (see colour bar for values). The thick red curved line indicates the relationship for a single OB zero-age main sequence star (from \citealt{martins2005}) and the light grey circles show the result of the Monte-Carlo simulation for clusters. The dashed black curve shows the ionizing photon flux that would be expected from a blackbody with the same radius and temperature as a ZAMS star (A.\,S\'anchez-Monge priv. comm.). Characteristic error bars for these parameters are shown in the lower right corner of the plot.} 

\end{center}
\end{figure}

Fig.\,\ref{fig:lum_ni_comparison} shows the Lyman photon flux as a function of bolometric luminosity for the \hii\ region sample. In cases where multiple \hii\ regions are within 10\arcsec\ of the RMS position we have integrated the Lyman flux and used the size of the largest and most evolved \hii\ region in the plot. We also show the luminosity-Lyman photon flux relation for OB stars from table of values given in \citet{davies2011}, the results of the Monte-Carlo cluster simulations discussed in the previous subsection (red curve and grey shaded region, respectively).

The distribution of compact and \uchii\ regions reveals a possible correlation between the bolometric luminosity and ionizing photon flux. However, this is effectively a luminosity-luminosity plot with both of these parameters having a distance dependence and we need to be careful that the observed correlation is not the result of the Malmquist bias. We have therefore used a partial Spearman correlation test to remove the dependence of these two parameters on distance (see Paper\,I for more details). This gives a correlation coefficient of 0.69 for the whole sample with a $p$-value $<1\times10^{-7}$, and so the correlation is significant and points to a strong correlation between these parameters. 

Fig.\,\ref{fig:lum_ni_comparison} reveals some interesting features.  First, the distribution of the measured parameters agrees well with the predictions of the Monte Carlo simulations for clusters, for which the single OB star locus forms an upper envelope.  But the agreement only holds for the high-luminosity sources. For \hii\ regions with luminosities equivalent to B stars, the observed distribution strongly disagrees with the theoretical relationship and the measured values lie well above the predicted upper envelope. This discrepancy has also been noted by \citet{lumsden2013}  and \citet{sanchez-monge2013} for the RMS sample of \hii\ regions (with which our sample shares a number of sources) and a sample of $\sim$200 IRAS selected sources, respectively. 
This excess Lyman continuum luminosity is unlikely to be the result of an overestimate of the radio flux as this is more likely to be underestimated due to the nature of interferometric observations, which filters out some of the larger spatial frequencies. This excess is, therefore, not easy to explain.

The sample of \citet{sanchez-monge2013} is split into three evolutionary types (millimetre source only (Type\,1), millimetre with associated infrared source (Type\,2), and infrared source only (Type\,3)).  They find that 70\,per\,cent of their Type\,1 and 2 sources have excess ionizing photon fluxes and suggest that this may preferentially occur for \emph{young} B type stars, a conclusion independently arrived at by \citet{lumsden2013} . Following \citet{sanchez-monge2013},  Fig.\,\ref{fig:lum_ni_comparison} also includes the locus of a blackbody with the same radius and temperature as ZAMS OB stars (dashed black curve), giving the theoretical upper limit to the Lyman continuum flux. The data for B-type \hii\ regions agree significantly better with the blackbody prediction than with the ZAMS stellar atmosphere models, albeit with considerable scatter in these data.

Another feature to note in Fig.\,\ref{fig:lum_ni_comparison} is in the O-star distribution, where the larger \hii\ regions appear closer to the Lyman luminosities expected from the single-star stellar atmosphere models. For late O type stars, the ionizing flux is still consistent with what we might expect if the bolometric luminosity were due to a cluster rather than a single star, however, this is not the case for the early O stars where there are clearly fewer ionizing photons than expected, even allowing that the luminosity is the result of a cluster. One explanation could be that the interferometer has filtered out some of the larger scale radio continuum flux, however, this is more likely to be an issue for the more extended \hii\ regions and most of the sources found to have an deficit of ionizing photons are the genuine \uchii\ regions. Another possibility is that some of these nebulae are actually optically thick which would result in the ionizing flux being underestimated and/or there could be significant absorption due to dust which would lower the overall number of ionizing photons in the \hii\ region.

\subsubsection{Clump mass-luminosity relationships}

\begin{figure}
\begin{center}

\includegraphics[width=0.49\textwidth, trim= 0 0 0 0]{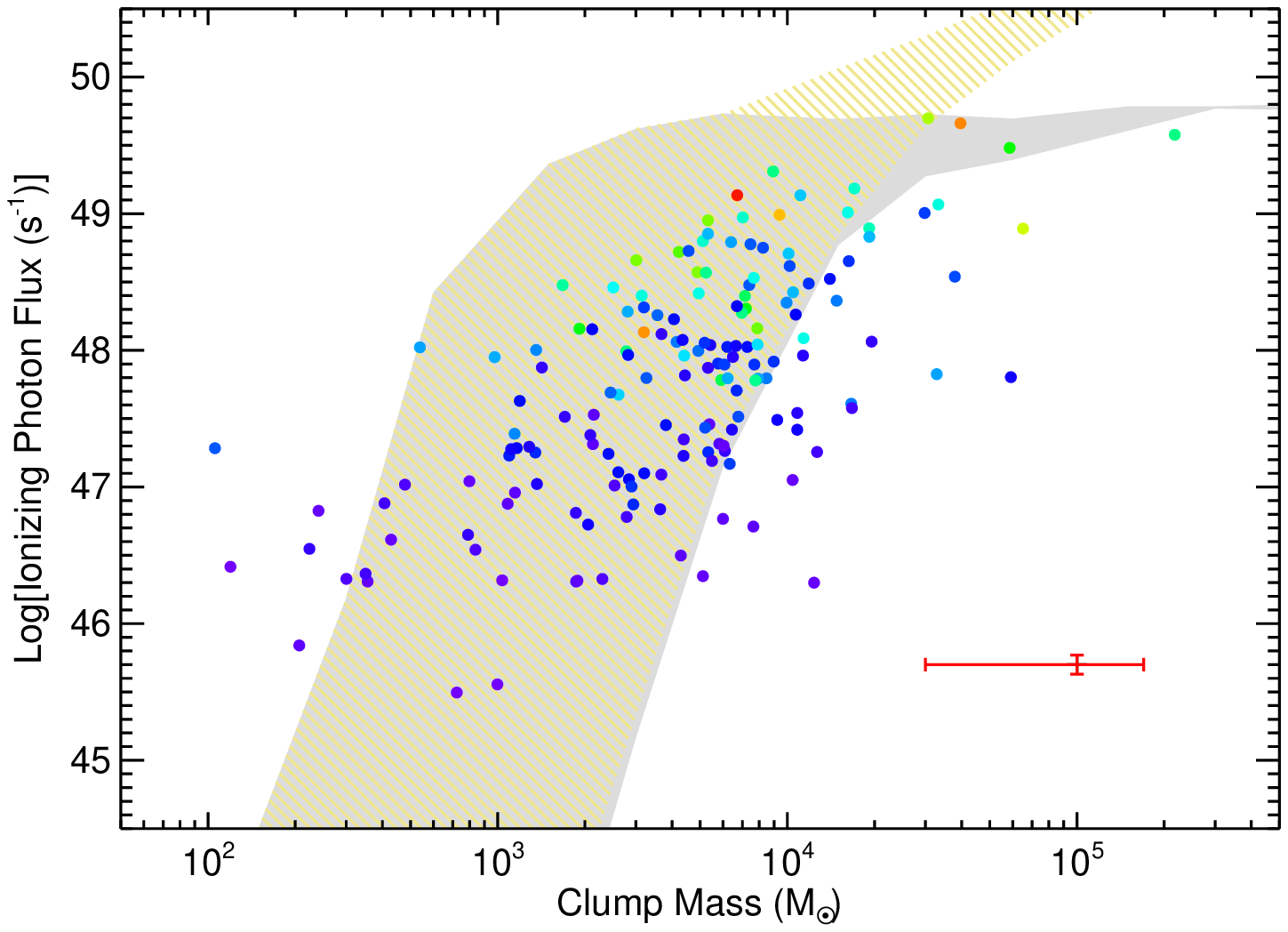}
\includegraphics[width=0.49\textwidth, trim= 0 0 0 0]{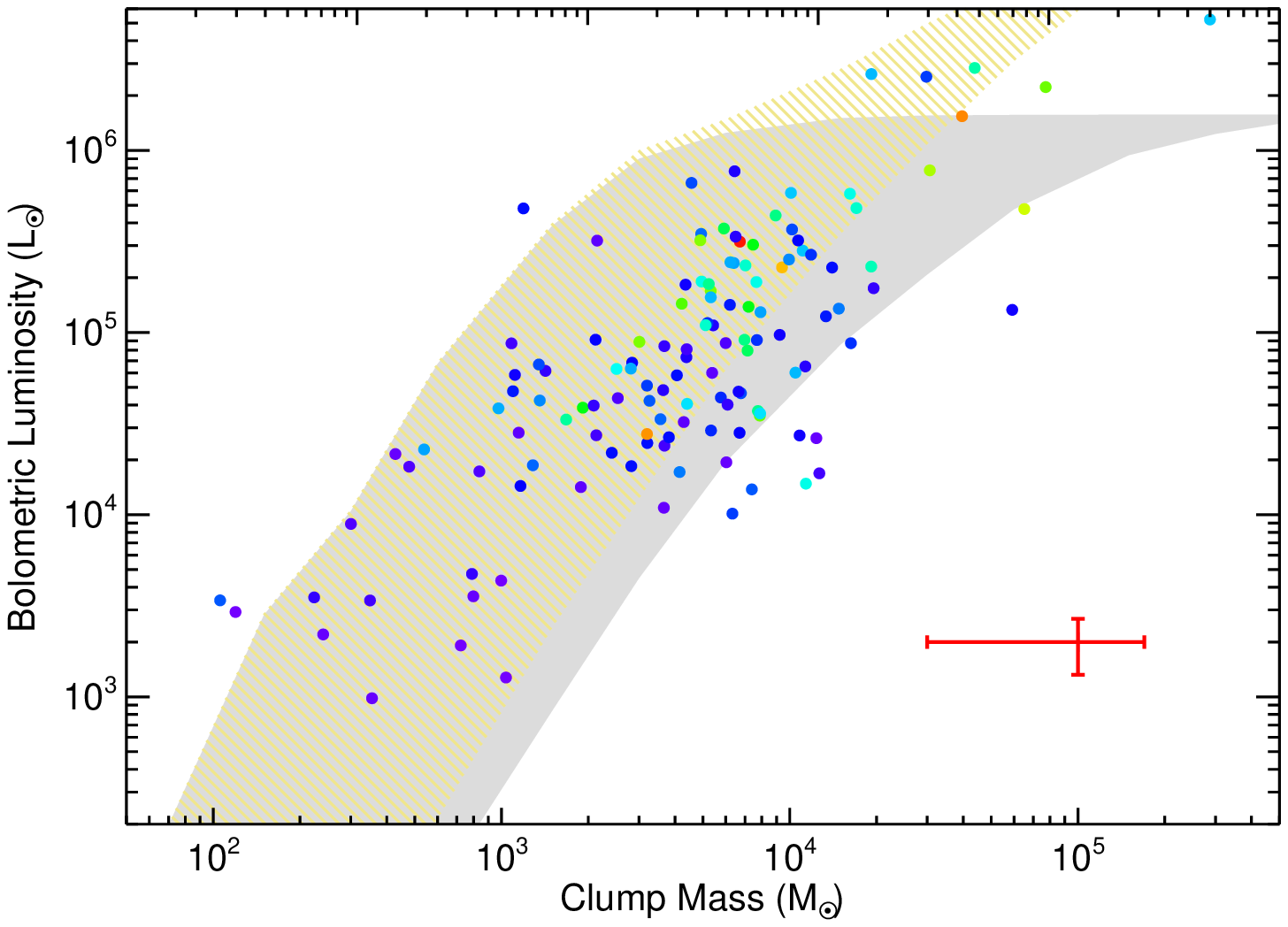}

\caption{\label{fig:mass_relations} Upper panel: Comparison of clump mass to ionising photon flux emitted from their embedded \uchii\ regions. Lower panel: Clump mass-bolometric luminosity relationship for the ATLASGAL-CORNISH sample of \hii\ regions. The colours are the same as for Fig.\,\ref{fig:lum_ni_comparison} and correspond to the physical radius of the brightest embedded \uchii\ region associated with each clump. The region of parameter space encompassed by the yellow hatching indicates the area in which 90\,per\,cent of the simulated clusters are located, while the region highlighted in grey indicates area where the most massive star of each cluster was located. Characteristic error bars for these parameters are shown in the lower right corner of each plot.} 

\end{center}
\end{figure}

In Fig.\,\ref{fig:mass_relations} we present the ionizing photon flux and bolometric luminosity as a function of clump mass. Since the main focus of this paper is the clump properties and, given that we are unable to subdivide the masses of clumps with multiple \hii\ regions, we simply integrate the radio flux for all embedded sources and their RMS counterparts to obtain the total bolometric luminosities and Lyman continuum flux for each clump. The yellow hatched region shown on these plots corresponds to the range of parameter values that includes 90\,per\,cent of the Monte-Carlo simulations, while the grey region indicates the region of parameter space covered by the most massive member of each cluster.

Inspecting these plots, it appears that the bolometric luminosity and ionizing photon flux are correlated with clump mass, even though there is a large scatter in the data. As above, we use the partial Spearman correlation test to remove any potential Malmquist bias. This gives a coefficient of 0.54 for the correlation between clump mass and Lyman continuum flux, and 0.53 for the clump mass and bolometric luminosity;\footnote{These correlation coefficients are slightly higher at 0.57 if we include only \hii\ regions with sizes less that 0.2\,pc in diameter in the mass-ionising flux relation and 0.58 for the mass-bolometric luminosity relation if we include only sources where an SED fit has been made to the data. The sample sizes drop to 97 and 108, respectively, for these correlations but the $p$-values remain unchanged.} the samples sizes are 164 and 130 and associated $p$-value for both correlations is $<1\times10^{-7}$.  The bolometric luminosity and Lyman continuum are therefore correlated with clump mass, with the most massive stars forming within the most massive clumps. Presumably the high accretion rate required to form the more massive stars can only be sustained by the most massive clumps. There is clearly an upper bound on cluster mass set by the clump mass, but the lower bound suggests that extremely low SFEs are rare (i.e., the SFE has a probability distribution that is not flat).

The dependence on clump mass of the Lyman continuum and bolometric luminosities derived from the data fit the Monte-Carlo simulation very well. If the SFE were higher than the 10\,per\,cent assumed in our simulations, the predicted Lyman and bolometric cluster luminosities (shaded regions) would be shifted to the left, worsening the match between measured and predicted values, requiring a further increase in cluster luminosity to restore the correspondence. The fact that the Lyman continuum flux and bolometric luminosities of the more evolved \hii\ regions are close to those prediction by the stellar atmosphere models suggests that their luminosities are unlikely to change significantly and that their evolution is at an end. 

Since the larger \hii\ regions effectively set the upper envelope to the Lyman continuum flux and this agrees with predictions derived using a SFE of 10\,per\,cent we would conclude that an average SFE is more likely. 

\section{Summary and conclusions}

We have combined the ATLASGAL and CORNISH surveys to identify a large sample of molecular clumps associated with compact and ultra-compact (UC) \hii\ regions. The ATLASGAL survey (\citealt{schuller2009}; $280\degr < \ell < 60\degr$ and $|b| < 1.5$\degr) traces the thermal emission from dense clumps at 870\,$\umu$m and is complete to all massive clumps above 1,000\,\msun\ to the far side of the inner Galaxy ($\sim$20\,kpc). In total the ATLASGAL survey has identified some 10,000 compact sources (\citealt{contreras2013} and Csengeri et al. in prep), many of which have the potential to form the next generation of massive stars. The CORNISH project has mapped the northern Galactic plane between $\ell=10\degr$ and 65\degr\ and $|b|<1$\degr\ for 5\,GHz radio-continuum emission (\citealt{hoare2012,purcell2013}). 

By comparing the positions of the $\sim$2,300 CORNISH radio sources with the dust emission maps and archival mid-infrared images extracted from the GLIMPSE database we have identified \uchiinum\ compact and \uchii\ regions between $\ell=10\degr$ and 60\degr\ and $|b|<1$\degr. These are associated with 170 clumps with multiple \uchii\ regions being associated with a single clump in 23 cases. Radial velocities have been assigned for every source, largely from values available in the literature, with the majority being drawn from high density tracers such as NH$_3$ (1,1) and HCO$^{+}$ (3-2) transitions. Where available, distances have been taken from maser parallax measurements, however, the majority are kinematically derived with distance ambiguity solutions provided from the literature or derived here from archival \hi\ data. 

These distances have been used to estimate clump masses and their effective radii, as well as the Lyman continuum fluxes and sizes of the embedded \hii\ regions. The clump masses range from 100 to a few 10$^5$\,\msun\ with a median value of $\sim$5,000\,\msun, with sizes from a couple of tenths of a pc to a few pc and so cover the full range of size scales (i.e., cores, clumps and clouds). The median clump radius is $\sim$1.4\,pc and so the majority would be considered clumps, however, classifications based on size seem somewhat arbitrary. We have estimated the virial masses from the effective radii and molecular-line width of a large fraction of the clumps and find, that in most clumps, these are lower than the derived clump masses. This may suggest that the majority of clumps are gravitationally unstable and therefore likely to be collapsing. This is still true if we assume an equipartition of thermal and magnetic energy within the clumps. 

We find that we are complete to optically thin \uchii\ regions driven by a zero age main sequence (ZAMS) star of spectral type B0 or earlier embedded within a 1,000\,\msun\ to the far side of the Galaxy ($\sim$20\,kpc and between $\ell=10$-60\degr). Assuming a standard initial mass function (IMF; i.e., \citealt{kroupa2001}) and a star formation efficiency of 10\,per\,cent, the minimum mass required to form at least one massive star is $\ge$1,000\,\msun. It is therefore unlikely any embedded B0 and earlier spectral type stars are associated with lower mass clumps. The \hii\ region Lyman photon fluxes range between a 10$^{45}$ to 10$^{50}$\,photon\,s$^{-1}$ and radii range from 0.01\,pc to 0.9\,pc, with median values of $\sim$10$^{48}$\,photon\,s$^{-1}$ and 0.11\,pc, respectively.  Bolometric luminosities have been obtained by matching the \hii\ regions with sources identified by the RMS survey; these have median value of $\sim$10$^5$\,\lsun, which roughly corresponds to a 30\,\msun\ star or ZAMS star of spectral type O7.

We use the derived parameters to investigate the correlation between the properties of these massive star forming clumps and the star formation taking place within them. We use these data to test several density criteria suggested in the literature for massive star formation and correlate the distribution of this sample of embedded massive stars with the various features of Galactic structure present in the First Quadrant. Our main findings are as follows:

\begin{enumerate}

\item We find that the lower envelope of the mass-radius relation for methanol and \hii\ region associated clumps is well modelled with a mean surface density of 0.05\,g\,cm$^{-2}$. As stated in Paper\,I this density threshold provides a better estimate of the lower envelope of the distribution for masses and radii greater than 500\,\msun\ and 0.5\,pc, respectively, than the \citet{kauffmann2010b} criterion. This surface density threshold is also approximately twice that determined by \citet{lada2010} and \citet{heiderman2010} for ``efficient'' star formation. So while the \citet{lada2010} and \citet{heiderman2010} criteria may be considered a minimum requirements for ``efficient'' star formation, the threshold determined for this sample of massive star forming clumps can be considered a minimum requirement for the ``efficient'' formation of \emph{massive} stars.

\item We find a clear correlation between the mass of the most massive embedded stars and the masses of their natal clumps, with the most massive stars forming in the most massive clumps. We find good agreement between the predicted Lyman and bolometric luminosities of an embedded cluster and those measured from our data assuming a standard initial mass function and a star formation efficiency of $\sim$10\,per\,cent. This is similar to the mean value of $7\pm 8$\,per\,cent reported by \citet{johnston2009} for a sample of 31 intermediate- and high-mass clumps, but significantly lower than the upper limit for cluster formation of 30\,per\,cent suggested by \citet{lada2003}.

\item Of the \uchiinum\ \hii\ regions we have identified we find only half of these satisfy the size criterion to be classified as \uchii\ regions. This proportion decreases to less than 20\,per\,cent when we consider \hii\ regions associated with O stars and to zero when considering only early O stars. %This would suggest that the \hii\ regions associated with the most massive stars evolve so rapidly that their \uchii\ region phase is effectively unobservable.

\item The Galactocentric distribution of our sample reveals strong peaks at approximately 4 and 6\,kpc, which corresponds to the expected radial positions of the Scutum-Centaurus arm, the end of the long bar, and the tangent of the Sagittarius and far side of the Perseus arms. We find another peak at 8.5\,kpc, however, this can be attributed solely to the W49A star forming complex.

\item We measure significantly different Galactic scale heights for the \hii\ regions and their host clumps (20.7$\pm$1.7\,pc and 29.1$\pm$3.0\,pc, respectively), with more extreme star forming environments (i.e., W31, W43 and W51) being located closer to the Galactic mid-plane. The star formation associated with these prominent regions is intense enough that they have a significant effect on the derived Galactic parameters. Using the surface density of \hii\ regions above our completeness limit (i.e, optically thin \hii\ regions associated with stars of spectral types of B0 or earlier) we estimate the Galactic population to be $\sim$400. However, we note that we are not complete to more extended \hii\ regions around nearby B0 stars, as these fall below the CORNISH surface brightness sensitivity.

\item We have found \hii\ regions associated with B stars have a higher Lyman continuum photon flux relative to that predicted from standard stellar atmosphere models compared to their bolometric luminosities. This is even more pronounced for later B stars. This would suggest that \emph{young} B type stars are significantly more luminous in the ultraviolet than predicted by current stellar atmosphere models. 

\item Although the region of the Galactic plane discussed in this paper includes the major star forming regions W31, W43 and W51 (the latter two of these have been labelled as mini-star bursts), it is W49A that stands out as being exceptional. It contains $\sim$7\,per\,cent of the \hii\ regions identified in this study and has an integrated luminosity of $\sim7\times10^6$\,\lsun, a mass of $\sim4\times 10^5$\,\msun\ and a very high star formation efficiency (15-32\,\lsun\,\msun$^{-1}$). This is one of the most massive and intense regions of star formation in the Galaxy.

\item The properties of the clumps associated with methanol masers and those associated with \hii\ regions are in excellent agreement despite being drawn from different quadrants. The masses, sizes and structure of the clumps are all very similar and the embedded young massive stars identified by these have a similar distribution with respect to the clump structure. We find that the massive stars traced by the methanol masers and \hii\ regions are found towards the centre of their host clumps, with their positions tightly correlated with the peaks in the dust emission. The similarities between the two samples and the fact that only 20\,per\,cent of the methanol masers are associated with the MYSO and \hii\ region stages provides strong support for the hypothesis that methanol masers are indeed tracing some of the earliest stages of massive star formation.

\item From the similarities between the \emph{envelope} structure of clumps associated with both the early and late embedded stages of massive star formation (as traced by the methanol masers and radio emission) we conclude that the structure does not change significantly during the massive star's or its associated cluster's evolution. It is likely then that the clumps are already in place before the  collapse  and the subsequent star formation begins, and that once the collapse has begun the clumps effectively decouple from any external influences.

\end{enumerate}

This is the second in a series of three papers planned to use the ATLASGAL survey to conduct a detailed and comprehensive investigation of  massive star formation environments. The main aim of these papers is to use the unbiased nature of the dust emission mapped by ATLASGAL over the inner Galactic plane to connect the results derived from samples selected using different high-mass star formation tracers. In the final paper in this series we will investigate the dust properties of a well-selected sample of massive young stellar objects identified by the RMS survey team.

\section*{Acknowledgments}

We would like to thank the anonymous referee for their many comments and suggestions that have helped to clarify and improve this work. The ATLASGAL project is a collaboration between the Max-Planck-Gesellschaft, the European Southern Observatory (ESO) and the Universidad de Chile.  The University of New South Wales Digital Filter Bank used for the observations with the Mopra Telescope was provided with support from the Australian Research Council.
This research has made use of the SIMBAD database operated at CDS, Strasbourg, France. This work was partially funded by the ERC Advanced Investigator Grant
  GLOSTAR (247078) and was partially carried out within the Collaborative
  Research Council 956, sub-project A6, funded by the Deutsche
  Forschungsgemeinschaft (DFG). This paper made use of information from the Red MSX Source survey database at www.ast.leeds.ac.uk/RMS which was constructed with support from the Science and Technology Facilities Council of the UK. L. B. acknowledges support from CONICYT project Basal PFB-06.

\bibliography{cornish}

\bibliographystyle{mn2e_new}

\end{document}